\documentclass[traditabstract]{aa}
\usepackage{txfonts}
\usepackage[english]{babel}
\usepackage{t1enc}
\usepackage{graphicx}
\usepackage{fixltx2e}
\usepackage{todonotes}
\usepackage{subfig}
\usepackage{lscape}
\usepackage{natbib}
\usepackage{footmisc}
\usepackage{url}
\usepackage{dingbat}
\usepackage{longtable}

\def \HII {\ion{H}{ii} region}
\def \HIIs {\ion{H}{ii} regions}
\def \kms {~km~s$^{-1}$}

\begin{document}

\title{The cometary \ion{H}{ii} regions of DR\,21: \\Bow shocks or champagne flows or both?}
\author{K. Immer\thanks{Member of the International Max Planck Research
School (IMPRS) for Astronomy and Astrophysics at the Universities
of Bonn and Cologne}\inst{1,2} \and C. Cyganowski\thanks{NSF Astronomy 
and Astrophysics Postdoctoral Fellow}\inst{2} \and M. J. Reid\inst{2} \and K. M. Menten\inst{1}}

\institute{
Max-Planck-Institut f\"ur Radioastronomie, Auf dem H\"ugel 69, D-53121 Bonn, Germany
\and
Harvard-Smithsonian Center for Astrophysics, 60 Garden Street, 02140, Cambridge, MA, USA}
\date{Received 19 April 2013 / Accepted 6 January 2014}

\abstract{We present deep Very Large Array H66$\alpha$ radio
recombination line (RRL) observations of the two cometary \ion{H}{ii}
regions in DR\,21.  With these sensitive data, we test the
``hybrid'' bow shock/champagne flow model previously proposed for the
DR\,21 \ion{H}{ii} regions.  The ionized gas down the tail of the
southern \HII \ is redshifted by up to $\sim$30 km s$^{-1}$ with
respect to the ambient molecular gas, as expected in the hybrid
scenario.  The RRL velocity structure, however,
reveals the presence of two velocity components in both the northern
and southern \HIIs.  This suggests that the ionized gas is flowing
along cone-like shells, swept-up by stellar winds.  The observed
velocity structure of the well-resolved southern \HII \ is most
consistent with a picture that combines a stellar wind with stellar
motion (as in bow shock models) along a density gradient (as in
champagne flow models).  The direction of the implied density gradient
is consistent with that suggested by maps of dust continuum
and molecular line emission in the DR\,21 region.}


\authorrunning{K. Immer et al.}
\titlerunning{Cometary \ion{H}{ii} region -- DR\,21}

\maketitle

\section{Introduction}
\label{intro}

\begin{figure*}[htbp]
\caption{(a): 22 GHz continuum image of the two
cometary \ion{H}{ii} regions in DR\,21. The contour levels are 0.01,
0.02, 0.05, 0.1, 0.2, 0.5, and 1 Jy beam$^{-1}$, chosen to match the
contour levels of Fig. 2 in Paper I as closely as possible. (The
lowest contour level in Fig. 2 of Paper I is 0.02 Jy beam$^{-1}$.)
(b) and (c): Same as (a). The boxes show
the areas over which the emission was summed to generate the ``long
slit'' (b) and the ``box'' (c) spectra in Figs. \ref{SpectrumEx} and
\ref{Spectra}.  The numbers in the boxes correspond to the position
numbers in Table \ref{FitResultsPaperI} and \ref{FitResults},
respectively. In each panel, the 3.4$\arcsec$ (FWHM) synthesized beam is shown in the
lower left corner and a scale bar of 0.1 pc is shown in the upper left corner.}
	\centering
	\subfloat[]{\label{DR21CONT}\includegraphics[width=9cm]{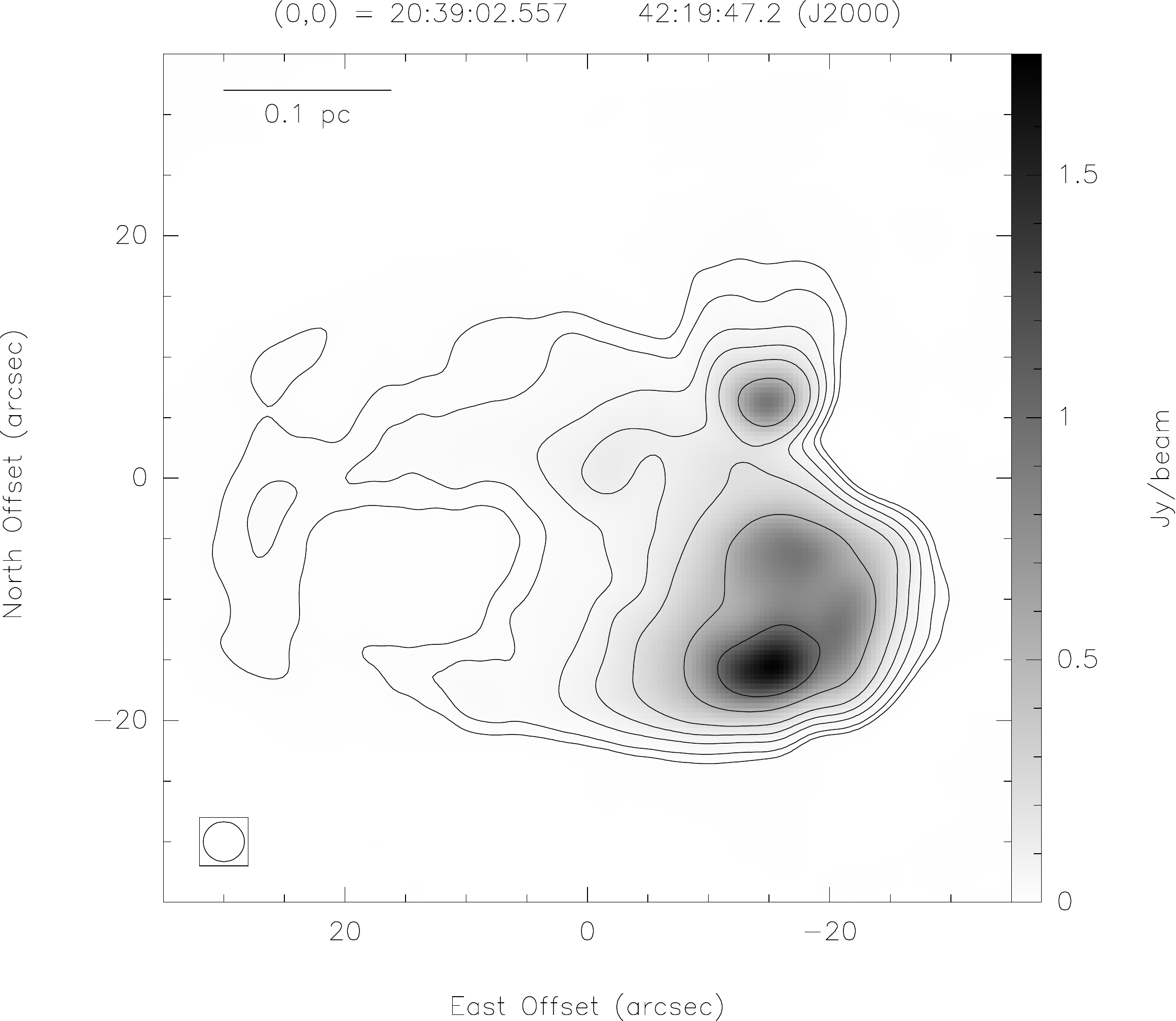}}
	\subfloat[]{\label{DR21CONTLS}\includegraphics[width=9cm]{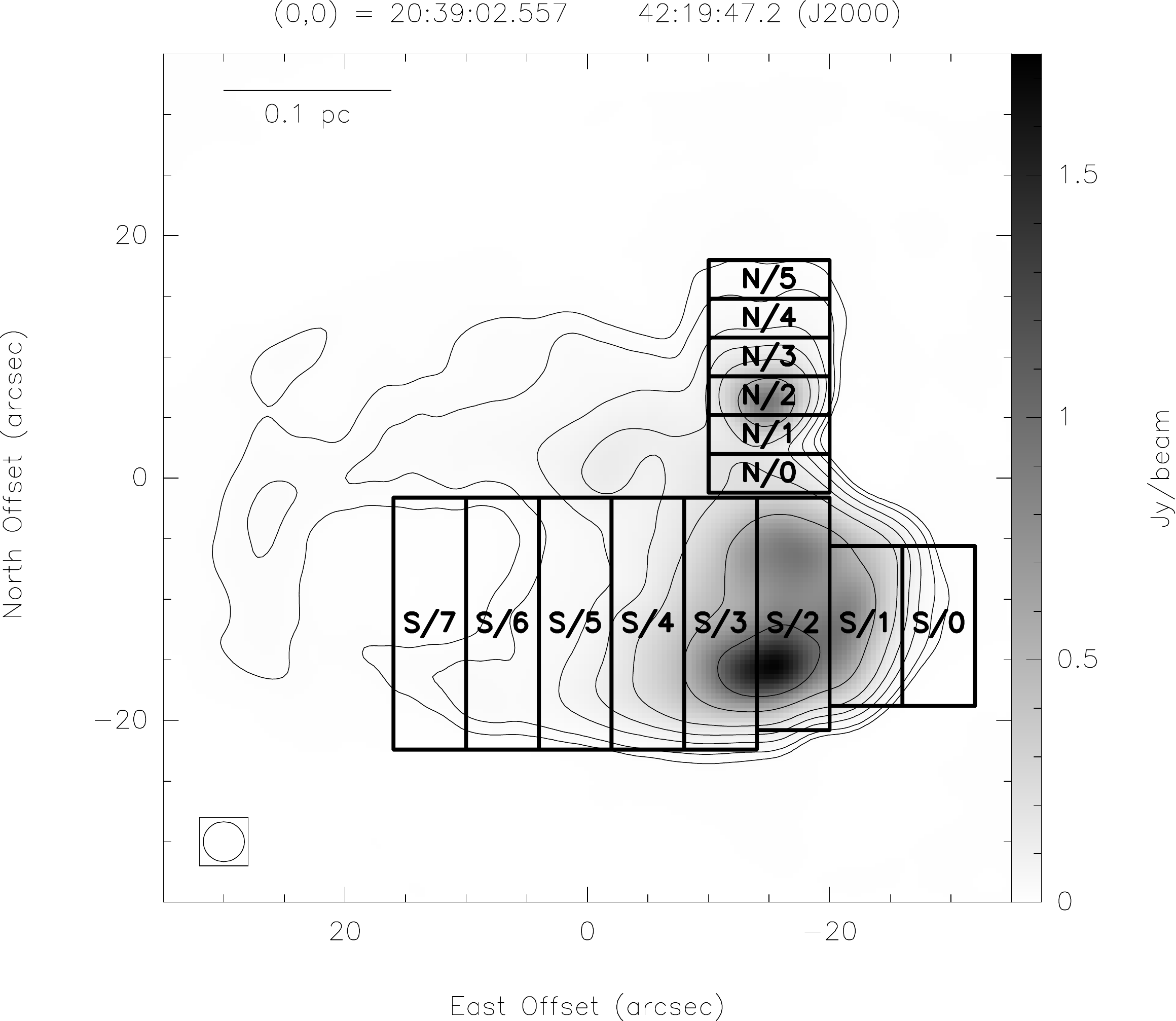}}\\
	\subfloat[]{\label{DR21CONTBOX}\includegraphics[width=9cm]{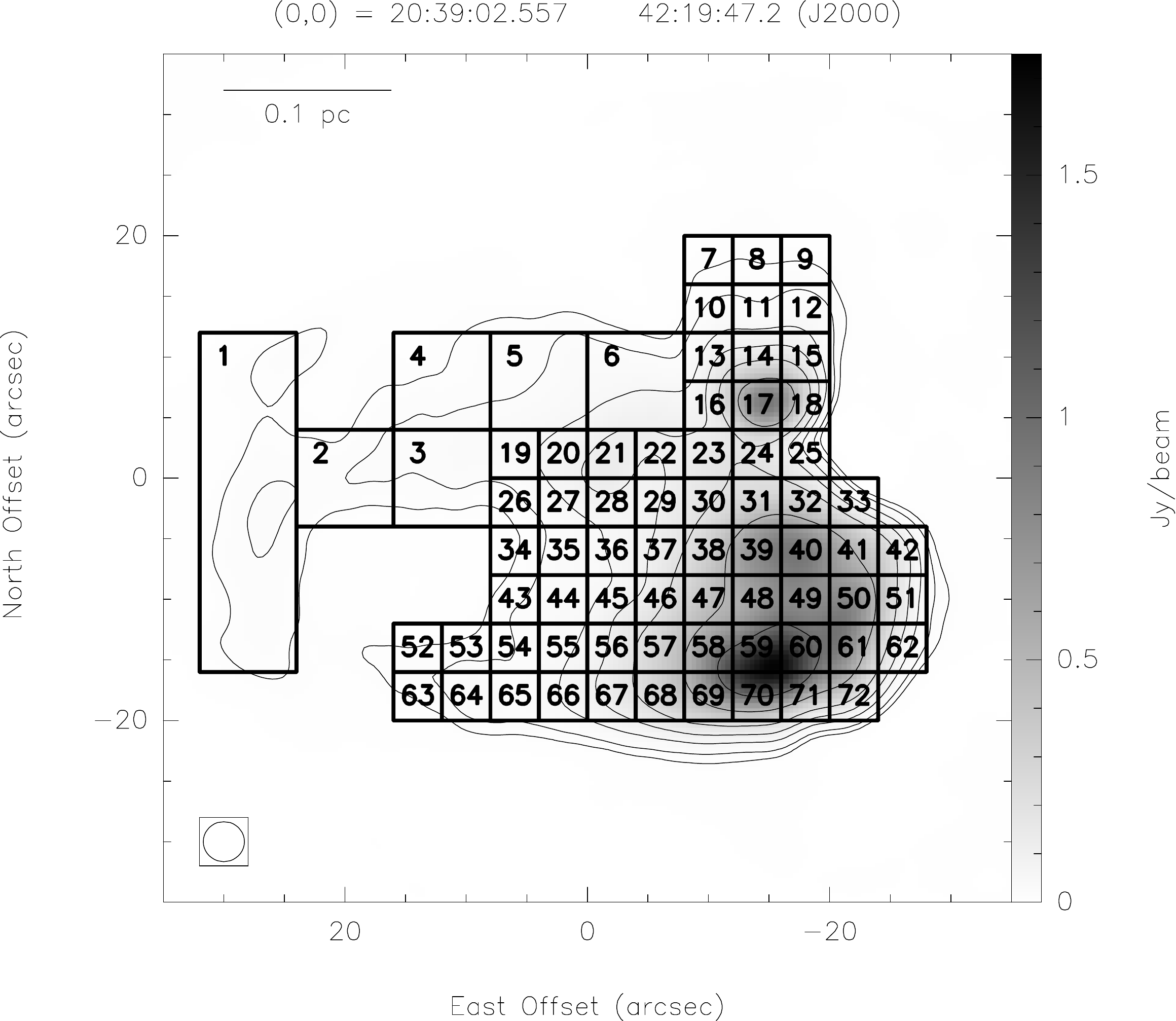}}
	\label{DR21-RadioCont}
\end{figure*}

There are several models that propose how an \ion{H}{ii} region may
develop a cometary shape.  In the ``blister'' or ``champagne flow''
models \citep[e.g.][]{Israel1978, Bodenheimer1979, Yorke1983}, the
ionized gas breaks into the diffuse intercloud medium at the edge of a
molecular cloud or flows down a density gradient in the birth cloud of
the star. In the bow shock model, the new-born star moves
supersonically through the surrounding medium \citep[e.g.][]{Reid1985,
MacLow1991, vanBuren1992}. In more recent works, the influence of a
strong stellar wind is incorporated into the models
\citep[e.g.][]{Gaume1994, Zhu2005, Arthur2006, Zhu2008}. The stellar
wind modifies the structure of the \ion{H}{ii} region by trapping it
in a swept-up shell and forcing ionized flows in that shell.
These models make different predictions for the velocity
structure of the ionized gas (see Appendix \ref{Models}), and so may
be tested with spatially resolved radio recombination line (RRL) observations of
cometary \ion{H}{ii} regions.

\citet{Cyganowski2003} (hereafter Paper I) observed the two cometary
\ion{H}{ii} regions in the well-studied DR\,21
region in NH$_{3}$ and the H53$\alpha$ and H66$\alpha$ radio
recombination lines with the Very Large Array (VLA) to investigate
whether the gas kinematics were consistent with bow shock or champagne
flow models.  They found supersonic velocity differences in the
ionized gas between the two cometary ``heads,'' and between each
``head'' and the molecular gas, consistent with simple bow shock
models.  However, indications of increasing ionized gas velocities
(relative to systemic) towards the tail of the southern DR\,21
\ion{H}{ii} region led the authors of Paper I to suggest a ``hybrid''
model, in which the ionized gas kinematics are bow-shock-like in the
cometary head, and champagne-flow-like in the cometary tail (their
Fig. 4).  More sensitive hydrogen recombination line data were,
however, required to probe the velocity field of the ionized gas
further down the cometary tail and test the hybrid model.  In this
paper, we present deep H66$\alpha$ observations of the cometary
\ion{H}{ii} regions in DR\,21, obtained with the VLA.  These new data
are $\sim$3 times more sensitive than the observations of Paper I,
allowing us to detect emission further down the tail of the southern
\HII .

The paper is structured as follows: in Section \ref{Obs}, we describe 
the observations and we present the results in Section \ref{Results}. 
In Section \ref{Discussion}, we compare the results with existing 
models for cometary \ion{H}{ii} regions. Complementary observations at 
infrared and submillimeter wavelengths are presented in Section \ref{DR21IRSubmm}. 
Section \ref{Conclusion} gives a short summary of the paper.

\section{Observations}
\label{Obs}

On July 21 2004, DR\,21 was observed in the hydrogen recombination line
H66$\alpha$ (rest frequency 22364.17 MHz) using the Very Large Array
(VLA)\footnote{The National Radio Astronomy Observatory operates the
VLA and is a facility of the National Science Foundation operated
under agreement by the Associated Universities, Inc.} in D configuration.  At this frequency,  
the VLA had a primary beam size of $\sim$2$\arcmin$. The pointing center of the observations was
$\alpha$~=~20$^{\rm h}$39$^{\rm m}$02$^{\rm s}$.0 
$\delta$~=~+42$^{\circ}$19$\arcmin$42.00$\arcsec$ (J2000). The total observing 
time was 9 hr with an on-source integration time of approximately 6.3 hr. 
The primary flux calibrator was J1331+305; J2015+371 was observed 
as the phase calibrator and used for bandpass calibration. The derived flux
density of J2015+371 was 2.92 Jy.  The bandwidth of the observations was
12.5 MHz, divided into 32 channels, resulting in a spectral resolution
of 390.625 kHz/channel (5.2~km~s$^{-1}$/channel). The band center was set 
to an LSR velocity of $-$2.0~km~s$^{-1}$. The full width at half maximum 
(FWHM) size of the synthesized beam was 3.4$\arcsec$. Since our 
observations are only sensitive on angular scales of less than 60$\arcsec$, 
smoothly distributed emission at scales larger than $\sim$0.4 pc was resolved out.

The data reduction was conducted with the NRAO Astronomical Image 
Processing System (AIPS). After calibration and flagging bad data, 
we inspected the line data set for spectral channels with no or almost no 
line emission. Since the radio recombination 
line is very broad, only a small number of line-free channels were 
available for spectral baseline determination 
on the low- (five channels) and high-frequency (two channels) 
ends of the spectrum.
The continuum in the line-free channels was fitted with a flat 
baseline which was then subtracted from the entire spectrum, 
yielding a continuum-free line database.  
Beside the continuum-free line database, a line-free continuum database 
was constructed from the baseline fit. 

The continuum data were self-calibrated, and the solutions transferred
to the line data.  A \emph{uv} taper was applied to baselines longer
than 50 k$\lambda$ to improve the signal-to-noise ratio and the data
were weighted with an AIPS ``robust'' parameter of 0.  The rms of the
resulting continuum image, away from sources, is 1~mJy~beam$^{-1}$, 
and of the line image cube 0.7~mJy~beam$^{-1}$ per channel. 
The continuum image does not reach the thermal
noise level because of dynamic range limitations
and the effects of poorly-represented large scale structure.  The
continuum peak is 1722 mJy~beam$^{-1}$, corresponding to a dynamic
range of $>$1700 (compared to a peak of 395 mJy~beam$^{-1}$ and
dynamic range of $\sim$560 for the H66$\alpha$ line image cube).

\section{Results}
\label{Results}

Our goal was to obtain more sensitive observations than those
presented in Paper I, and in particular to probe the velocity
structure of the ionized gas further down the cometary tails to
distinguish between the different classes of models for cometary
\ion{H}{ii} regions.  A comparison of our continuum image
(Fig. \ref{DR21CONT}) with that in Paper I (their Fig. 2) illustrates
the increase in sensitivity. 
We detect continuum emission $\sim$0.14 pc
further down the tail of the southern \ion{H}{ii}
region\footnote{Recent trigonometric parallax observations of 6.7 GHz
methanol masers revise the distance to DR21 to 1.5 kpc
\citep{Rygl2012}, which we adopt.  For a distance of 1.5 kpc, 0.1 pc
corresponds to $13\rlap{.}''8$.}, increasing its observed
east-west extent by $\sim$50\%. The observed head-tail (north-south)
extent of the northern \ion{H}{ii} region is also increased, by
about 0.04 pc ($\sim$33\%).  In addition to tracing ionized gas further down the cometary tails, the sensitivity of the new H66$\alpha$ data also allows the ionized gas kinematics to be studied on smaller spatial scales.

\begin{table*}
\caption{Fitting results for the ``long slit'' spectra. The values obtained in Paper 
I are shown in the last three columns for comparison.}
\begin{tabular}{|l|ccc|ccc|ccc|} \hline
              & \multicolumn{3}{|c|}{Gauss 1} & \multicolumn{3}{c|}{Gauss 2} & \multicolumn{3}{c|}{Paper I}\\ \hline
Position & Amplitude & Velocity$$ & FWHM & Amplitude & Velocity & FWHM  & Amplitude & Velocity & FWHM \\ 
             & (mJy) & (km s$^{-1}$) & (km s$^{-1}$) & (mJy) & (km s$^{-1}$) & (km s$^{-1}$) & (mJy) & (km s$^{-1}$) & (km s$^{-1}$)\\ \hline
S/0 & 23.8 & 4.8 & 19.2 & 22.3 & $-$16.5 & 32.5 & & &\\
S/1 & 612.6 & $-$3.3 & 32.3 & & & & 398 & $-$2.6 & 30.9\\
S/2 & 1611.1 & $-$2.9 & 31.6 & & & & 1132 & $-$2.5 & 30.9\\
S/3 & 1110.9 & $-$0.5 & 32.2 & & & & 770 & $-$0.8 & 31.3\\
S/4 & 282.0 & 0.4 & 38.7 & & & &168 & 0.2 & 35.4\\
S/5 & 59.3 & 21.6 & 34.2 & 37.2 & $-$10.1 & 28.2 & & &\\
S/6 & 23.7 & 27.1 & 30.4 & & & & & &\\
S/7 & 14.3 & 26.3 & 22.3 & & & & & &\\ \hline
N/0 & 106.9 & 8.9 & 27.8 & & & & & &\\
N/1 & 145.4 & 6.5 & 26.9 & & & &145 & 5.9& 23.9\\
N/2 & 280.1 & 4.9 & 27.6 & & & & 186 & 5.1 & 26.7\\
N/3 & 92.6 & 4.2 & 30.4 & & & & 38 & 4.1 &29.3\\
N/4 & 24.1 & 5.5 & 32.7 & & & &9 & $-$0.5 & 29.3\\
N/5 & 7.0 & 19.5 & 24.3 & 7.7 & 0.4 & 18.5 & & &\\
\end{tabular}
\label{FitResultsPaperI}
\end{table*}

\begin{figure*}[htbp]
\caption{H66$\alpha$ spectra of the ``long slits'' (see Fig. \ref{DR21CONTLS}). The vertical 
line marks the systemic velocity of the molecular gas in DR\,21, measured from 
NH$_{3}$ ($-$1.5 km s$^{-1}$). Positions S/0, S/5, and N/5 clearly show two 
velocity components, and are fit with two Gaussians.  ``Gaussian 1'', the component with the higher velocity, is shown in red, and ``Gaussian 2'' in blue (see Table~\ref{FitResultsPaperI}).  Note that in N/5, both Gaussian 1 and 2 are redshifted with respect to the systemic molecular gas velocity.}
	\centering
	\includegraphics[width=16cm]{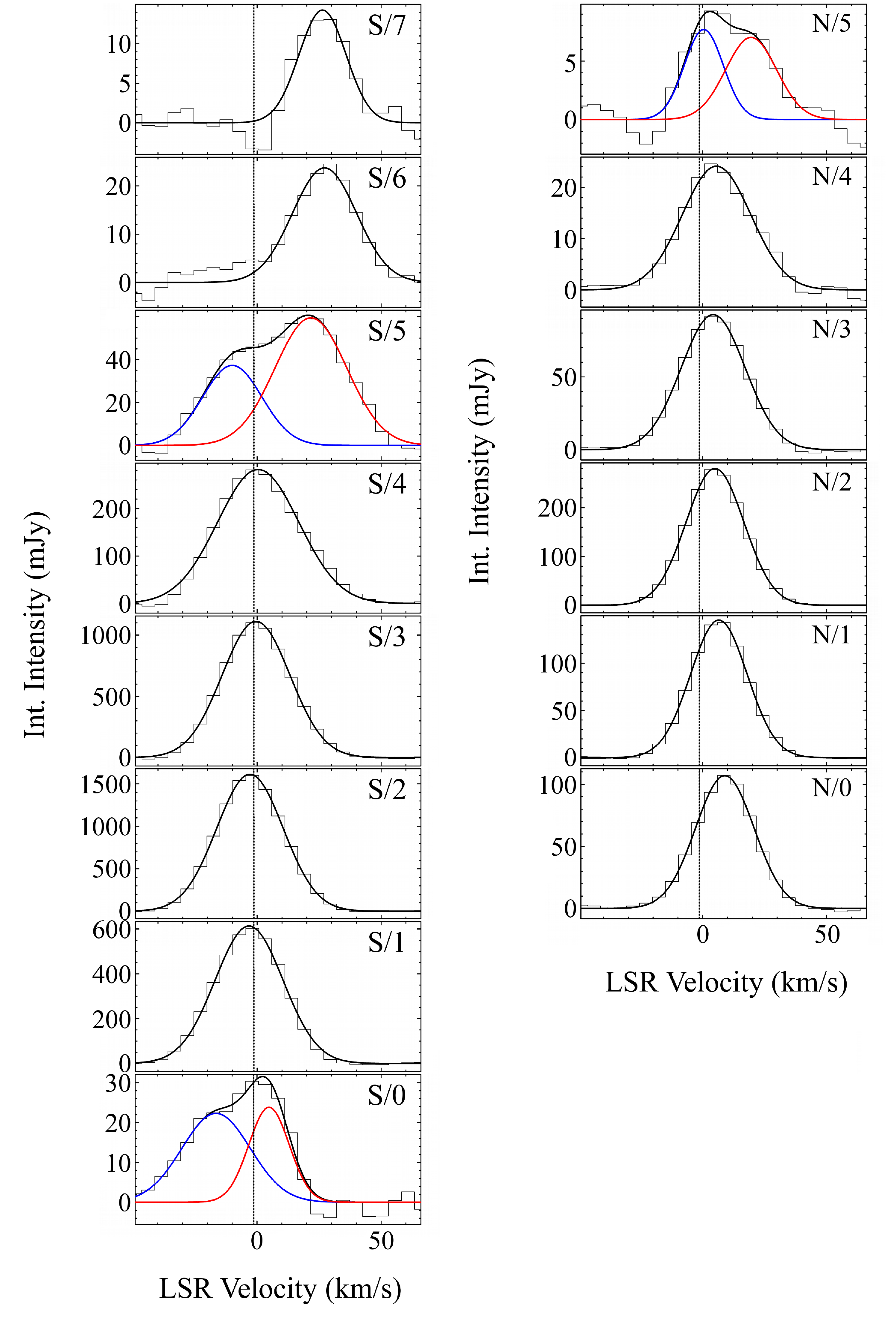}
	\label{Slits_S}
\end{figure*}

\begin{figure}[htbp]
	\caption{Example H66$\alpha$ spectrum (position 36).   ``Gaussian 1'', the component with the higher velocity, is shown in red, and ``Gaussian 2'' in blue (Table~\ref{FitResults}).  The dashed 
black line presents the sum of the two Gaussian fits. The vertical black line marks the 
systemic molecular gas velocity of $-$1.5\kms.}
	\centering
	\includegraphics[angle=270,width=8cm]{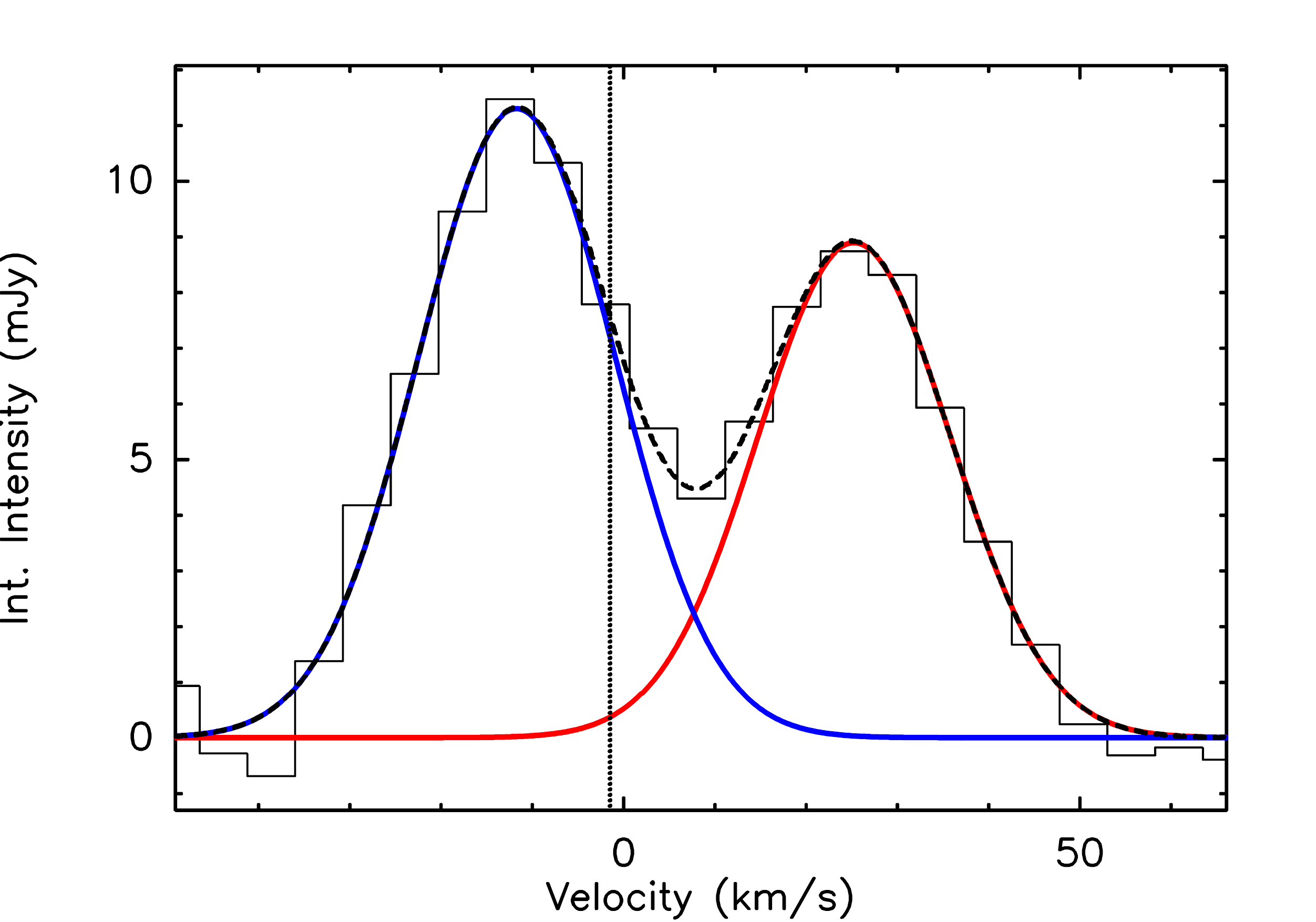}
	\label{SpectrumEx}
\end{figure}

\subsection{``Long slit'' Analysis}\label{results_longslit}

With our increased sensitivity, we first extend the ``long
slit''-style analysis of Paper I to positions further down the tails of the
two \ion{H}{ii} regions. We generated spectra by summing the emission
over eight slits in the southern \ion{H}{ii} region and six slits in
the northern \ion{H}{ii} region (for the positions of the slits, see
Fig. \ref{DR21CONTLS}).  For a better comparison with the results of
Paper I, we chose similar sizes and positions for the slits S/1 to S/4
and N/1 to N/4 as in Paper I. Then, we added the slits S/0 (with the
same size as S/1, 6$\arcsec$~x~13.2$\arcsec$) at the head of the
southern \ion{H}{ii} region and S/5--S/8 (with the same size as S/4,
6$\arcsec$~x~20.8$\arcsec$) along the tail of the southern \ion{H}{ii}
region. Similarly, we added slits N/0 (with the same size as N/1,
10$\arcsec$~x~3.2$\arcsec$) at the head of the northern \ion{H}{ii}
region, and N/5 (with the same size as N/4,
10$\arcsec$~x~3.2$\arcsec$) at the end of the tail of the northern
\ion{H}{ii} region. Each spectrum was fitted with a Gaussian line
profile, yielding the amplitude, the central velocity, and the FWHM of
the line. The spectra and the fitting results are shown in
Fig. \ref{Slits_S} and Table \ref{FitResultsPaperI}, respectively.
Our new measurements of the ionized gas velocities are
consistent with those of Paper I for seven of the eight common
``slits'' (S/1--S/4 and N/1--N/3).  For slit N/4, we find a central 
velocity of 5.5 $\pm$ 0.4 km s$^{-1}$, while Paper I 
reported $-$0.5 $\pm$ 1.6 km s$^{-1}$.
The most likely reason for this discrepancy is the improved
sensitivity of the new data.  The N/4 H66$\alpha$ detection in Paper I
was marginal, with the lowest S/N of the reported fits (Table
\ref{FitResultsPaperI} and compare their Fig. 3 to
Fig. \ref{Slits_S}).

The molecular gas in DR\,21 -- traced by NH$_{3}$ -- shows only a
small velocity dispersion, with an average systemic velocity of
$\approx$$-$1.5~km~s$^{-1}$ (Paper I).  
In the northern \ion{H}{ii}
region, the central velocity of the ionized gas at the head is $\sim$9
km s$^{-1}$ (position N/0). Due to the proximity of the southern
\ion{H}{ii} region, the spectrum at N/0 might be confused by a
contribution from the southern \ion{H}{ii} region.  Moving along the
tail, the Gaussian shifts blueward until position N/3, then shifts
redward at position N/4, where the broader wing at higher positive
velocities indicates the presence of a second velocity component. In
the spectrum at position N/5, two velocity components can clearly be
distinguished.
In the southern \ion{H}{ii} region, the spectrum at the head (S/0) shows two 
velocity components, one blueshifted at $\sim-$17 km s$^{-1}$ and 
one redshifted at $\sim$5 km s$^{-1}$. 
In spectra S/1--S/4, we detect single Gaussians whose 
central velocities increase (shift redward) along the tail. 
%
%
The spectrum at S/4 shows a broader wing at positive velocities, indicative of a second velocity component;  
this is supported by the spectrum at S/5, in which the two components can be distinguished.
In the spectra at positions S/6 and S/7 we only detect the redshifted velocity component, 
at large positive velocities ($\sim$27 km s$^{-1}$). 
%

\subsection{Small-scale Kinematics: Two velocity components}\label{results_boxes}

``Long slit'' analysis, such as that employed in Paper I and
Section \ref{results_longslit} above, assumes symmetry about the
cometary axis.  With the higher signal-to-noise ratio of our new data,
we can dispense with this assumption and investigate the velocity
structure on smaller spatial scales.  To do so, we construct a grid of
boxes parallel and perpendicular to the major axes of both cometary
\ion{H}{ii} regions.  We generated 66 spectra by summing the emission
over 4$\arcsec$~x~4$\arcsec$ boxes, chosen to approximately correspond
to the area of the synthesized beam.  Further along the tail of the
southern \ion{H}{ii} region the emission is weaker; thus, we chose
larger boxes (five boxes of 8$\arcsec$~x~8$\arcsec$ and one box of
28$\arcsec$~x~48$\arcsec$) to increase the signal-to-noise ratio. The
positions and sizes of the boxes are shown in Fig. \ref{DR21CONTBOX}.
Analyzed on these spatial scales, many of the H66$\alpha$ spectra
along the tail of the southern \ion{H}{ii} region (positions 3, 4, 5,
19, 20, 27, 28, 35, 36, 44, 45, 55, 63, and 64) clearly show two
velocity components.
An example (for position 36) is shown in Fig. \ref{SpectrumEx}; spectra for 
all positions are included as online material (Fig. \ref{Spectra}).   

At the cometary heads, the spectra appear as single Gaussians
(Fig. \ref{Spectra}); however, a single broad Gaussian can approximate
a blend of two narrower Gaussians.  The observed linewidths are
consistent with two velocity components being present throughout the
northern and southern \ion{H}{ii} regions.  Where two clearly
separated velocity components are observed, their linewidths are
generally close to the thermal linewidth (expected for 8000 K plasma, 
FWHM $\sim$ 20 km s$^{-1}$).
In contrast, where the observed RRL could be interpreted as a single
line, the linewidths are broader, consistent with the presence of two,
blended velocity components (Fig. \ref{Spectra}).  We thus fitted all
spectra with two Gaussian line profiles to study the distribution of
both velocity components over the entire \ion{H}{ii} regions.
However, in ten spectra at the end of the tail of the southern
\ion{H}{ii} region (positions 1, 2, 26, 34, 43, 52, 53, 54, 65, and
67), the second velocity component is too weak to be a 3$\sigma$
detection.  Therefore, we fitted these spectra with only one
Gaussian. The fitting results are summarized in Table
\ref{FitResults}, and the Gaussian fits are overlaid on the spectra in
Figs. \ref{SpectrumEx} and \ref{Spectra}.

\begin{figure*}[htbp]
\caption{22 GHz continuum map of DR\,21 (contours; same as
Fig. \ref{DR21-RadioCont}).  The dots represent the Gaussian fitting
results; their positions correspond to the centers of the
boxes.  The systemic velocity of the molecular material is
$\approx$$-$1.5~km~s$^{-1}$ (green in the upper panels).  (a-b): The
dot size scales logarithmically with the amplitude of the Gaussians. 
(c): The color of the dots represents the velocity difference between the two
Gaussian components. The figure shows that the velocity
difference between the two components increases from the head down the
tail in both \ion{H}{ii} regions.  (d): The color of the dots represents $|\textnormal{v}_{\textnormal{Gauss1}}-\textnormal{v}_{\textnormal{sys}}|-|\textnormal{v}_{\textnormal{Gauss2}}-\textnormal{v}_{\textnormal{sys}}|$, an indication of asymmetry in the velocity structure of the ionized gas (see also Sect. \ref{Discussion}). In each panel, the synthesized beam is shown
in the lower left corner, and a
scale of 0.1 pc is indicated in the upper left corner.}
	\centering
	\subfloat[Higher velocity component (Gaussian 1)]{\label{Gauss1}\includegraphics[width=9cm]{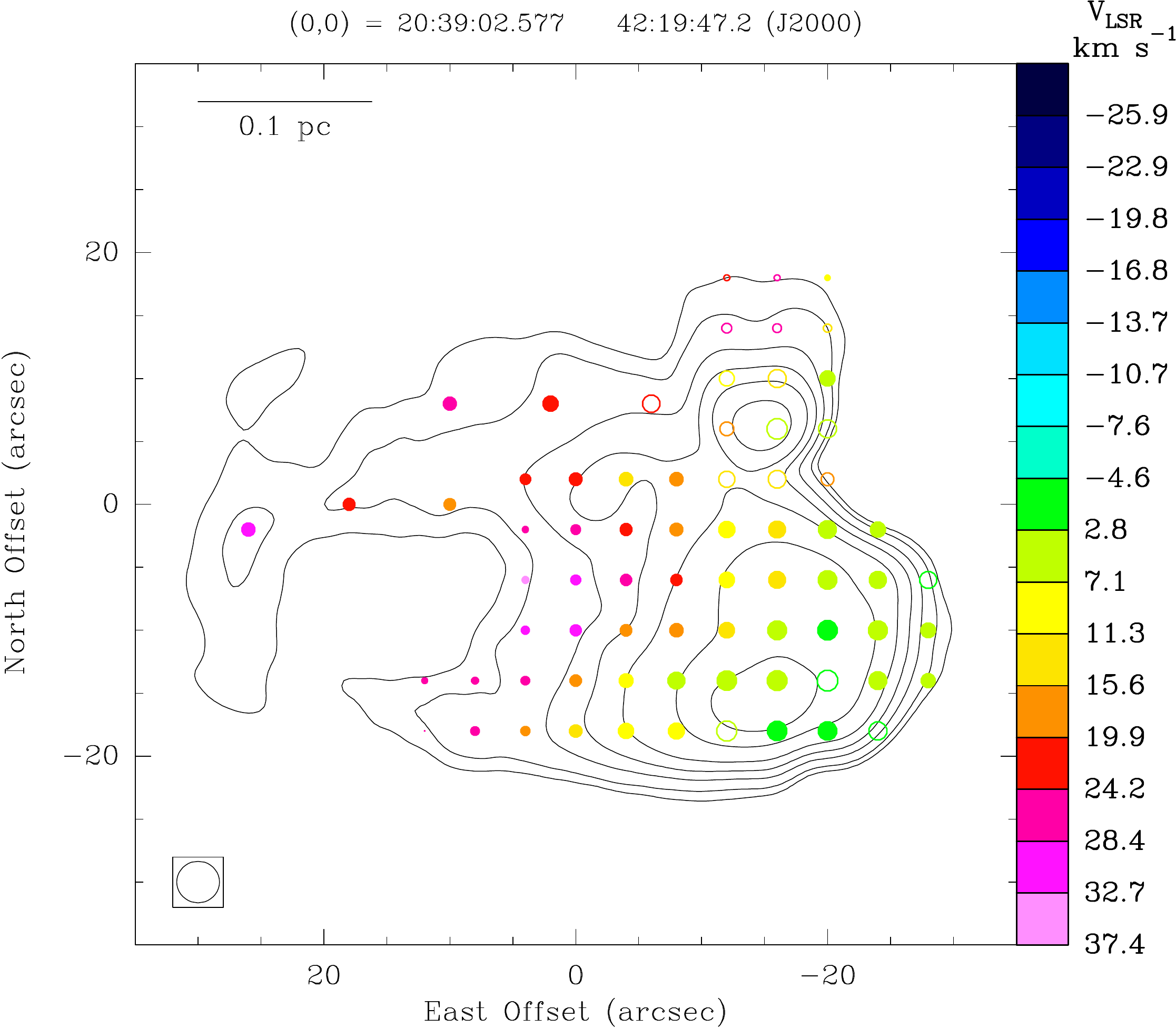}}
	\subfloat[Lower velocity component (Gaussian 2)]{\label{Gauss2}\includegraphics[width=9cm]{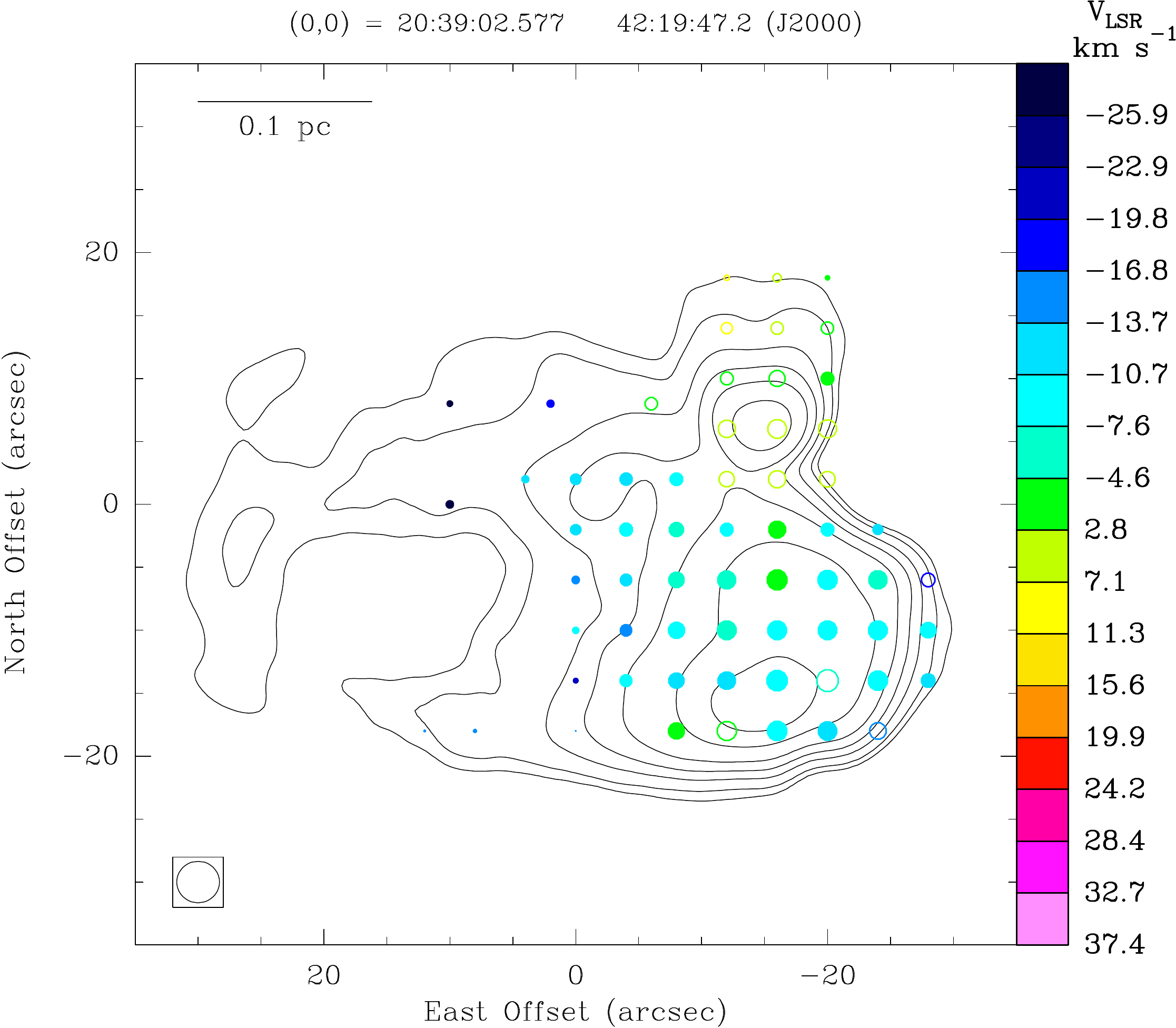}}\\
	\subfloat[Velocity difference of the two fitted Gaussians 
($\textnormal{v}_{\textnormal{Gauss1}}-\textnormal{v}_{\textnormal{Gauss2}}$)]
{\label{Gaussdiff}\includegraphics[width=9cm]{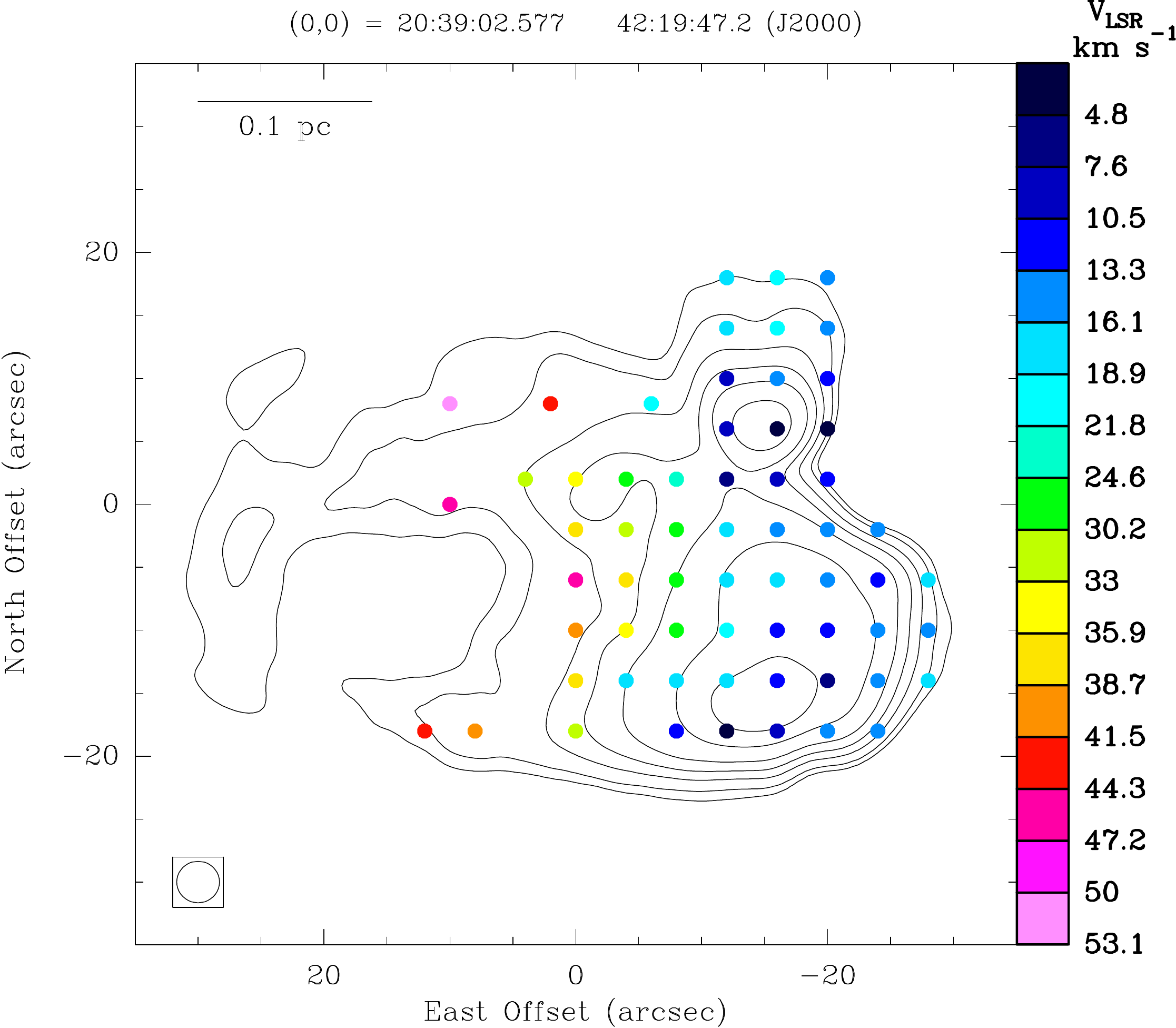}}
	\subfloat[Difference in the offset from the systemic velocity of the two fitted Gaussians 
($|\textnormal{v}_{\textnormal{Gauss1}}-\textnormal{v}_{\textnormal{sys}}|-|\textnormal{v}_{\textnormal{Gauss2}}-\textnormal{v}_{\textnormal{sys}}|$)]
{\label{Gaussdiffsym}\includegraphics[width=9cm]{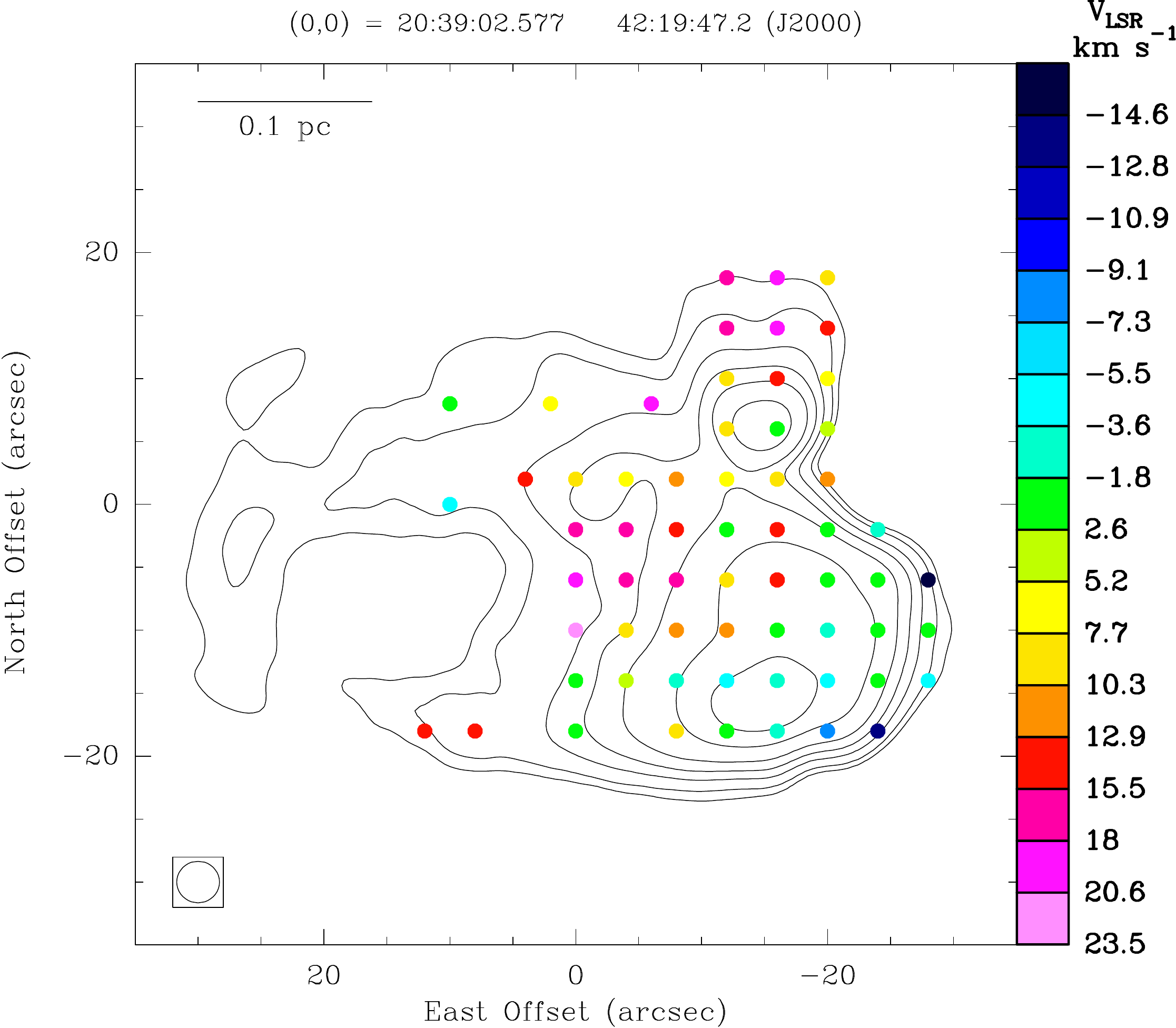}}
	\label{GaussFitting+Cont}
\end{figure*}

The central velocities of the two fitted Gaussian components
are shown in Fig. \ref{GaussFitting+Cont}.  Throughout this paper, we
refer to the higher-velocity component as ``Gaussian 1''
(Fig. \ref{GaussFitting+Cont}a) and the lower-velocity component as
``Gaussian 2'' (Fig. \ref{GaussFitting+Cont}b).\footnote{Note that while
Gaussians 1 and 2 are plotted in red and blue, respectively, in
Fig. \ref{Spectra}, both components may be red- or blue-shifted with
respect to the systemic molecular gas velocity: see
Fig. \ref{GaussFitting+Cont}a,b and Table \ref{FitResults}.} The
positions of the dots in Fig. \ref{GaussFitting+Cont} correspond to
the centers of the 72 boxes described above; the dot size scales
logarithmically with the amplitude of the Gaussian. The velocity of
the ionized material spans a range of $\sim$39~km~s$^{-1}$ ($-$2.4 to
36.6~km~s$^{-1}$) in Gaussian 1 and $\sim$36~km~s$^{-1}$
($-$28.1 to 7.5~km~s$^{-1}$) in Gaussian 2.
Fig. \ref{GaussFitting+Cont}c plots the velocity difference
between the two Gaussian components
($\textnormal{v}_{\textnormal{Gauss1}}-\textnormal{v}_{\textnormal{Gauss2}}$), 
showing that the two velocity components are more separated
further down the tails of both \ion{H}{ii} regions.
Fig. \ref{GaussFitting+Cont}d illustrates asymmetries in the
velocity structure: in this panel, the color scale represents
$|\textnormal{v}_{\textnormal{Gauss1}}-\textnormal{v}_{\textnormal{sys}}|-|\textnormal{v}_{\textnormal{Gauss2}}-\textnormal{v}_{\textnormal{sys}}|$
(see also Sect. \ref{Discussion}).

In the southern \HII, emission at velocities close to the systemic
velocity is detected in the vicinity of the continuum emission peak
for Gaussian 1 (Fig. \ref{GaussFitting+Cont}a and
Table \ref{FitResults}, e.g. positions 49, 60, 71). Single-peaked
Gaussian line profiles at the head indicate that the velocity
difference between the two components is small (see
Fig. \ref{Spectra}). Down the tail of the southern \HII, the velocity
difference increases, with the velocities of Gaussian 1 and 2 reaching $+$37\kms \ and $-$28\kms, respectively
(Fig. \ref{Gaussdiff}). The maximum velocities of the two components near the head 
are 7 (Pos. 51) and $-$18\kms \ (Pos. 42) for Gaussian 1 and 2, respectively. 

In the northern \HII, velocity components close to the systemic
velocity are also detected around the continuum emission peak, near
the head of the cometary \HII. Here again, the spectra show single
broad Gaussians, approximating a blend of two narrower Gaussians close
in velocity.  Down the tail, the velocity of Gaussian 1
increases to $\sim$25\kms \ (positions 8 and 11).
Notably, our high spatial resolution analysis reveals a
velocity gradient of $\gtrsim$8\kms \ \emph{across} the flow, seen in
both Gaussian components (e.g. positions 10--12 and 7--9).
This velocity gradient is also evident in Fig. \ref{DR21-NH},
which presents the moment 1 map of the northern cometary \HII.
Fig. \ref{DR21-NH} shows that the northeastern part of the northern
\HII \ is dominated by emission at large positive velocities, while
emission at negative velocities peaks to the northwest.


\section{Discussion}
\label{Discussion}

\subsection{Kinematics of the DR21 Cometary \HIIs}
\label{dis_models}


In our H66$\alpha$ observations, we clearly detect two separate
velocity components in the radio recombination line in the tails of
the northern and southern DR\,21 \HIIs \ (Section \ref{results_boxes}).  To our knowledge,
this is the first time two velocity components have been detected in
RRLs over large areas of cometary \HIIs . The two velocity
components suggest the presence of strong stellar winds, causing the
ionized gas to flow along cone-like shells (with the interior of the
cone being filled with low-density gas or completely devoid of
material; see also Appendix \ref{Models}). Thus, the two
velocity components of the radio recombination line are emitted by
denser gas at the front and the back of the cone.

Our H66$\alpha$ data provide a direct test of the ``hybrid''
bow shock/champagne flow model, proposed in Paper I.  Long-slit
analysis of the well-resolved southern \ion{H}{ii} region (Section
\ref{results_longslit}) shows that the velocity of the ionized gas:
(i) is blueshifted with respect to the ambient molecular material near
the cometary head (S/1), (ii) transitions to being redshifted relative
to the ambient molecular gas around positions S/3-S/4, and (iii) is
increasingly redshifted relative to the ambient molecular gas further
down the cometary tail (S/5-S/7).  Our grid-based kinematic analysis
(Section \ref{results_boxes}) further shows (iv) that the ionized gas
at the easternmost extent of the southern \ion{H}{ii} region's
cometary tail is highly redshifted, by $\sim$ 30 km s$^{-1}$ with
respect to the ambient molecular gas.  Our data thus confirm that the
ionized gas velocity ``drifts past'' the molecular gas velocity around
position S/4 (i and ii, above), as suggested in Paper I.  The
detection of increasingly redshifted gas further down the tail (iii
and iv) -- beyond the region studied in Paper I -- is strong evidence in
support of a hybrid model, which predicts champagne-flow-like
kinematics in the cometary tail.

The schematic ``hybrid'' model proposed in Paper I is
consistent with the overall velocity structure of our data, as
outlined above.  Importantly, however, the hybrid model does not explain the two
detected velocity components, since it does not include a confining mechanism 
such as provided by a stellar wind.  
In the following, we will assume that the stellar wind scenario is the best 
explanation for the detected velocity structure in DR\,21.
The widespread detection of two velocity components in RRLs is
an important observational constraint for cometary \HII \ modeling, as
it excludes models that do not incorporate a stellar wind
\citep[e.g. the ``pure'' champagne flow case considered
by][see also Appendix \ref{Models}]{Arthur2006,Zhu2008}.
In the southern DR21 \HII, from
the continuum emission peak of the ionized gas down to the end of the
tail, the absolute velocities of the ionized gas increase. A
champagne flow fits best to explain this velocity distribution,
with the flow of the ionized gas being accelerated down the density
gradient of the surrounding material. Molecular line observations of DR\,21 indicate a
density gradient in the southern \ion{H}{ii} region's surrounding environment, as discussed below (Sect. \ref{DR21IRSubmm}).  Consistent with Paper I, we find lower
velocities than the systemic velocity in the head of the \HII,
consistent with a bow shock evoked by the movement of the exciting
star through dense material.  In sum, the velocity structure of the southern \HII \ is best
explained by a champagne flow + stellar wind + bow shock
model \citep[such as models G, H, and I of][see Appendix \ref{Models} for more details]{Arthur2006}.
%


\subsubsection{Using kinematics to constrain orientation: Southern \ion{H}{ii} region}

Detailed analysis of the two velocity components also allows us to
constrain the orientation of the southern \HII. If the southern \HII
\ was seen side-on, i.e. with a cometary axis perpendicular
to the line-of-sight, we would expect the velocities of the
two Gaussian components to be symmetric with respect to the systemic
velocity of $-$1.5\kms \ \citep[Figs. 46 and 47 of][]{Zhu2008}. This
is shown schematically in the left panel of
Fig. \ref{Orientation_SHII}, which presents three lines of sight
through a cometary \ion{H}{ii} region which is shaped by a bow shock,
champagne flow, and stellar wind and is viewed side-on. For each line
of sight, the expected velocity profile is presented (comprised of a
Gaussian component from the front and from the back of the swept-up
shell). Since for this inclination the velocities of the two
components are symmetric relative to the systemic velocity (dashed
line), the difference
($|\textnormal{v}_{\textnormal{Gauss1}}-\textnormal{v}_{\textnormal{sys}}|-|\textnormal{v}_{\textnormal{Gauss2}}-\textnormal{v}_{\textnormal{sys}}|$)
would be approximately zero at each position along the \HII.  This
velocity difference for the two velocity components in the southern
\ion{H}{ii} region is 
plotted in Fig. \ref{Gaussdiffsym}. We see that the difference is not close to zero at most positions in the \ion{H}{ii} region. Close to the continuum emission peak of the ionized gas 
and in the head of the \HII, the velocity difference is slightly negative, 
i.e. the velocity difference between Gaussian 2 and the
systemic velocity is larger than the difference between 
Gaussian 1 and the systemic velocity. Down the champagne flow, the
velocity difference is positive at most positions, i.e. the velocity
offset of Gaussian 1 relative to the systemic velocity is here
larger than the velocity offset of Gaussian 2.  

Since, at the head, the blueshifted spectral line seems to be emitted
closer to the ionization front than the redshifted spectral line and,
in the tail, we observe emission further down the flow for the
redshifted Gaussian (i.e. at higher velocities) than for the
blueshifted Gaussian, we conclude that the cometary axis of 
the southern \ion{H}{ii} region has an angle of <90$\degr$ to 
the line-of-sight. This is shown schematically in the right panel of 
Fig. \ref{Orientation_SHII} for a cometary \ion{H}{ii} region whose 
cometary axis has an angle of 75$\degr$ to the line-of-sight. Again, 
the two velocity components from the two sides of the shell are shown 
for three lines of sight. From the central velocities of the two Gaussian components,
we estimate an inclination range of 50$\degr$$-$80$\degr$ of the cometary axis to the line-of-sight (see Figs. \ref{Gauss1}, \ref{Gauss2}, and \ref{Gaussdiffsym}).
%
%
These suggested inclinations for the \ion{H}{ii} region
indicate that the exciting star is moving towards us, consistent with
the blueshifted ionized gas velocities near the cometary head.


\begin{figure}[htbp]
   \caption{H66$\alpha$ moment 1 map of the northern \ion{H}{ii} region in DR\,21. A velocity 
gradient from the eastern to the western side of the cometary \ion{H}{ii} region is detected.}
	\centering
    \includegraphics[width=9cm]{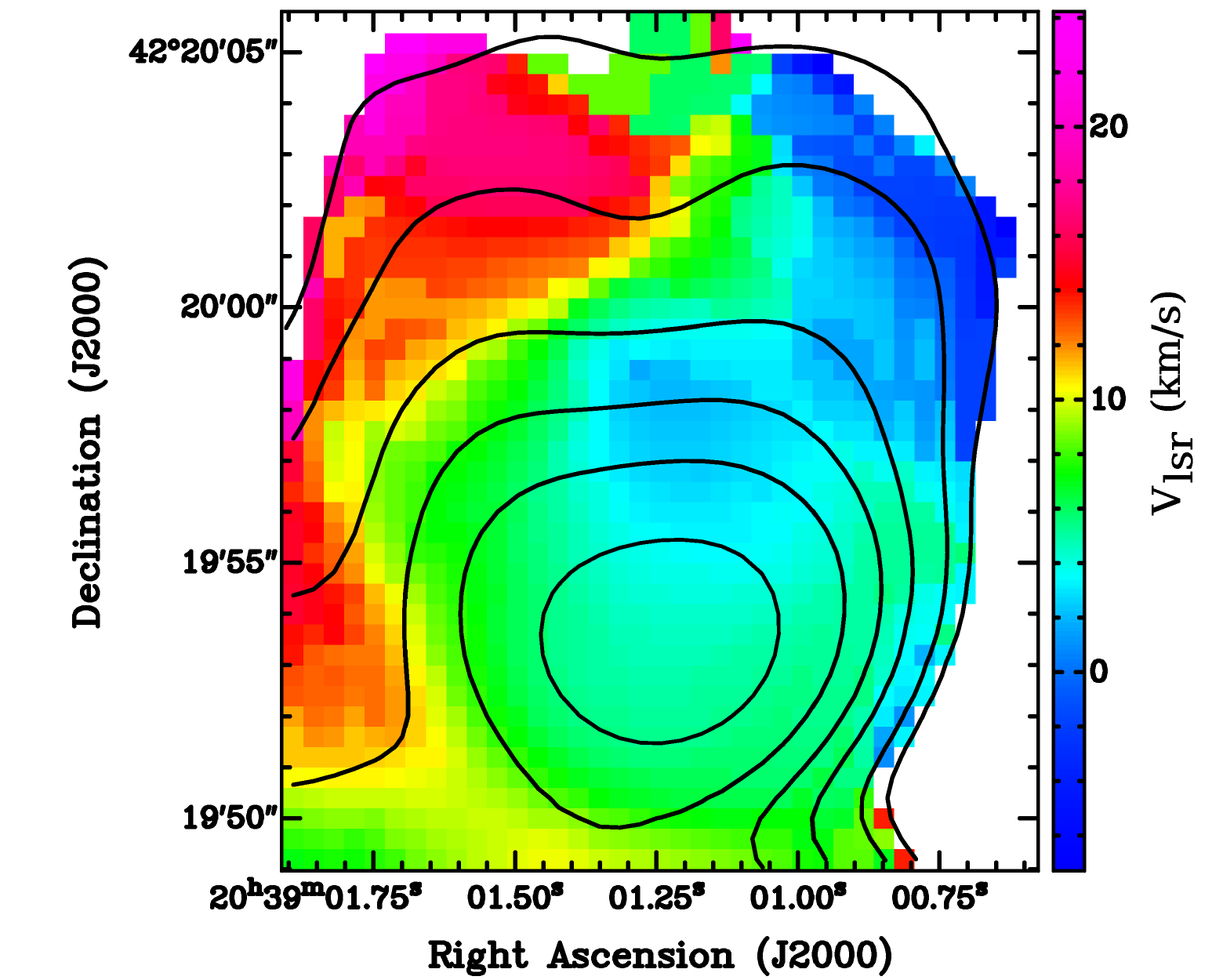}
	\label{DR21-NH}
\end{figure}

\begin{figure*}[htbp]
   \caption{Schematic diagram showing the effect of orientation on observed ionized gas velocities for a cometary \ion{H}{ii} region shaped by a bow shock, champagne flow, and strong stellar wind.  Due to the stellar wind, the ionized gas is confined to a swept-up shell, which gives rise to two velocity components. The two expected Gaussians components (from the near and far sides of the shell) are sketched for three different lines of sight. In the case of 90$\degr$ inclination (w.r.t. the line of sight), the two velocity components are emitted at corresponding positions on the sides of the shell, so the velocities are symmetric with respect to the systemic velocity (left panel). However, in the case of 75$\degr$ inclination (w.r.t the line of sight), the two components come from shifted positions along the shell and thus, the offset from the systemic velocity is not the same for the two components (right panel). The second case reflects the velocity structure of the southern \ion{H}{ii} region, suggesting that the exciting star is moving towards us.}
	\centering
    \includegraphics[width=15cm]{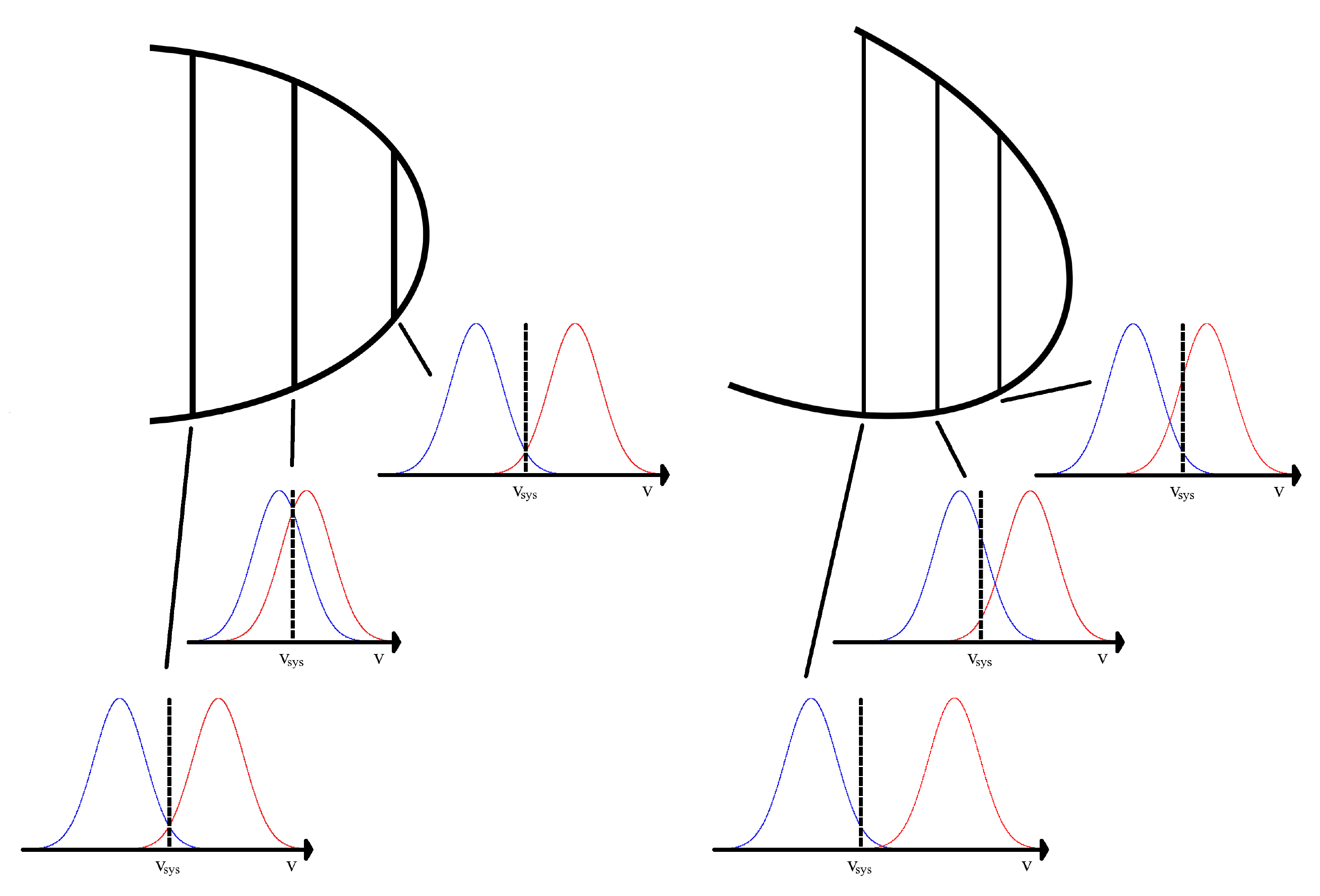}
	\label{Orientation_SHII}
\end{figure*}

\subsubsection{A rotating ionized flow?: Northern \ion{H}{ii} region kinematics}

The absolute velocities of the ionized material in the northern \HII \
appear to increase down the tail, showing evidence of a
champagne flow. The ionized gas velocity at the cometary head is
redshifted relative to the systemic velocity of the molecular gas, by
up to $\sim$10 km s$^{-1}$ (see also Section \ref{Results} and Table
\ref{FitResults}).  This is evidence for a bow-shock contribution to
the velocity structure near the cometary head (consistent with the
results of Paper I).  We note, however, that the identification of
velocity components at the head of the northern \ion{H}{ii} region is
complicated by the superposition of the velocity components from the
two \ion{H}{ii} regions. At positions 16--18 and 23--25, we
attempted to fit the spectra with four Gaussians.  The quality of the
fits was not encouraging due to the low signal-to-noise ratio; thus,
we only show the two Gaussian fits for these positions.

The detection of two velocity components again indicates the presence of a
conical ionized shell, confined by a stellar wind, as for the southern \HII. In addition, we
detected a velocity gradient \textit{across} the ionized flow, from the eastern
to the western side of the northern \HII \ (Fig. \ref{DR21-NH}). A
rotating star with a rotating stellar wind might force the ionized gas
in the wind swept-up shell to follow the rotation.  This interesting
possibility warrants further investigation; however, higher spatial 
resolution data are required to better constrain the velocity structure.

\subsection{Evidence for a density gradient in DR\,21}
\label{DR21IRSubmm}

\begin{figure}[htbp]
	\caption{SCUBA 870 $\mu$m image of DR\,21 with a HPBW of 
15$\arcsec$ \citep{Davis2007}. The contours show the 22 GHz continuum 
emission with the contour levels corresponding to the contours in 
Fig. \ref{DR21-RadioCont}. The offset between the submillimeter and radio 
continuum emission peaks is $\sim$0.04 pc.}
	\centering
	\includegraphics[width=9cm]{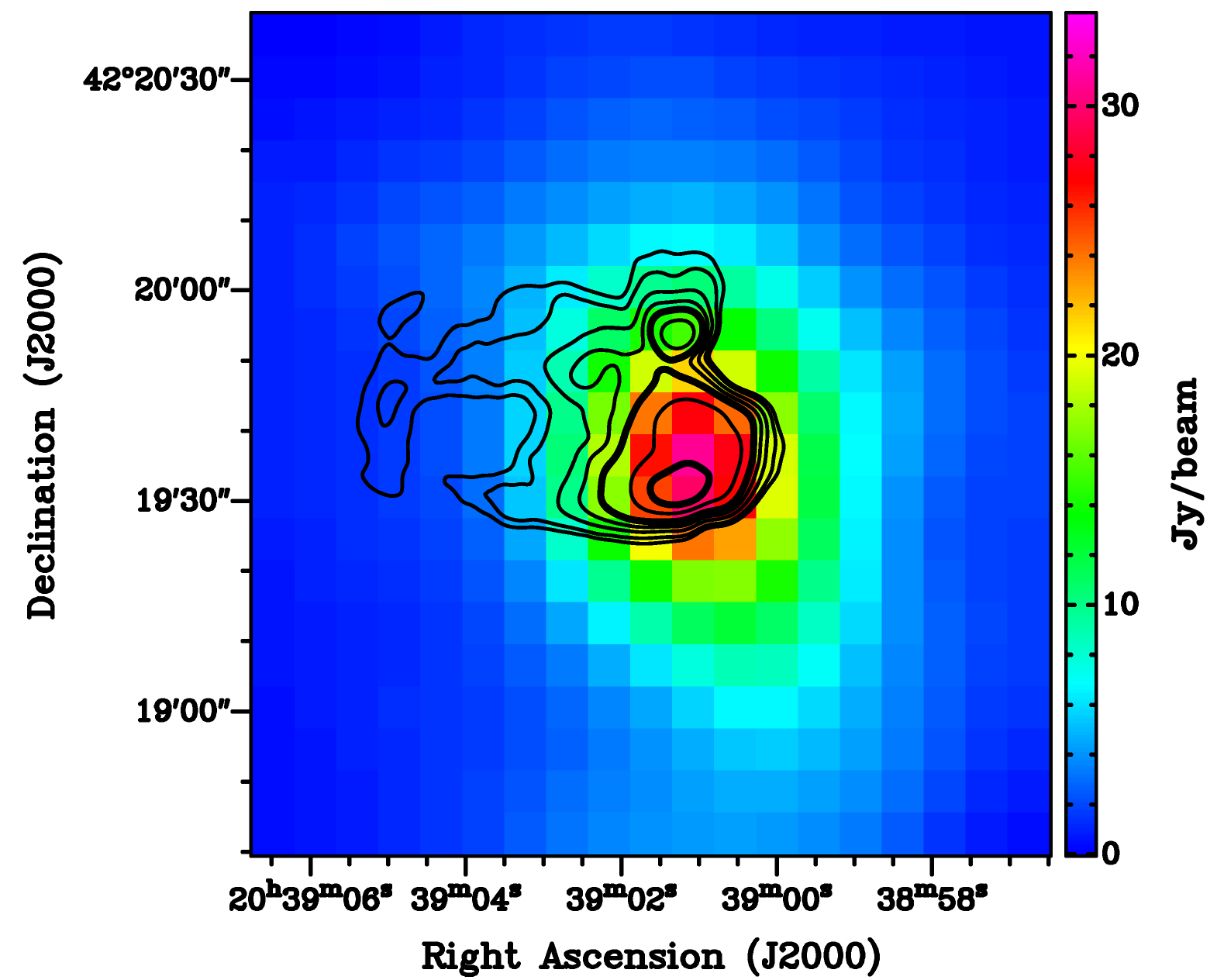}
	\label{DR21+Submm}
\end{figure}


As discussed in Section \ref{Discussion}, the
ionized gas kinematics of the DR\,21 \HIIs \ indicate a champagne flow
contribution to the cometary tails.  Champagne flows require a density
gradient in the surrounding medium. Recent large-scale studies of the DR\,21 region in (sub)millimeter continuum and molecular line emission
provide insight into the interaction of the \HIIs \ with
their environment.   
\citet{Davis2007} mapped the DR\,21 region in 870 $\mu$m dust continuum
emission with the Submillimeter Common User Bolometer Array (SCUBA) on
the James Clerk Maxwell Telescope (JCMT) (Fig. \ref{DR21+Submm}).  The
half power beam width (HPBW) of the SCUBA observations is 15$\arcsec$.
The 870 $\mu$m image shows that the head of the southern \HII \ is
embedded in a dust core with a size of $\sim$0.2 $\times$ 0.12 pc
\citep[27$\arcsec$~$\times$~16$\arcsec$,][]{Davis2007, diFrancesco2008}. The peak of the
submillimeter emission is slightly offset (by $\sim$0.04 pc in the
north-west direction) from the peak of the ionized emission.

In \citet{Schneider2010}, the emission of several molecular lines ($^{12}$CO(2--1), 
$^{13}$CO(2--1), HCO$^{+}$(1--0), H$^{13}$CO$^{+}$(1--0), 
H$_{2}$CO(3$_{1,2}$--2$_{1,1}$), C$^{34}$S (2--1), and 
N$_{2}$H$^{+}$(1--0)) was observed with the IRAM 30 m telescope towards DR\,21 
with a HPBW between 11$\arcsec$ and 29$\arcsec$. Channel maps of $^{13}$CO, 
H$^{13}$CO$^{+}$, and C$^{34}$S confirm the small velocity dispersion of 
the molecular gas in DR\,21 (their Fig. 7). Velocity-integrated line maps of HCO$^{+}$ and 
$^{13}$CO show emission of low-density gas from the whole DR\,21 region and 
the associated outflow. In contrast, C$^{34}$S and N$_{2}$H$^{+}$ emission, which trace
higher density gas, are only observed towards the central part of DR\,21, roughly 
coincident with the SCUBA core.
(There is also some weak C$^{34}$S emission extending to the west.)
Neither C$^{34}$S nor N$_{2}$H$^{+}$ is detected at the end of the
tail of the southern \HII, consistent with a density gradient in
the molecular material surrounding the southern \HII \ with density
decreasing from W (cometary head) to E (tail).

\section{Conclusion}
\label{Conclusion}

We have obtained deep H66$\alpha$ radio recombination
line observations of the two cometary \HIIs \ in DR\,21, with the aim of 
testing proposed models for cometary \HII \ kinematics, in particular 
the ``hybrid'' bow shock/champagne flow
model proposed in Paper I.  
Our analysis of the velocities of the
cometary heads is consistent with the results of Paper I: we find
offsets between the velocities of the ionized gas and the ambient
molecular material, indicative of stellar motion as in bow shock
models.
With the increased sensitivity of the new data, we study the kinematics of the ionized gas along the cometary tails, and find velocities increasing with distance from the cometary heads.
In the new data, we detect RRL emission at the extreme
eastern end of the tail of the southern \HII, at velocities
redshifted by up to $\sim$30 km s$^{-1}$ with respect to the ambient
molecular gas.  This velocity structure is consistent with a champagne
flow contribution in the tail, as in the hybrid scenario.

The sensitivity of the new data allows us to move beyond ``long slit''
analysis (which assumes symmetry about the cometary axis) and probe
the velocity structure on $\sim$4$\arcsec$ scales. A detailed analysis
of the ionized gas kinematics clearly shows that two velocity
components are widespread in both the northern and southern \HIIs.
To our knowledge, these are the first observations of cometary \HIIs \
in which two velocity components, belonging to the same region, have
been detected in RRLs.  

The two velocity components indicate that the ionized gas is likely confined
to a thin conical or paraboloidal shell, as predicted by models that
include strong stellar winds.  Our data suggest that the combined
effects of a stellar wind, stellar motion (as in bow shock models) and
an ambient density gradient (as in champagne flow models) are
necessary to explain the observed ionized gas kinematics.  Recent
observations of dense gas tracers in the DR\,21 region are consistent
with the ambient density gradient inferred from the ionized gas
kinematics: the cometary head of the southern \HII \ is embedded
in a dense core, while no emission from dense gas tracers is detected
towards the eastern end of the cometary tail.
Intriguingly, in the northern \HII \ we also
find tentative evidence for a velocity gradient \emph{across} the flow of the
ionized material, which might indicate a rotation of the flow.  Higher
spatial resolution data are required to investigate this
possibility.

\begin{acknowledgements} 
K.I. would like to thank M. Hoare for fruitful discussions.
This research has made use of NASA's Astrophysics Data System
Bibliographic Services.  C.J.C. is supported by an NSF Astronomy and
Astrophysics Postdoctoral Fellowship under award AST-1003134.
\end{acknowledgements}

\clearpage \onecolumn 
\begin{longtable}{|l|ccc|ccc|}
\caption{\label{FitResults} Gaussian fitting results}\\ \hline
              & \multicolumn{3}{|c|}{Gauss 1} & \multicolumn{3}{|c|}{Gauss 2}\\ \hline
Position & Amplitude & Velocity & FWHM & Amplitude & Velocity & FWHM \\ 
             & (mJy) & (km s$^{-1}$) & (km s$^{-1}$) & (mJy) & (km s$^{-1}$) & (km s$^{-1}$)\\ \hline
\endfirsthead
\caption{continued.}\\ \hline
              & \multicolumn{3}{|c|}{Gauss 1} & \multicolumn{3}{c}{Gauss 2}\\ \hline
Position & Amplitude & Velocity & FWHM & Amplitude & Velocity & FWHM \\ 
             & (mJy) & (km s$^{-1}$) & (km s$^{-1}$) & (mJy) & (km s$^{-1}$) & (km s$^{-1}$)\\ \hline
\endhead \hline
\endfoot \hline
POS 1   				& 16.0   & 29.0 & 37.8 &           &           & \\			
POS 2   				& 10.9   & 23.4 & 32.6 &           &           & \\			
POS 3\tablefootmark{a}   	& 10.2   & 19.3 & 43.9 & 4.6     & $-$26.5 & 9.9 \\
POS 4\tablefootmark{a}   	& 15.8   & 24.7 & 24.8 & 3.4     & $-$28.1 & 24.7 \\
POS 5\tablefootmark{a}   	& 32.1   & 22.3 & 26.5 & 4.3     & $-$19.7 & 22.6 \\
POS 6   				& 41.3   & 21.6 & 28.2 & 9.9     & 1.8     & 34.6 \\
POS 7\tablefootmark{a}  	& 2.4     & 24.1 & 13.1 & 2.8     & 7.1     & 22.4\\
POS 8\tablefootmark{a}   	& 2.4     & 25.2 & 14.0 & 4.3     & 5.1     & 19.9\\
POS 9\tablefootmark{b}   	& 2.7     & 10.1 & 24.2 & 2.9     & $-$4.5   & 12.5\\
POS 10 				& 4.8     & 24.3 & 17.0 & 7.7     & 7.4     & 24.4\\
POS 11\tablefootmark{b} 	& 3.8     & 25.4 & 13.7 & 9.5     & 5.0     & 27.5\\
POS 12 				& 3.4     & 13.9 & 32.1 & 9.7     & $-$1.0   & 24.0\\
POS 13 				& 18.7   & 11.2 & 28.3 & 10.9   & 2.6     & 25.9\\
POS 14 				& 53.9   & 13.6 & 23.6 & 27.1   & $-$0.5   & 26.5\\
POS 15 				& 32.5   & 7.0   & 26.7 & 16.0   & $-$4.4   & 25.1\\
POS 16 				& 13.3   & 16.3 & 28.2 & 40.9   & 6.1     & 26.1\\
POS 17 				& 151.1 & 5.7   & 28.5 & 69.7   & 3.4     & 24.4\\
POS 18 				& 48.2   & 6.6   & 29.0 & 50.2   & 2.9     & 24.3\\
POS 19\tablefootmark{a} 	& 7.4     & 21.1 & 21.4 & 4.3     & $-$11.2 & 40.3\\
POS 20\tablefootmark{a} 	& 12.8   & 20.1 & 28.7 & 8.6     & $-$13.0 & 33.1\\
POS 21\tablefootmark{b} 	& 17.4   & 14.1 & 39.1 & 13.5   & $-$11.1 & 25.5\\
POS 22\tablefootmark{b} 	& 17.2   & 16.4 & 34.7 & 15.2   & $-$7.8   & 26.3\\
POS 23 				& 26.2   & 12.1 & 27.2 & 21.7   & 5.2     & 34.6\\
POS 24 				& 44.2   & 13.0 & 24.0 & 41.0   & 3.4     & 23.3\\
POS 25 				& 9.7     & 16.2 & 20.4 & 23.6   & 4.1     & 24.8\\
POS 26 				& 3.0     & 27.2 & 23.2 &           &           &\\			
POS 27\tablefootmark{a} 	& 6.1     & 25.5 & 28.4 & 8.1     & $-$12.2 & 26.9\\
POS 28\tablefootmark{a} 	& 11.1   & 21.6 & 29.3 & 16.5   & $-$8.7   & 26.1\\
POS 29\tablefootmark{b} 	& 14.0   & 18.3 & 27.0 & 23.9   & $-$7.1   & 23.9\\
POS 30 				& 49.0   & 8.0   & 33.0 & 16.2   & $-$9.5   & 22.8\\
POS 31 				& 57.7   & 12.4 & 25.7 & 62.5   & $-$1.6   & 29.0\\
POS 32 				& 90.1   & 5.7   & 27.2 & 16.1   & $-$7.9   & 33.8\\
POS 33 				& 31.0   & 4.5   & 24.3 & 7.7     & $-$10.8 & 32.2\\
POS 34 				& 3.4     & 36.4 & 20.0 &           &           &\\			
POS 35\tablefootmark{a} 	& 6.3     & 31.0 & 20.9 & 4.6     & $-$14.1 & 20.7\\
POS 36\tablefootmark{a} 	& 8.9     & 25.2 & 24.9 & 11.3   & $-$11.7 & 25.4\\
POS 37 				& 9.0     & 19.9 & 25.0 & 37.2   & $-$7.2   & 26.7\\
POS 38 				& 33.0   & 11.2 & 27.2 & 90.6   & $-$5.2   & 26.3\\
POS 39 				& 58.4   & 14.2 & 23.5 & 203.0 & $-$3.5   & 27.6\\
POS 40 				& 134.6 & 5.4   & 27.4 & 153.5 & $-$8.4   & 27.2\\
POS 41 				& 66.3   & 5.5   & 22.9 & 104.4 & $-$7.2   & 29.2\\
POS 42 				& 37.9   & $-$2.0 & 26.1 & 13.6   & $-$18.4 & 28.1\\
POS 43 				& 4.4     & 29.6 & 22.0 &           &           & \\			
POS 44\tablefootmark{a} 	& 8.6     & 29.6 & 26.0 & 3.8     & $-$9.4   & 29.3\\
POS 45\tablefootmark{a} 	& 9.8     & 19.6 & 37.5 & 11.1   & $-$14.3 & 28.4\\
POS 46 				& 17.4   & 18.0 & 31.6 & 46.7   & $-$9.0   & 28.7\\
POS 47 				& 34.8   & 14.7 & 28.6 & 122.3 & $-$6.0   & 26.8\\
POS 48 				& 137.4 & 4.4   & 28.2 & 124.3 & $-$8.2   & 25.4\\
POS 49 				& 190.9 & 2.5   & 26.3 & 125.1 & $-$8.9   & 26.0\\
POS 50 				& 144.7 & 5.0   & 24.0 & 129.0 & $-$9.7   & 26.6\\
POS 51 				& 27.4   & 7.0   & 22.7 & 37.5   & $-$8.5   & 30.6\\
POS 52 				& 2.9     & 27.1 & 22.3 &           &           & \\			
POS 53 				& 3.4     & 25.4 & 21.0 &           &           & \\			
POS 54 				& 4.8     & 24.4 & 30.4 &           &           & \\			
POS 55\tablefootmark{a} 	& 9.6     & 17.0 & 41.7 & 3.0     & $-$20.8 & 20.2\\
POS 56 				& 19.6   & 9.3   & 40.5 & 12.7   & $-$9.5   & 32.3\\
POS 57 				& 59.6   & 5.8   & 36.1 & 33.7   & $-$11.3 & 27.6\\
POS 58 				& 181.4 & 5.4   & 29.9 & 75.5   & $-$12.8 & 23.0\\
POS 59 				& 218.5 & 2.8   & 28.0 & 258.5 & $-$9.4   & 26.4\\
POS 60 				& 160.8 & $-$2.0 & 30.5 & 222.4 & $-$7.0   & 29.9\\
POS 61 				& 90.9   & 6.1   & 23.0 & 158.9 & $-$9.0   & 28.0\\
POS 62 				& 22.0   & 5.5   & 22.3 & 21.6   & $-$12.3 & 27.7\\
POS 63\tablefootmark{a} 	& 1.7     & 27.6 & 26.4 & 2.3     & $-$15.8 & 15.6\\
POS 64\tablefootmark{a} 	& 5.0     & 24.7 & 44.3 & 2.6     & $-$14.5 & 11.5\\
POS 65 				& 5.4     & 18.6 & 43.6 &           &           & \\			
POS 66 				& 12.7   & 14.8 & 36.2 & 2.2     & $-$16.0 & 17.8\\
POS 67 				& 31.7   & 7.5   & 36.2 &           &           &\\			
POS 68 				& 43.3   & 8.1   & 30.3 & 46.3   & $-$2.7   & 27.3\\
POS 69 				& 121.1 & 3.0   & 27.5 & 80.1   & 1.0     & 31.8\\
POS 70 				& 176.8 & 2.1   & 28.1 & 161.2 & $-$7.7   & 28.2\\
POS 71 				& 115.3 & 1.5   & 29.6 & 109.3 & $-$12.6 & 25.2\\
POS 72 				& 57.6   & $-$2.2 & 26.7 & 33.1   & $-$16.3 & 22.2\\
\end{longtable}
\tablefoot{
\tablefoottext{a}{Two velocity components visible.}
\tablefoottext{b}{Asymmetric line profile indicates two velocity components.}
}

\clearpage \twocolumn

\Online

\begin{appendix} 



\section{Cometary \ion{H}{ii} region models}
\label{Models}
Since the publication of Paper I, the state of the art in compact \HII \ modeling has advanced considerably. 
Of particular relevance, \citet{Arthur2006} present four groups of 
radiation-hydrodynamic simulations of cometary \ion{H}{ii} regions: 
\begin{itemize}
\item[i)]   Pure bow shock: stellar motion + stellar wind, no density gradient
\item[ii)]  Pure champagne flow: density gradient, no stellar wind, no stellar motion
\item[iii)] Champagne flow + stellar wind: density gradient, stellar wind, no stellar motion
\item[iv)] Champagne flow + stellar wind + stellar motion: density gradient, stellar wind, stellar motion
\end{itemize}
We note that the schematic ``hybrid'' model proposed in Paper I (champagne flow and bow shock) 
is similar to (iv), but did not include a stellar wind.

For each of their models, \citet{Arthur2006} determine line intensity, mean velocity, and velocity 
dispersion at different positions in the \ion{H}{ii} region.
For the bow shock model (model i), the velocities of the ionized gas at the head of the 
\ion{H}{ii} region are similar to the stellar velocity. Since the influence of the 
stellar motion on the gas is small in the tail, the velocities quickly converge to 
the systemic velocity of the surrounding material in this region. 
In the pure champagne flow model (model ii), the velocity of the ionized material 
at the position of the star is similar to the velocity of the 
ambient molecular material. Down the tail, the velocity of the ionized gas 
increases due to the drop in pressure. Including a stellar wind in the champagne flow 
model (model iii) causes the ionized gas to flow around the stellar wind bubble in a thin shell. 
The density gradient in the ambient material then leads to large accelerations at the 
sides of the flow. If additionally the star is moving up a density gradient (model iv), the 
direction of the flow between the ionization front and the stellar position changes 
(compared to the pure champagne flow model), being now in the 
direction of stellar motion \citep[compare Figs. 3, 4, 5, and 12 of][]{Arthur2006}. 

In all cases, the observed morphology and velocity structure will be
affected by viewing angle.  \citet{Zhu2008} consider viewing angle
effects in some detail, in the context of [Ne II] 12.8 $\mu$m
observations of a sample of compact and UC \HIIs.  
%
While \citet{Zhu2008} emphasize that their models
are qualitative, their schematic position-velocity diagrams (their Figs. 46--48)
are illustrative.
%
%
In particular, for a cometary \HII \ viewed side-on (e.g., cometary axis
in the plane of the sky), two velocity components are expected from a
wind-confined ionized flow--one from the near and one from the far
side of the paraboloidal shell \citep[Fig. 46--47 of][]{Zhu2008}.  
%
%
At all positions, the two velocity components
are symmetric about the velocity of the ambient gas.  Notably, for
this viewing angle it is not possible to distinguish between the
bow-shock and pressure-driven stellar wind cases based on a
position-velocity diagram. For intermediate viewing angles, two
velocity components are expected near the head, while down the tail
one velocity component (side of the shell) becomes dominant
\citep[Figs. 46-47 of][]{Zhu2008}.  The line-of-sight ionized gas velocities depend both
on the precise viewing angle and on the model type.  In contrast to
the stellar-wind cases, in the wind-free champagne flow the ionized
gas fills the ``blister'' cavity, limited by the ionization front.  In
this case, a single broad line
is expected at all positions \citep[Fig. 48 of][]{Zhu2008}.

\section{General kinematics}
\begin{figure*}
\centering
\includegraphics[width=18cm]{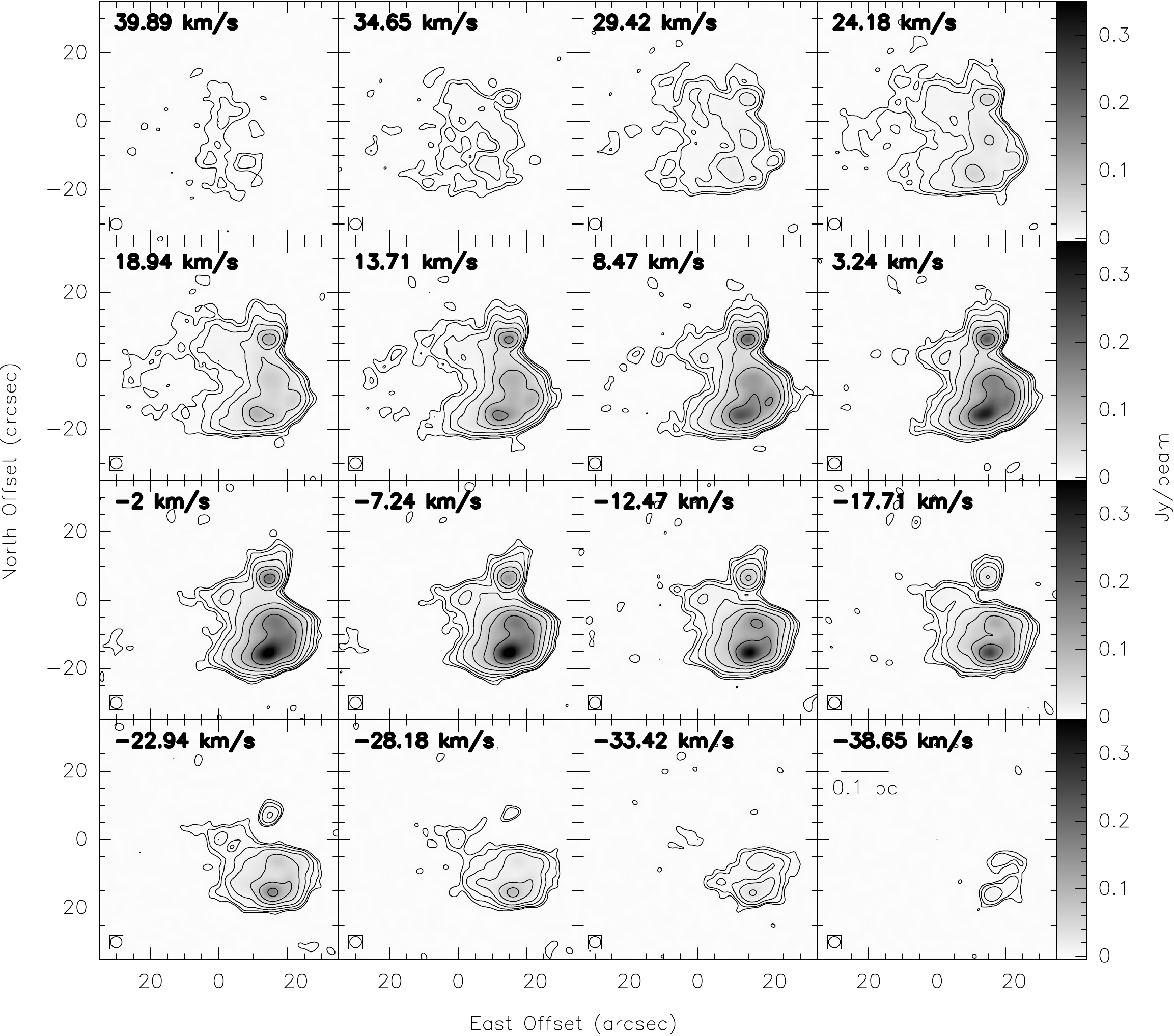}
\caption{Individual channel maps of the H66$\alpha$ line 
emission, showing every channel between 40 and $-$39~km~s$^{-1}$ . 
The velocity of each channel is given in the upper 
left corner of each plot. The contour levels are at 3$\sigma$, 5$\sigma$, 10$\sigma$, 
20$\sigma$, 50$\sigma$, 100$\sigma$, 200$\sigma$, and 500$\sigma$ 
(1$\sigma$~=~0.7~mJy~beam$^{-1}$). 
The image shows the inner (1$\arcmin$)$^2$ of the primary beam.
The 3.4$\arcsec$ synthesized beam is shown in
the lower left corner of each channel map. A scale of 0.1 pc is shown in the lower right 
channel map. The (0,0) position is at R.A.(J2000)~=~20h~39m~02.557s and Dec(J2000)~=~42d~19$\arcmin$~47.2$\arcsec$.}
\label{DR21-AllVel}
\end{figure*}

To study the general kinematics of the DR\,21 \HIIs , we made
individual channel maps of the continuum-free emission of the
H66$\alpha$ line (Fig. \ref{DR21-AllVel}), showing emission
between $\pm$~40~km~s$^{-1}$. The RRL emission from the northern
\ion{H}{ii} region generally covers a smaller velocity range than that
from the southern \ion{H}{ii} region.  Emission from the northern
\ion{H}{ii} region peaks at $\sim$3 km s$^{-1}$, while the emission of
the southern \ion{H}{ii} region peaks at $\sim$$-$7 km s$^{-1}$. At
high positive velocities the north-east part of the southern
\ion{H}{ii} region is more prominent whereas its south-west part is
more prominent at high negative velocities.

\begin{figure*}
	\caption{H66$\alpha$ radio recombination line spectra.  ``Gaussian 1'', the component with the higher velocity, is shown in red, and ``Gaussian 2'' in blue (Table~\ref{FitResults}).  The dashed black line presents the sum 
of the two Gaussian fits. The vertical black line marks the systemic molecular gas 
velocity of $-$1.5\kms.}
	\centering
	\subfloat[POS01]{\includegraphics[angle=270,width=7.5cm]{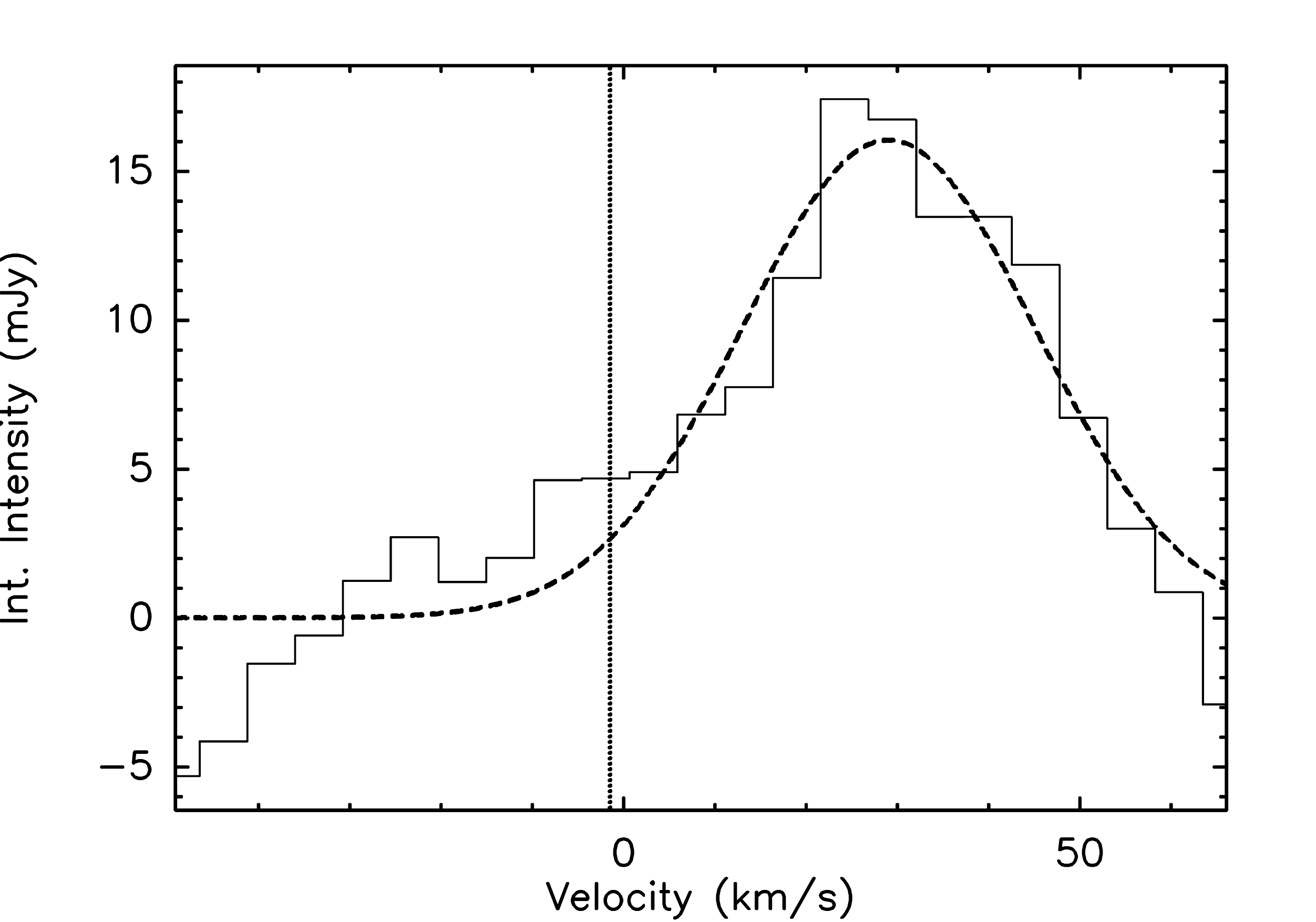}}	
	\subfloat[POS02]{\includegraphics[angle=270,width=7.5cm]{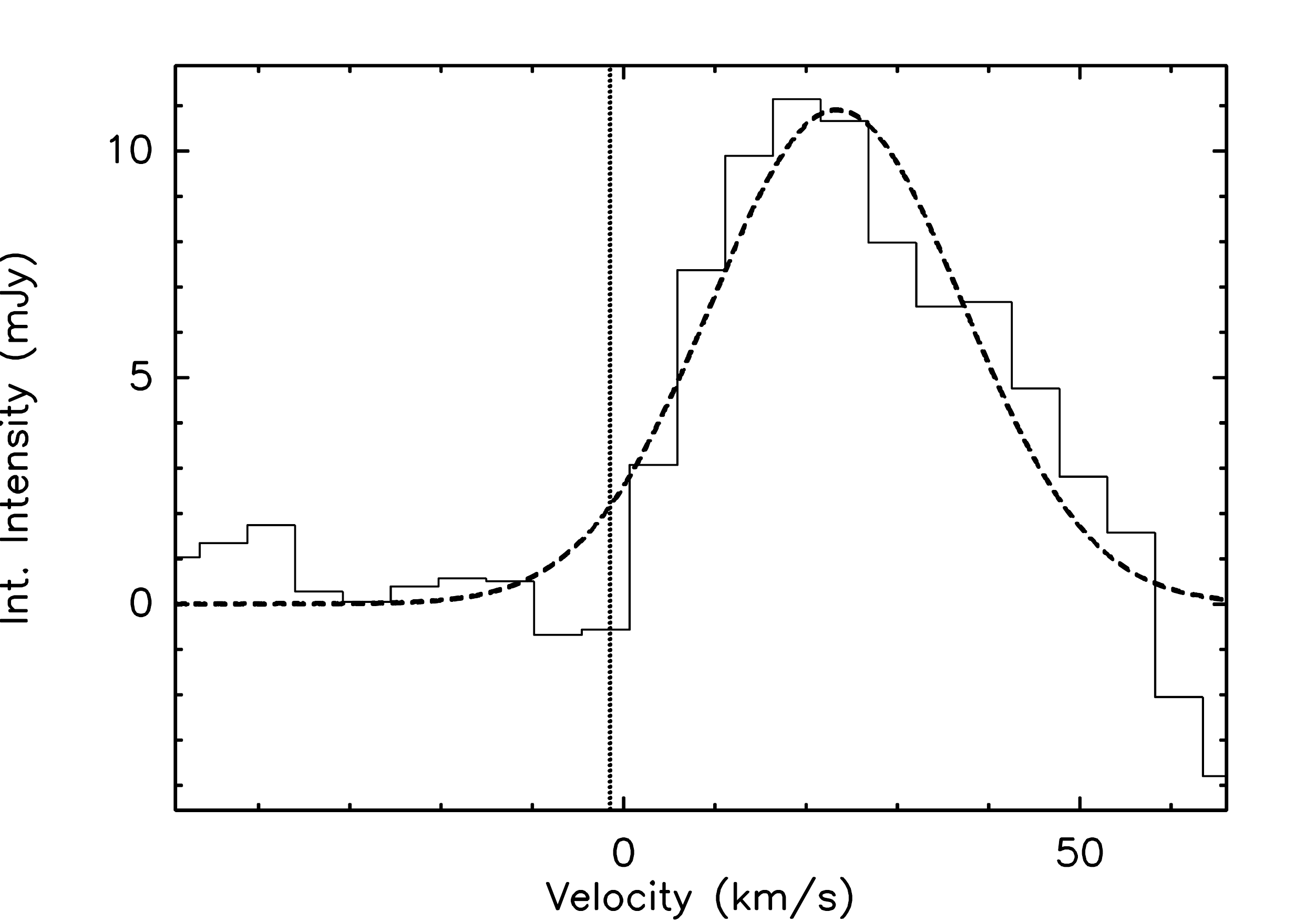}}\\
	\subfloat[POS03]{\includegraphics[angle=270,width=7.5cm]{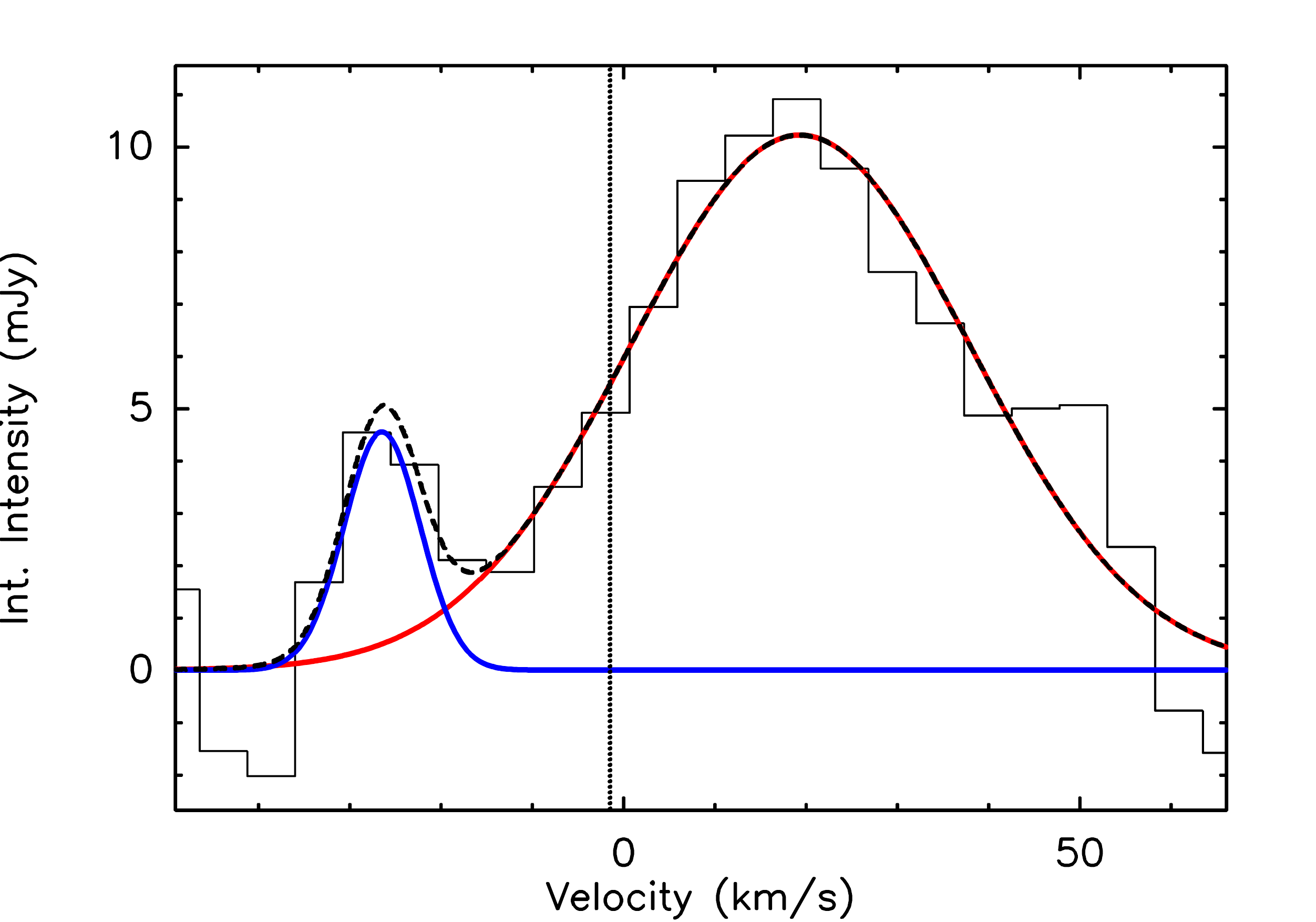}}
	\subfloat[POS04]{\includegraphics[angle=270,width=7.5cm]{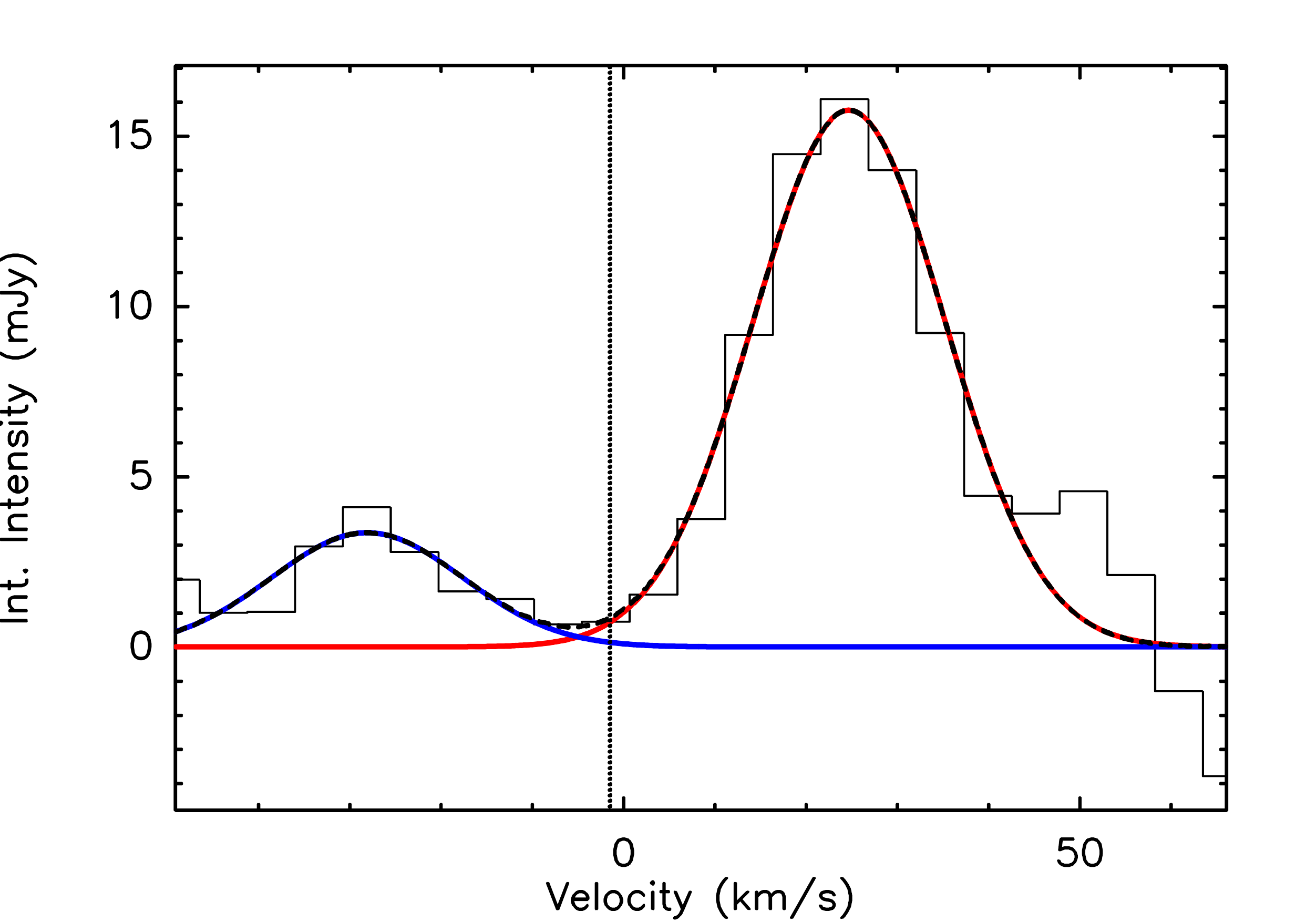}}\\	
	\subfloat[POS05]{\includegraphics[angle=270,width=7.5cm]{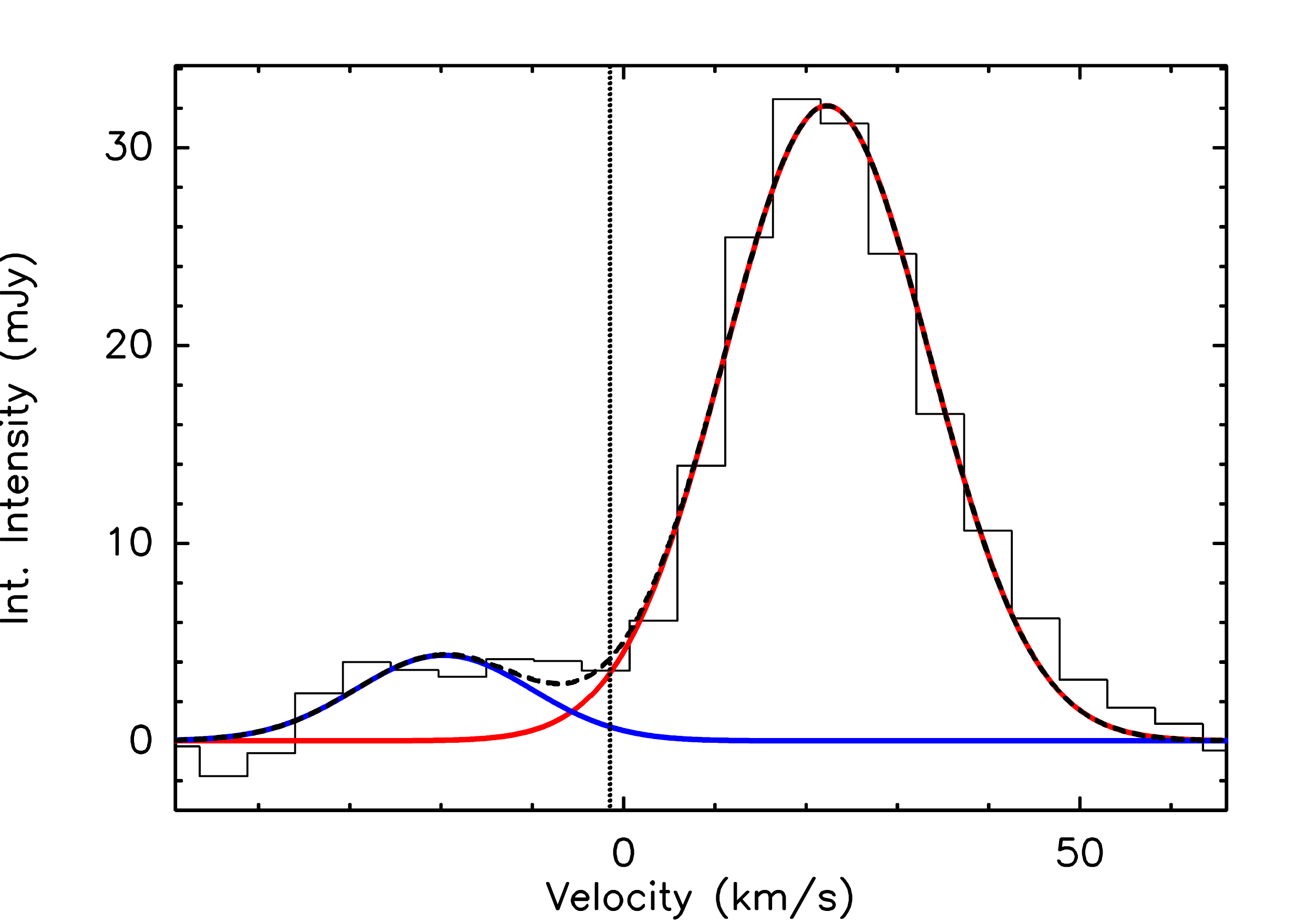}}
	\subfloat[POS06]{\includegraphics[angle=270,width=7.5cm]{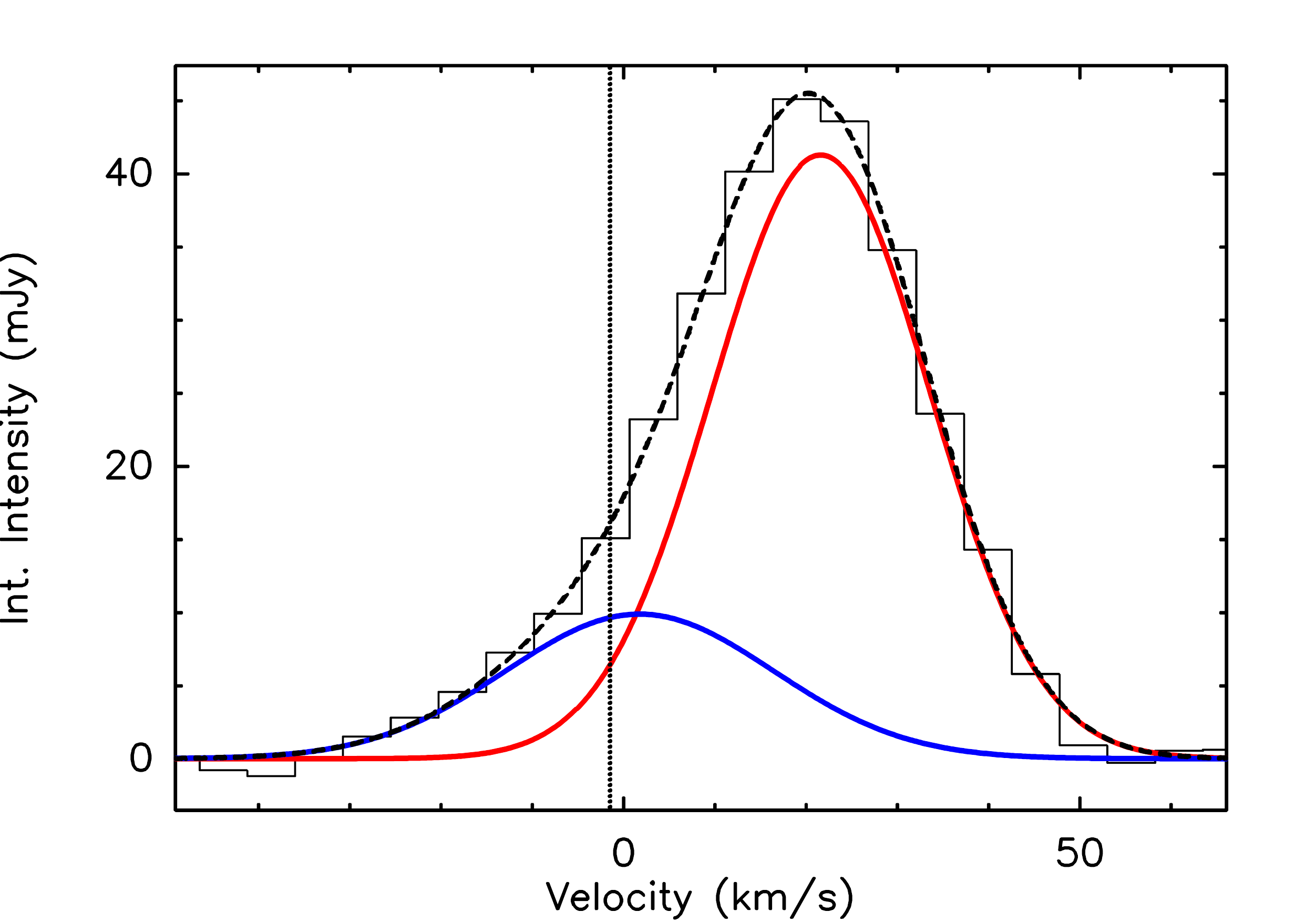}}\\	
	\subfloat[POS07]{\includegraphics[angle=270,width=7.5cm]{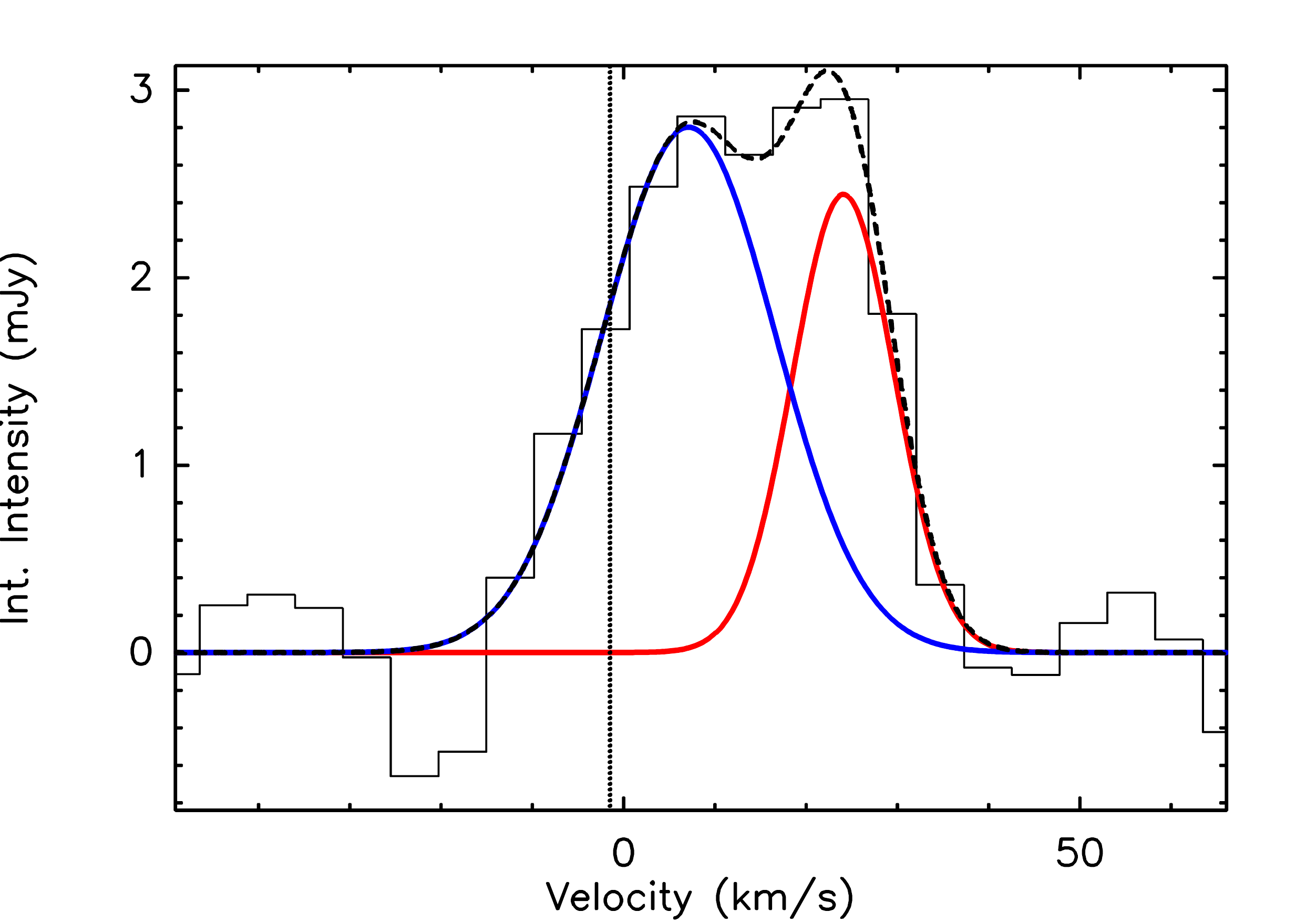}}
	\subfloat[POS08]{\includegraphics[angle=270,width=7.5cm]{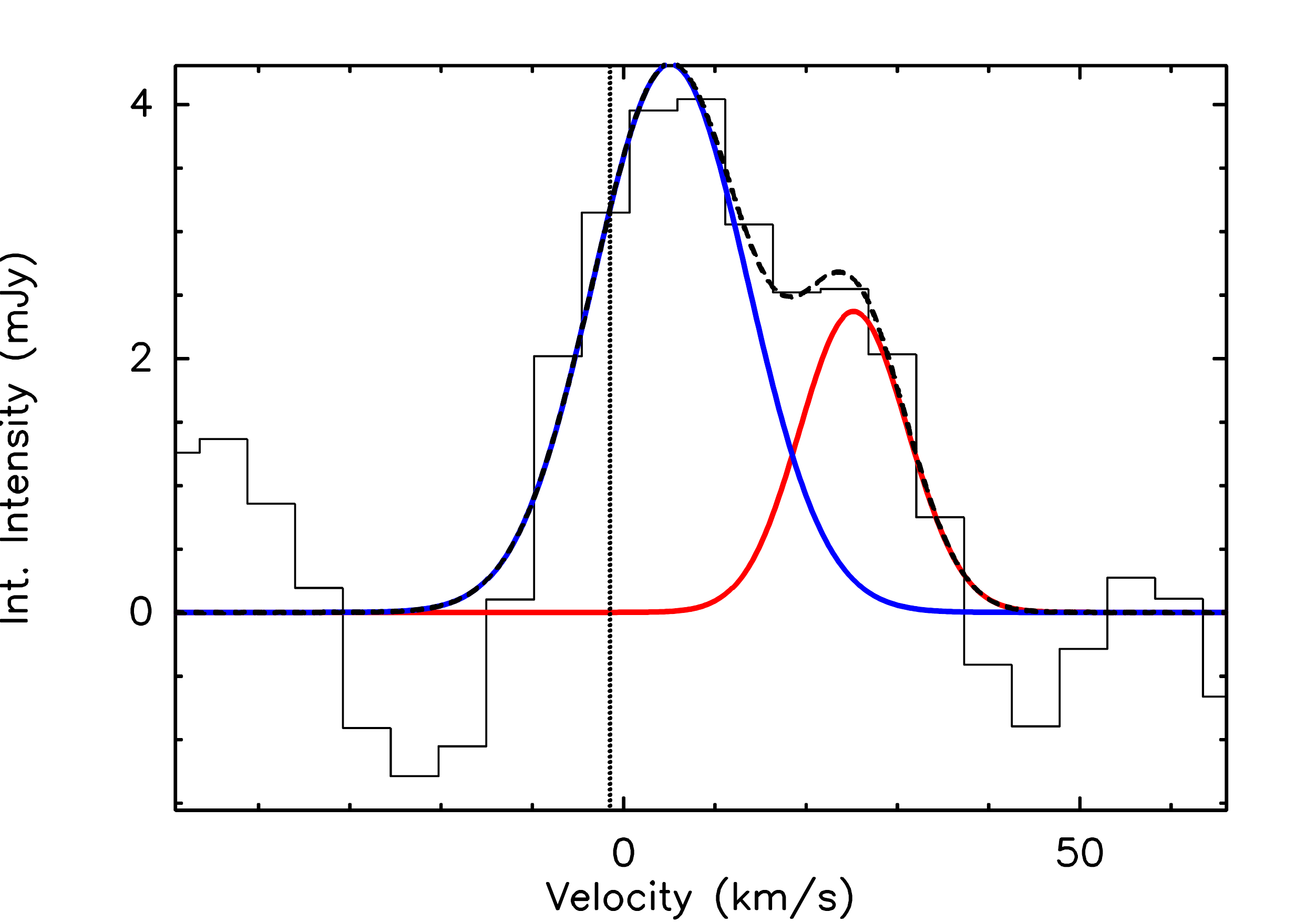}}
	\label{Spectra}
\end{figure*}

\addtocounter{figure}{-1}
\begin{figure*}
	\centering
	\caption{Continued}
	\subfloat[POS09]{\includegraphics[angle=270,width=7.5cm]{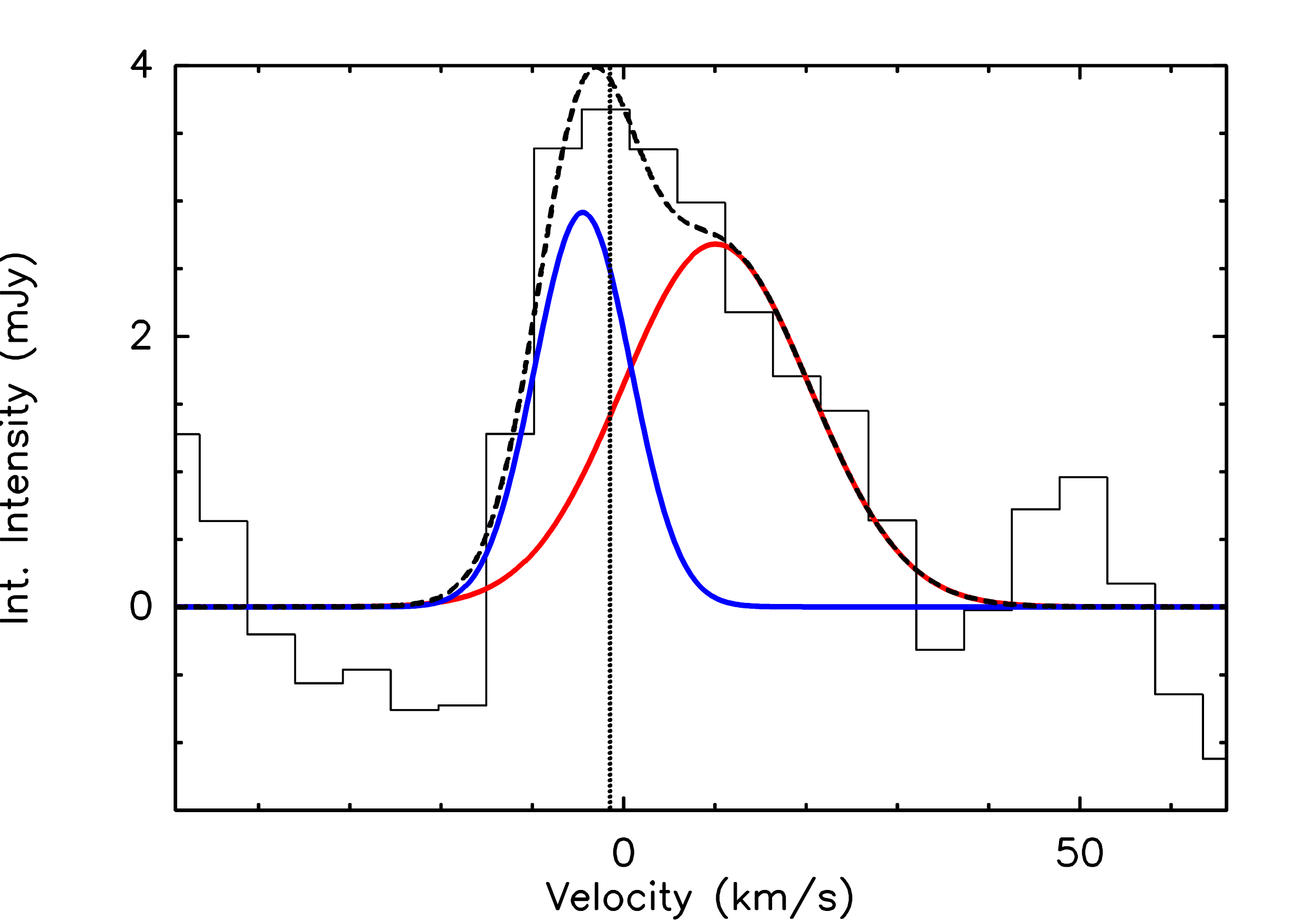}}	
	\subfloat[POS10]{\includegraphics[angle=270,width=7.5cm]{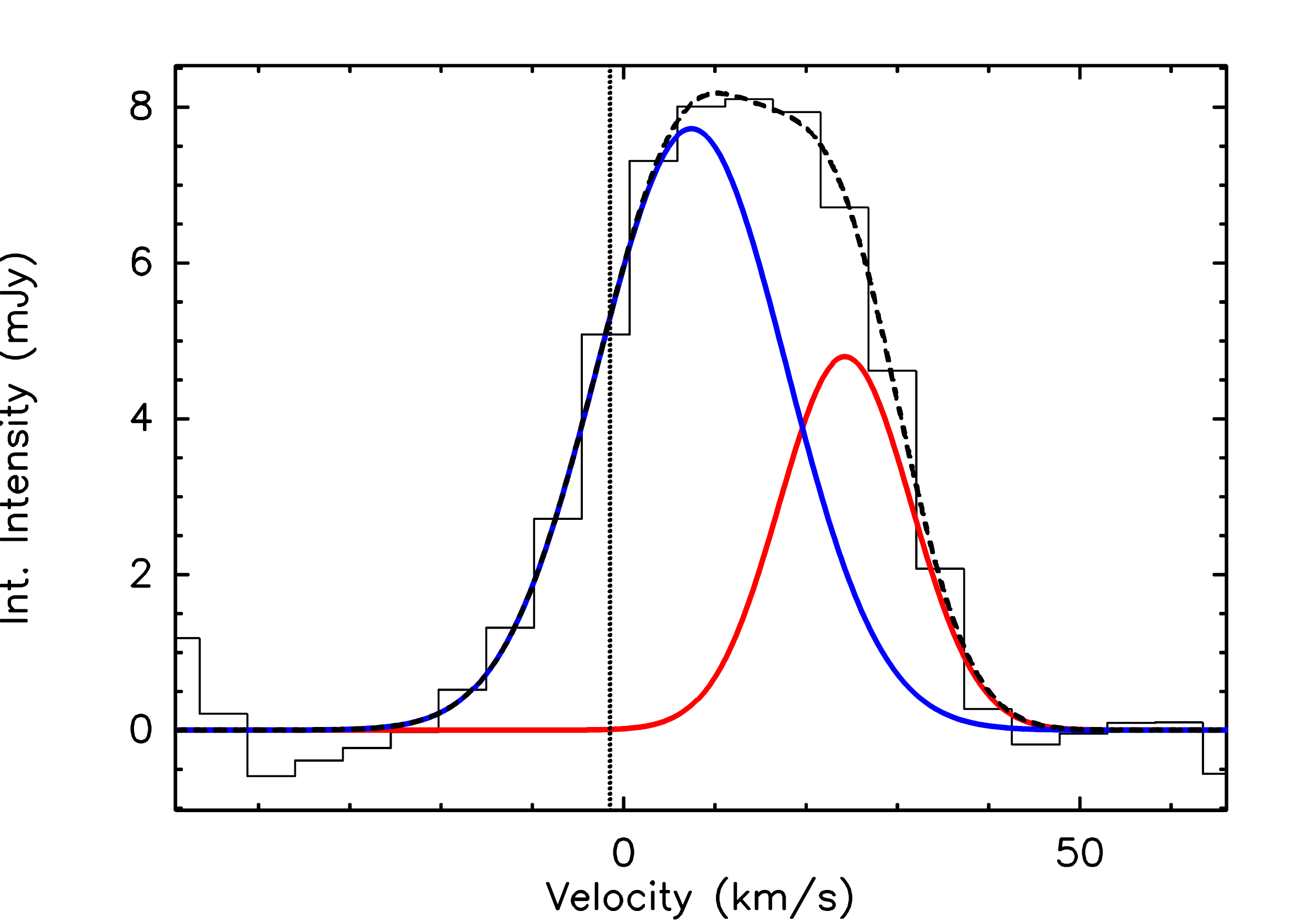}}\\
	\subfloat[POS11]{\includegraphics[angle=270,width=7.5cm]{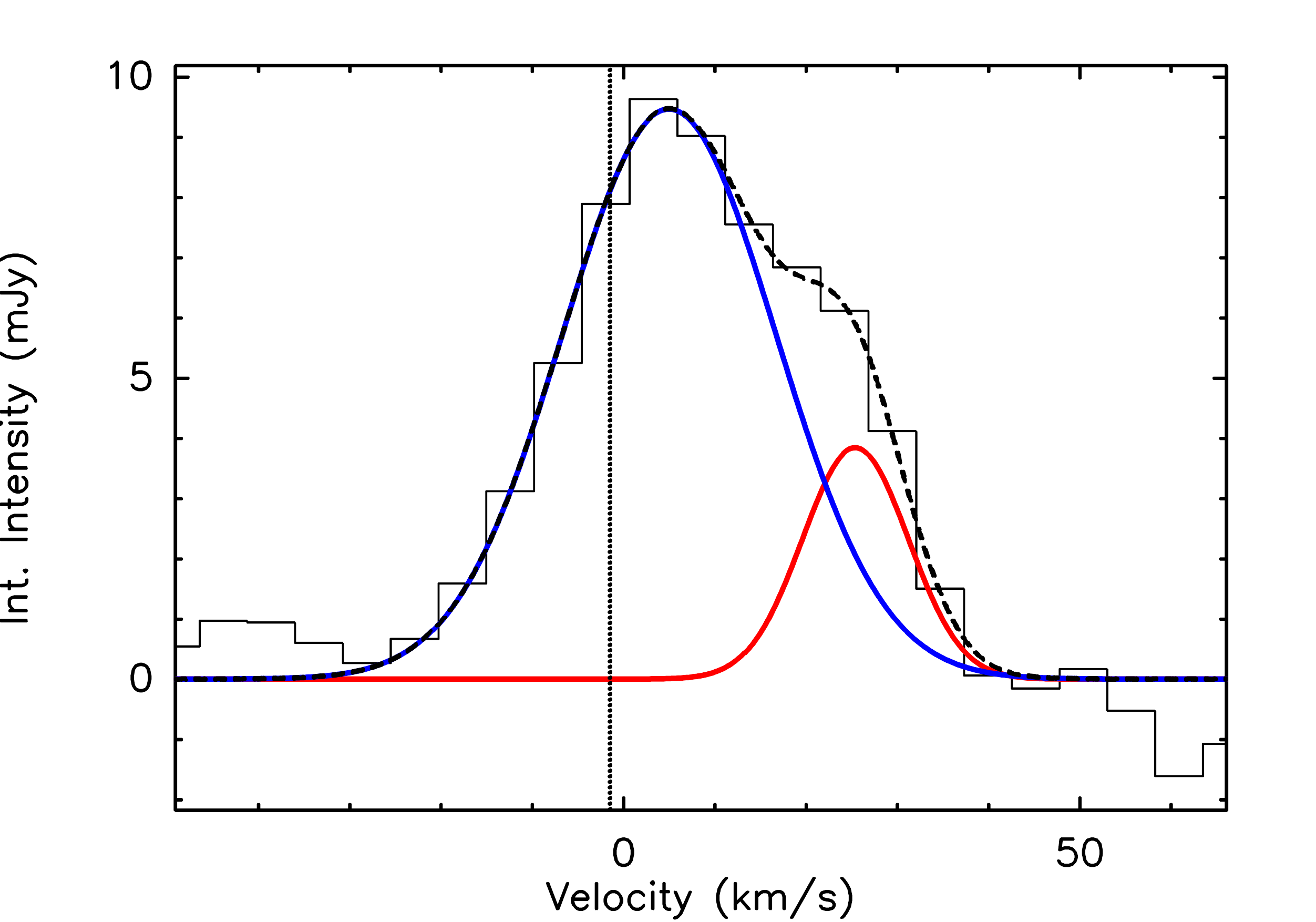}}
	\subfloat[POS12]{\includegraphics[angle=270,width=7.5cm]{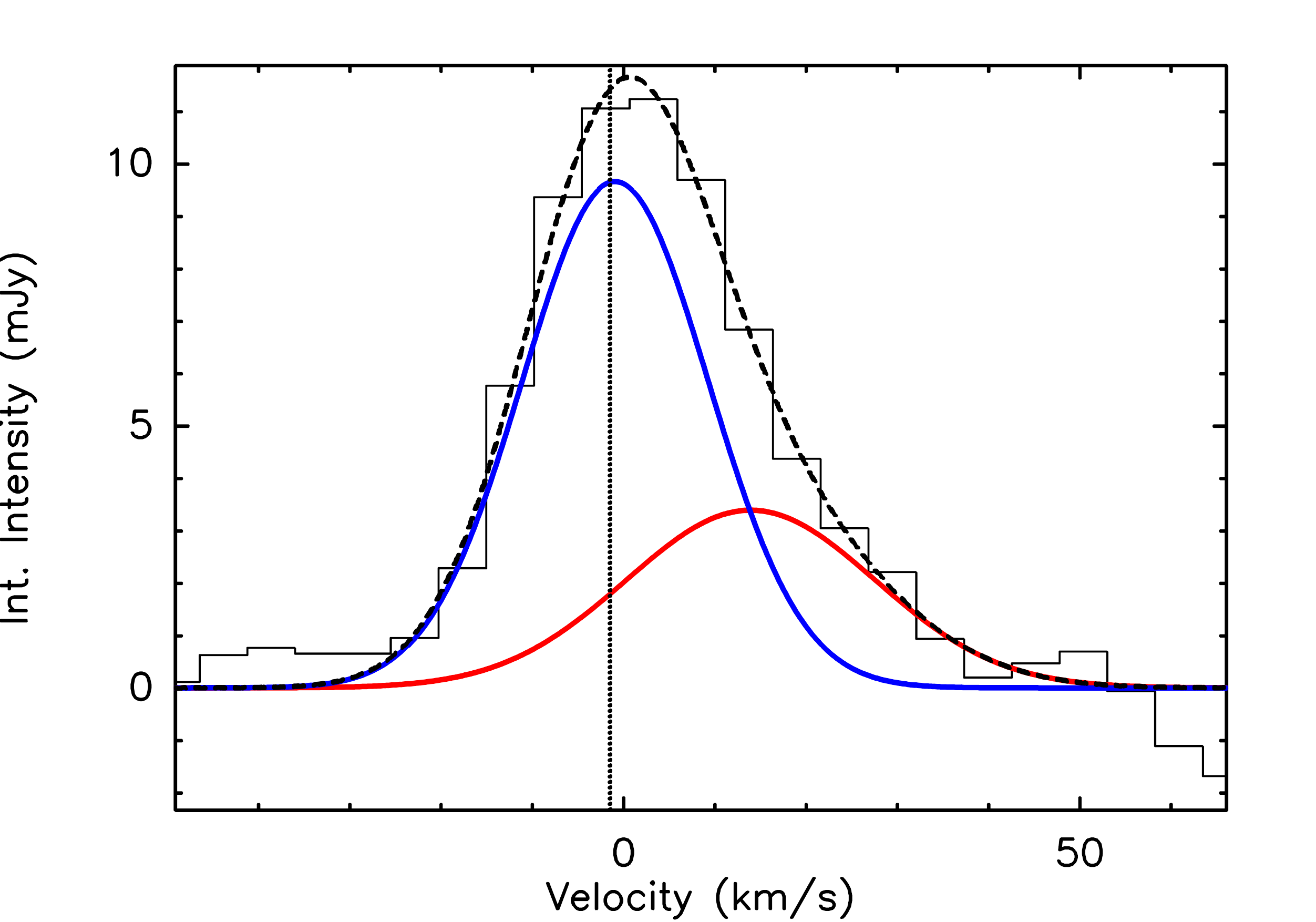}}\\	
	\subfloat[POS13]{\includegraphics[angle=270,width=7.5cm]{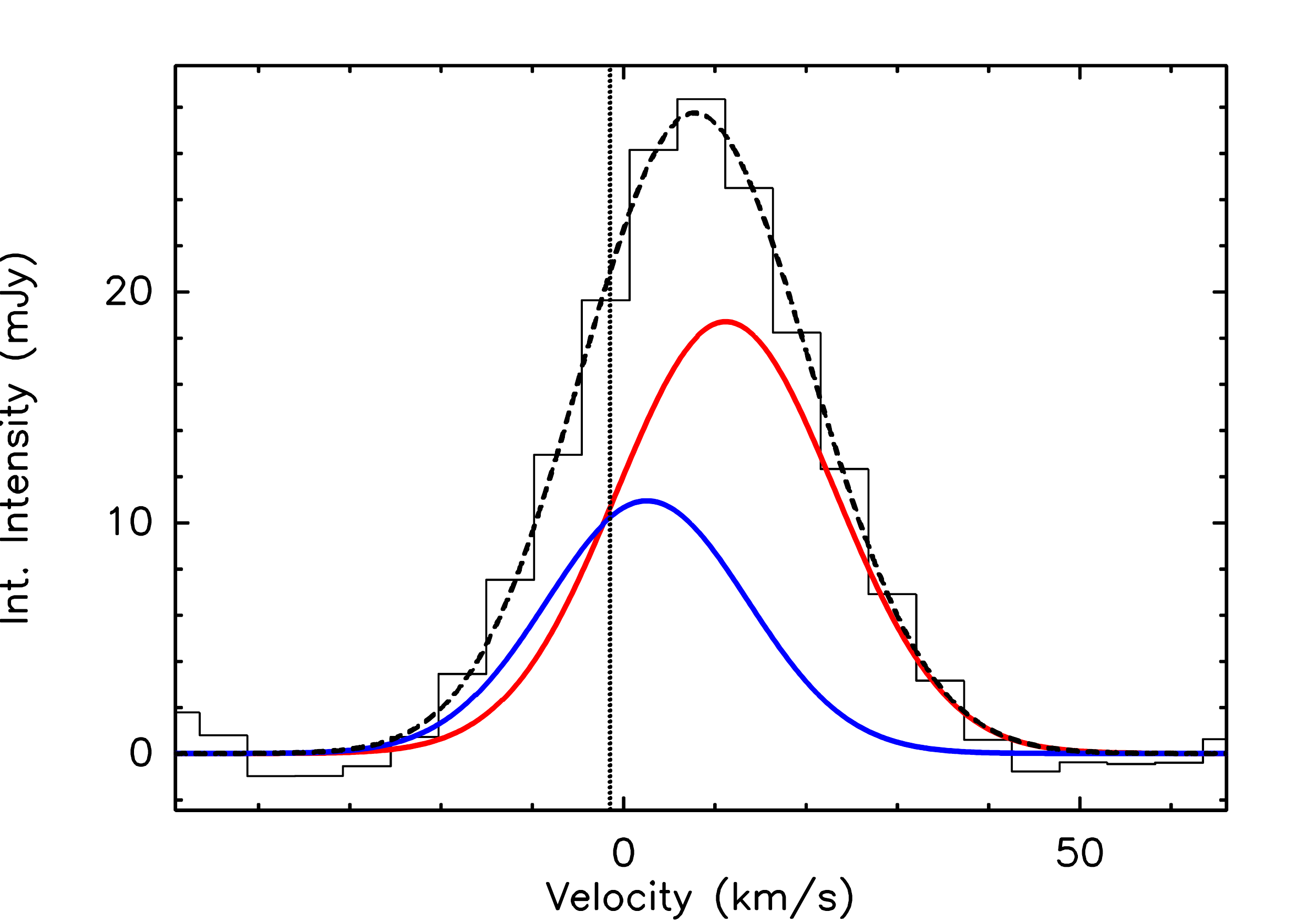}}
	\subfloat[POS14]{\includegraphics[angle=270,width=7.5cm]{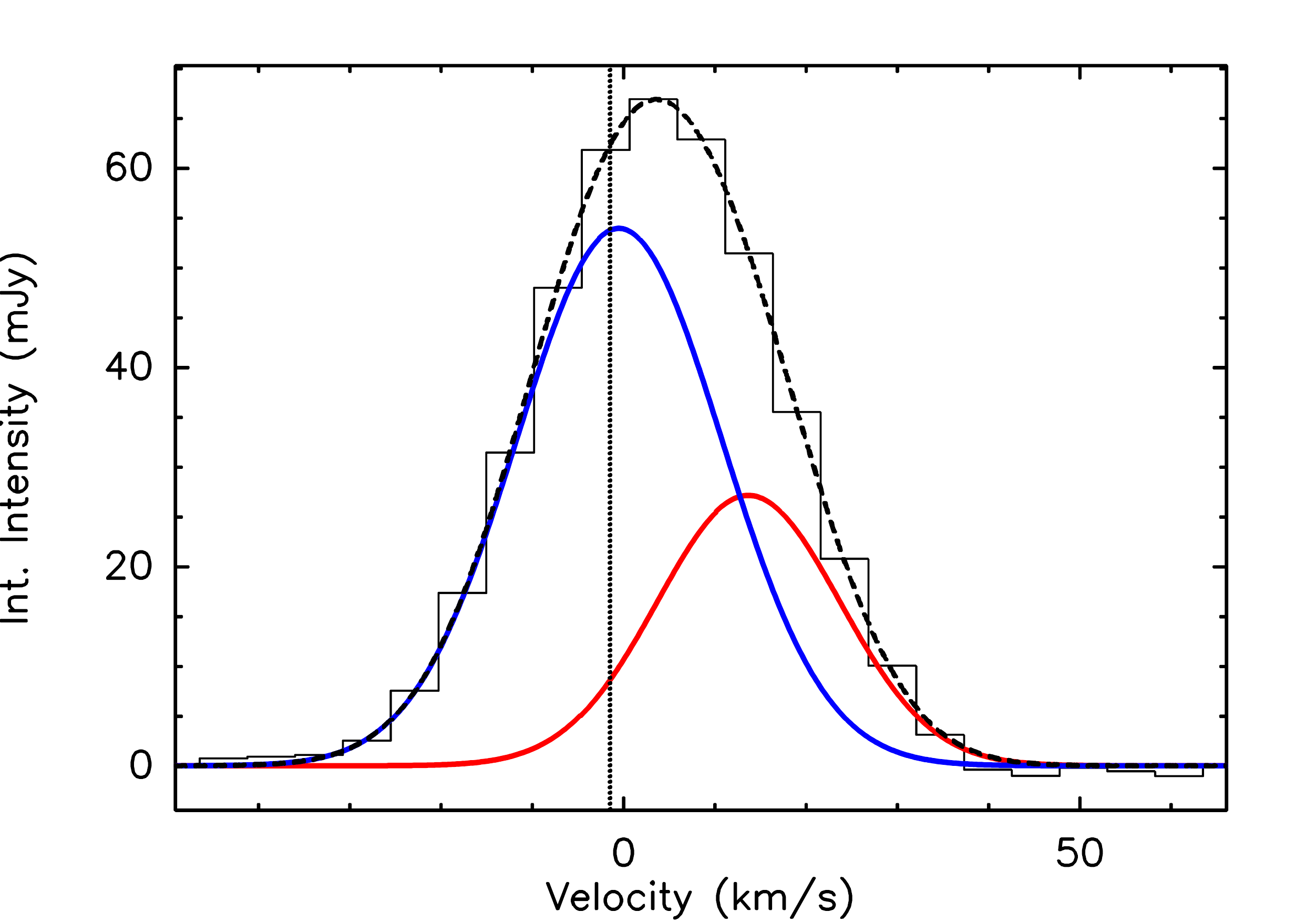}}\\	
	\subfloat[POS15]{\includegraphics[angle=270,width=7.5cm]{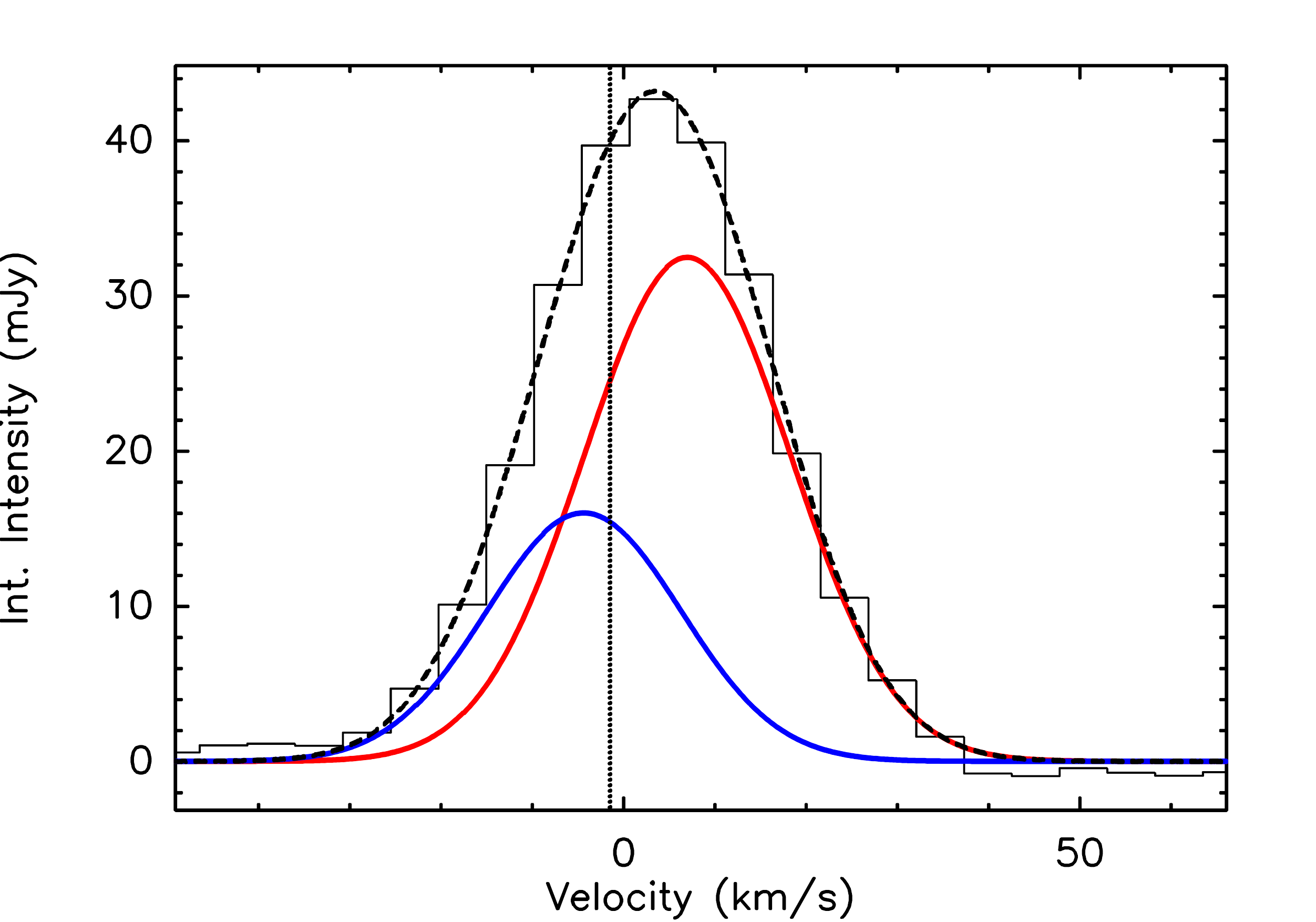}}
	\subfloat[POS16]{\includegraphics[angle=270,width=7.5cm]{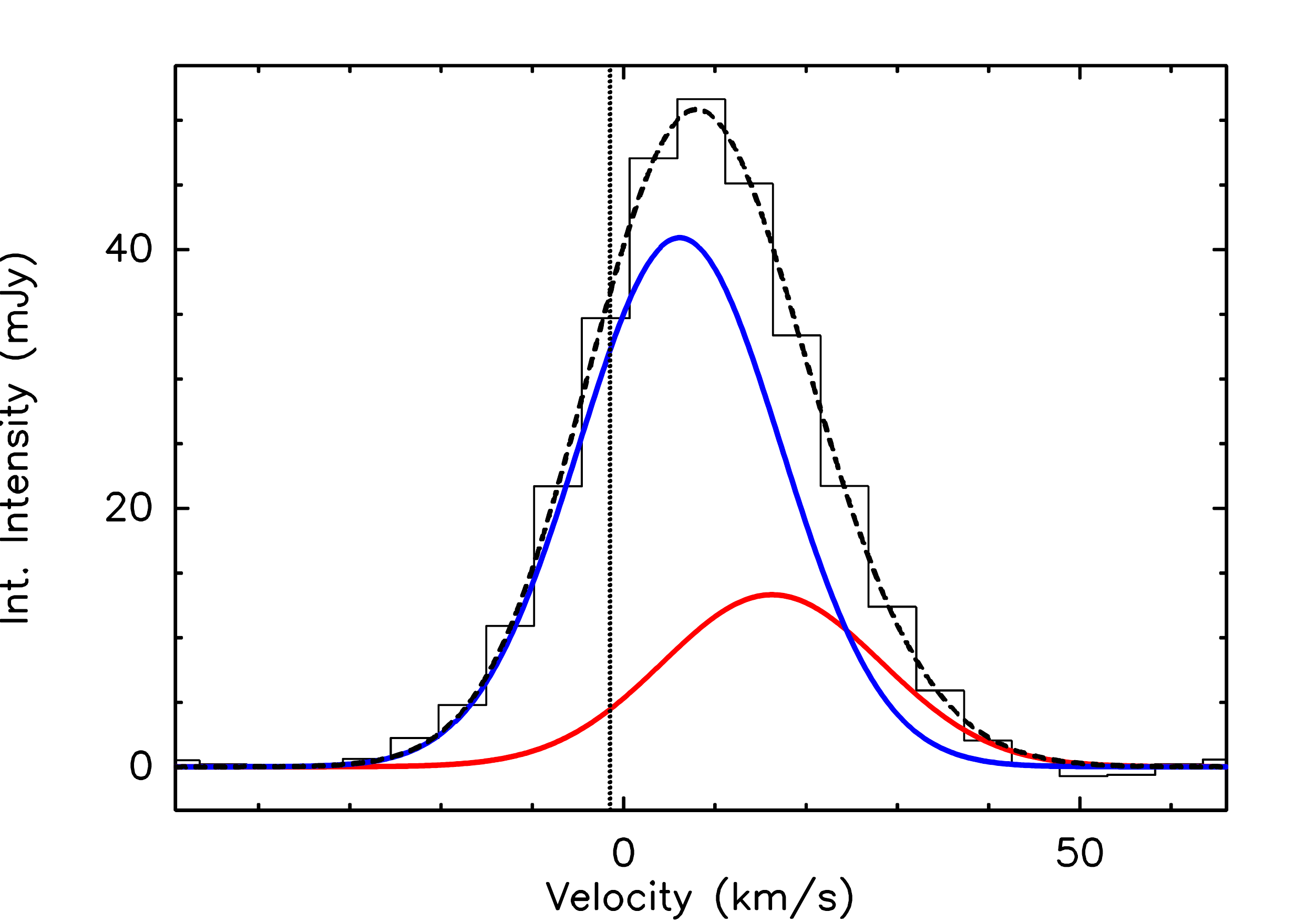}}
\end{figure*}

\addtocounter{figure}{-1}
\begin{figure*}
	\centering	
	\caption{Continued}
	\subfloat[POS17]{\includegraphics[angle=270,width=7.5cm]{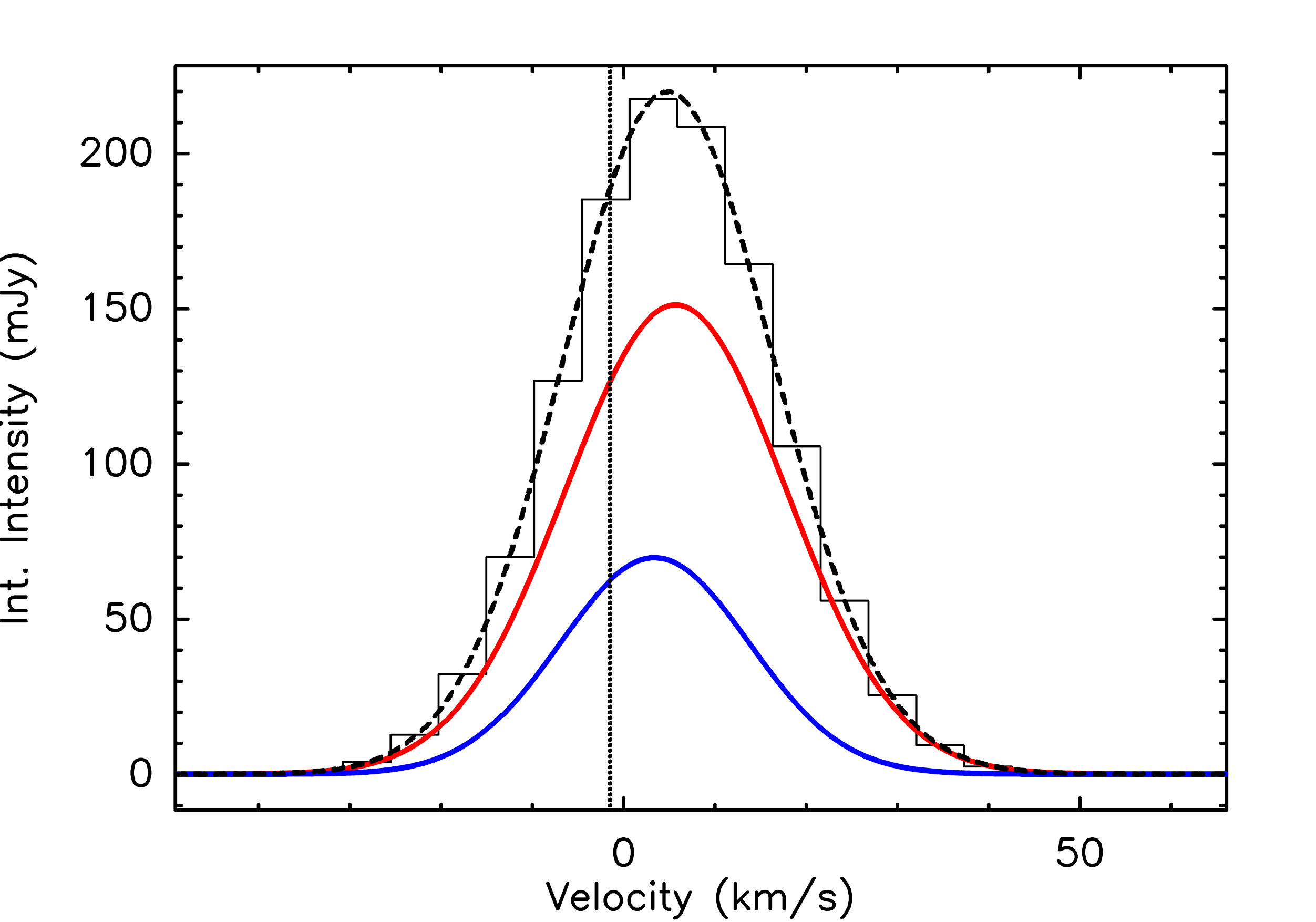}}	
	\subfloat[POS18]{\includegraphics[angle=270,width=7.5cm]{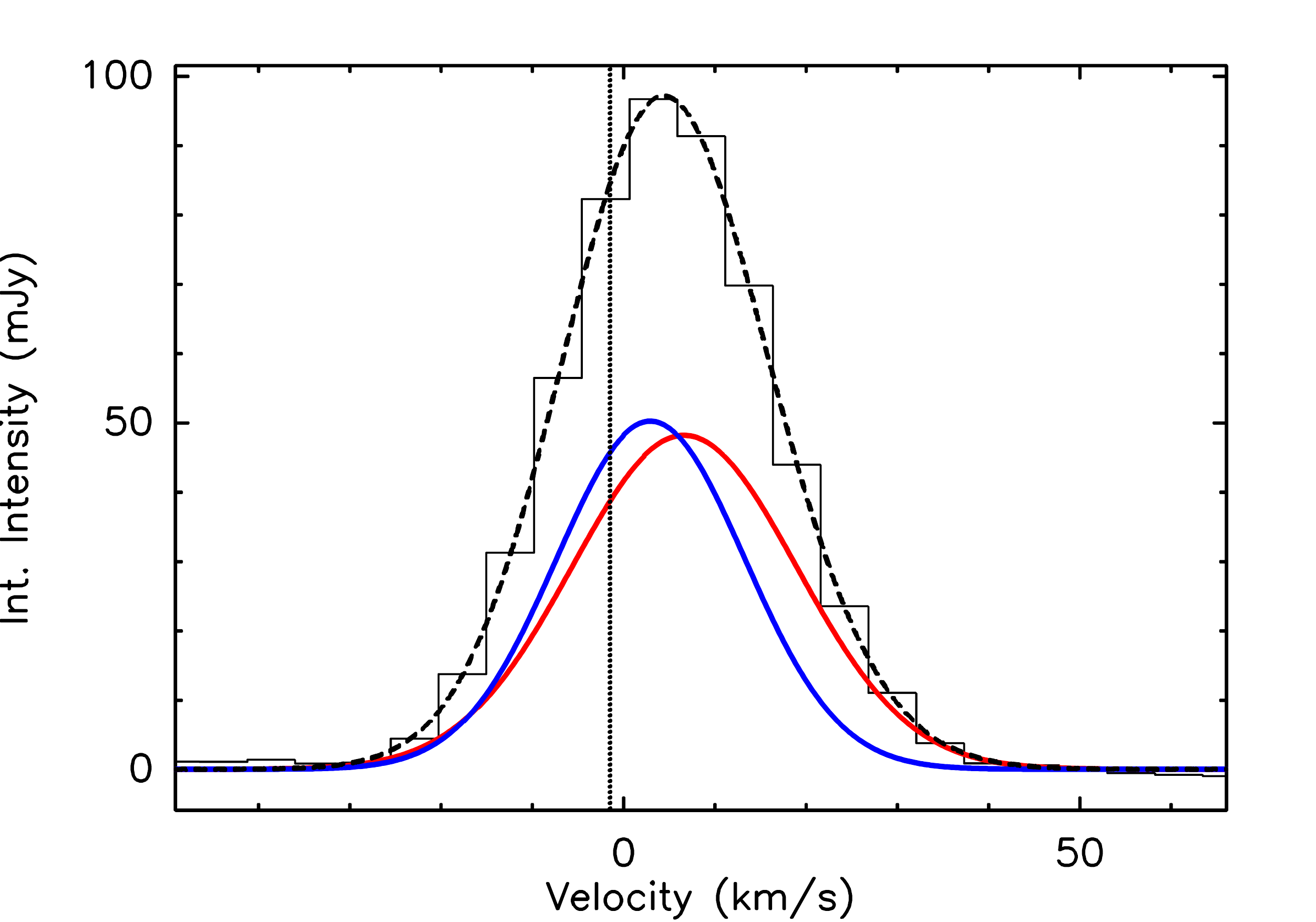}}\\
	\subfloat[POS19]{\includegraphics[angle=270,width=7.5cm]{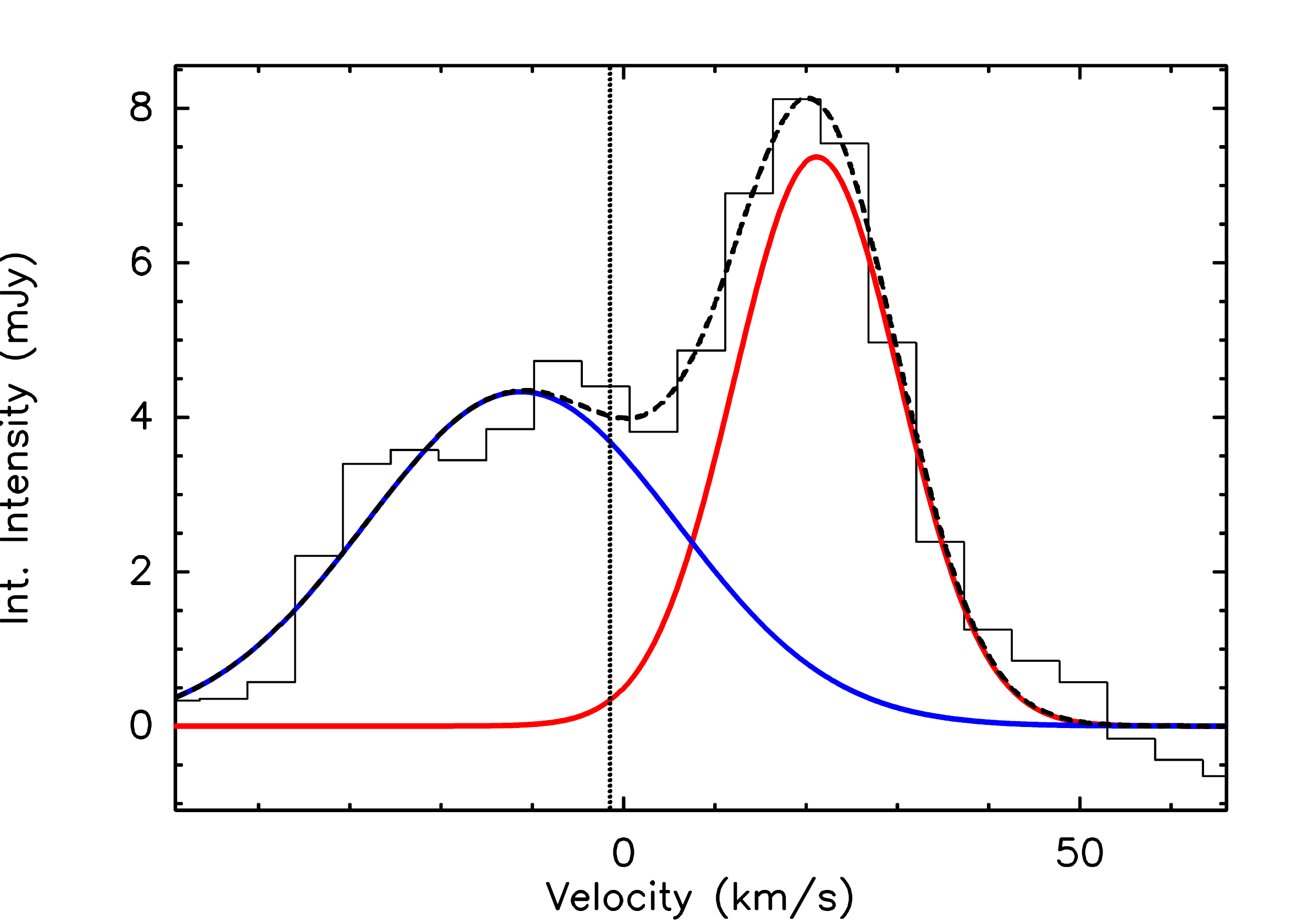}}
	\subfloat[POS20]{\includegraphics[angle=270,width=7.5cm]{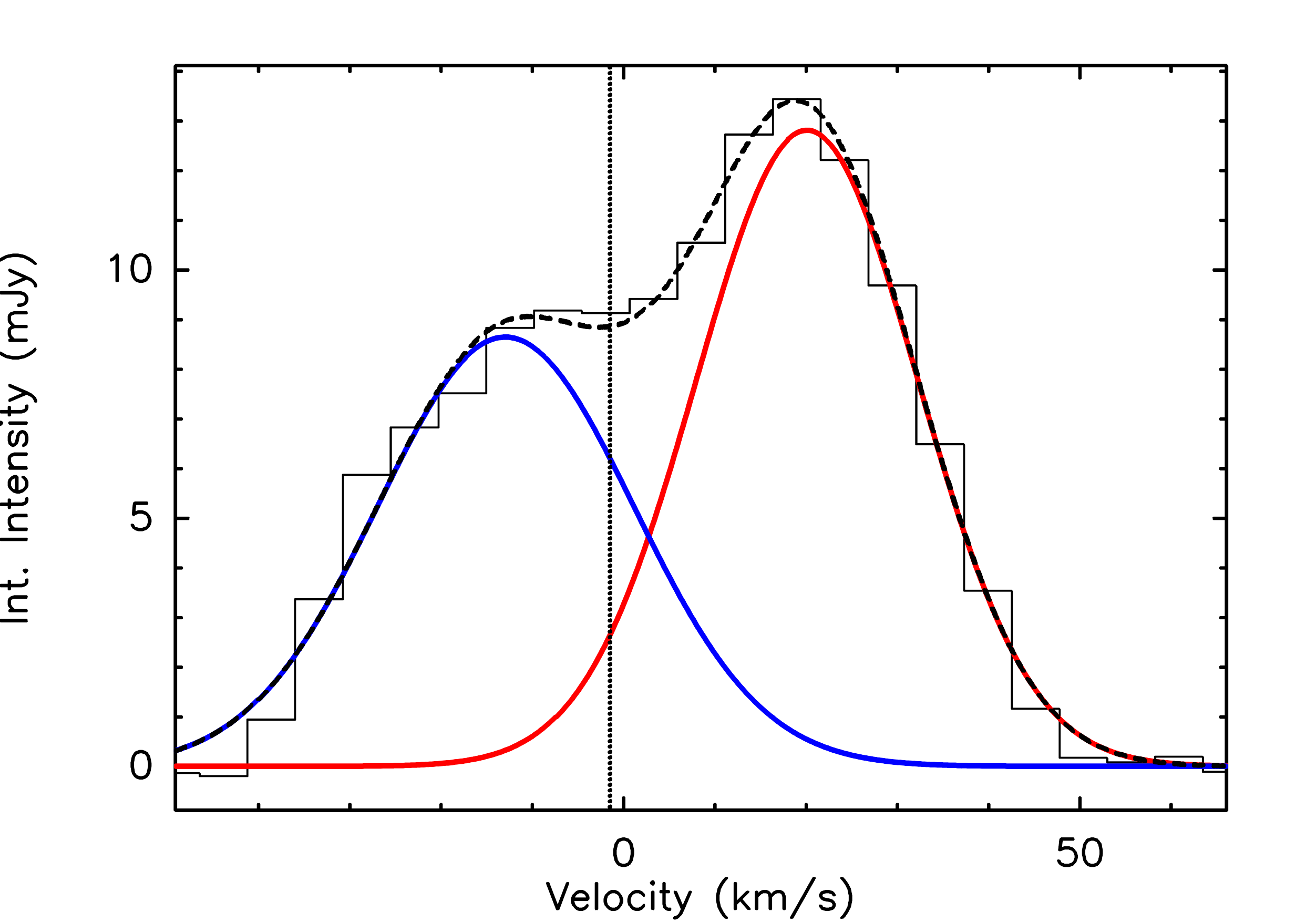}}\\	
	\subfloat[POS21]{\includegraphics[angle=270,width=7.5cm]{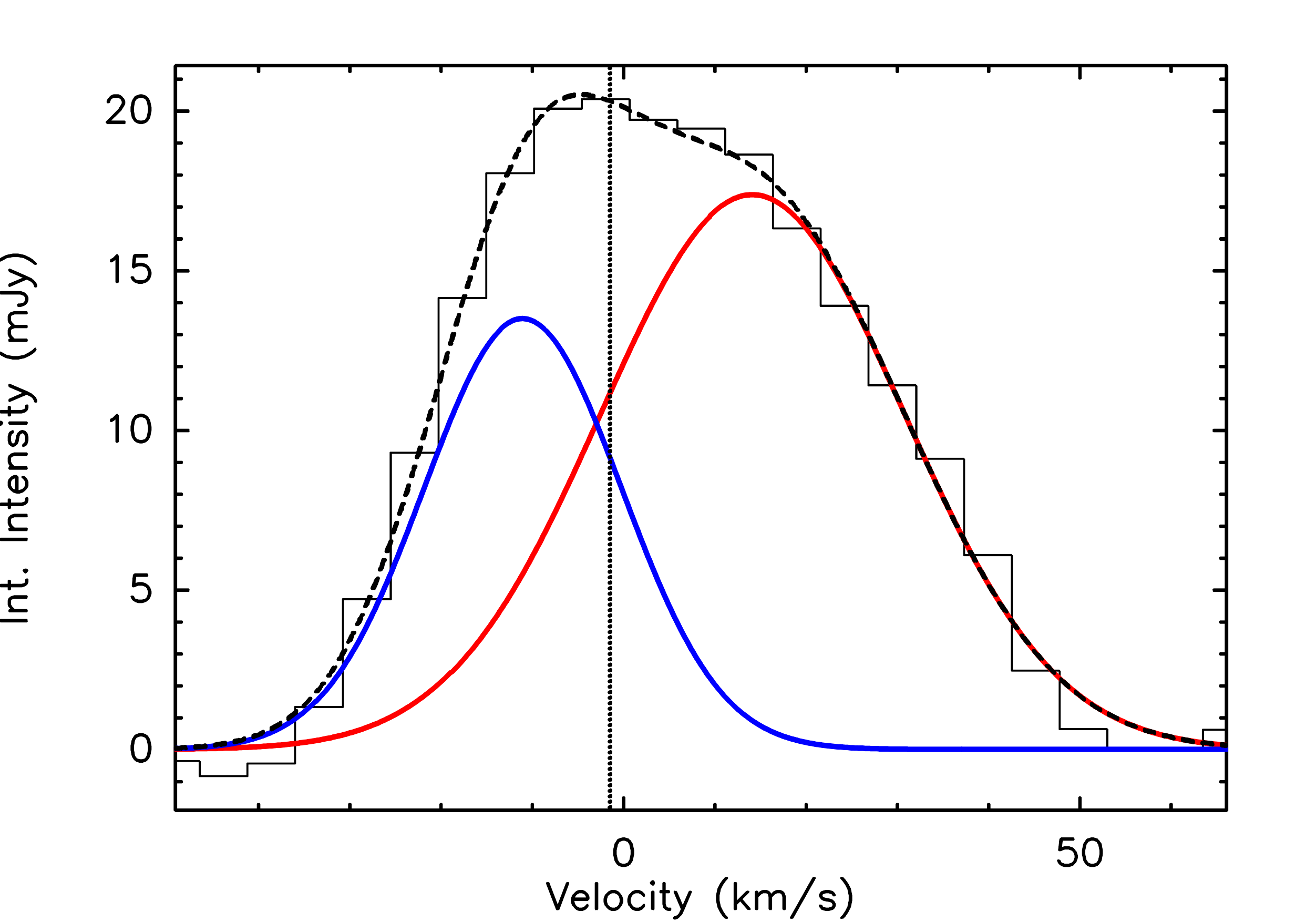}}
	\subfloat[POS22]{\includegraphics[angle=270,width=7.5cm]{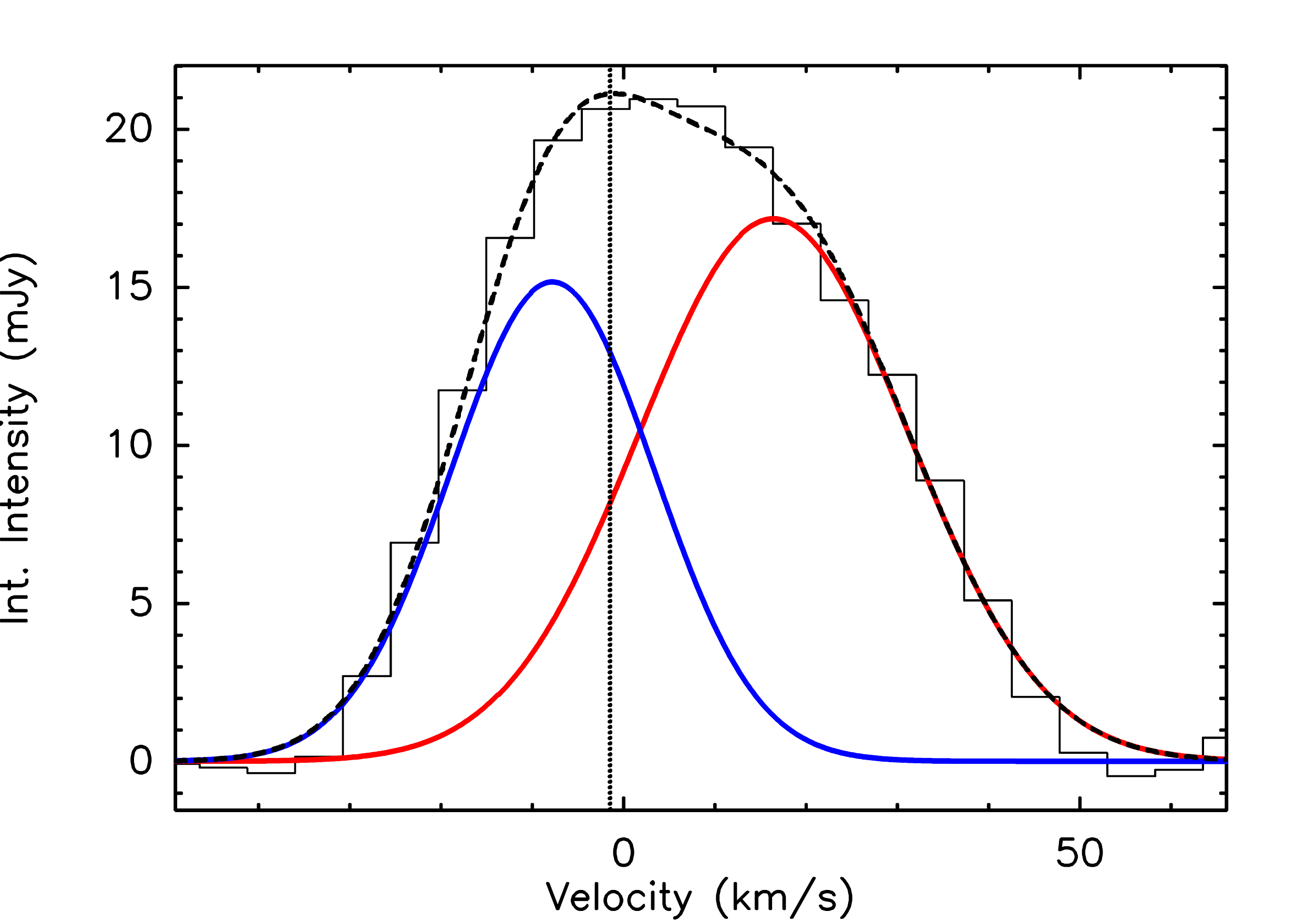}}\\	
	\subfloat[POS23]{\includegraphics[angle=270,width=7.5cm]{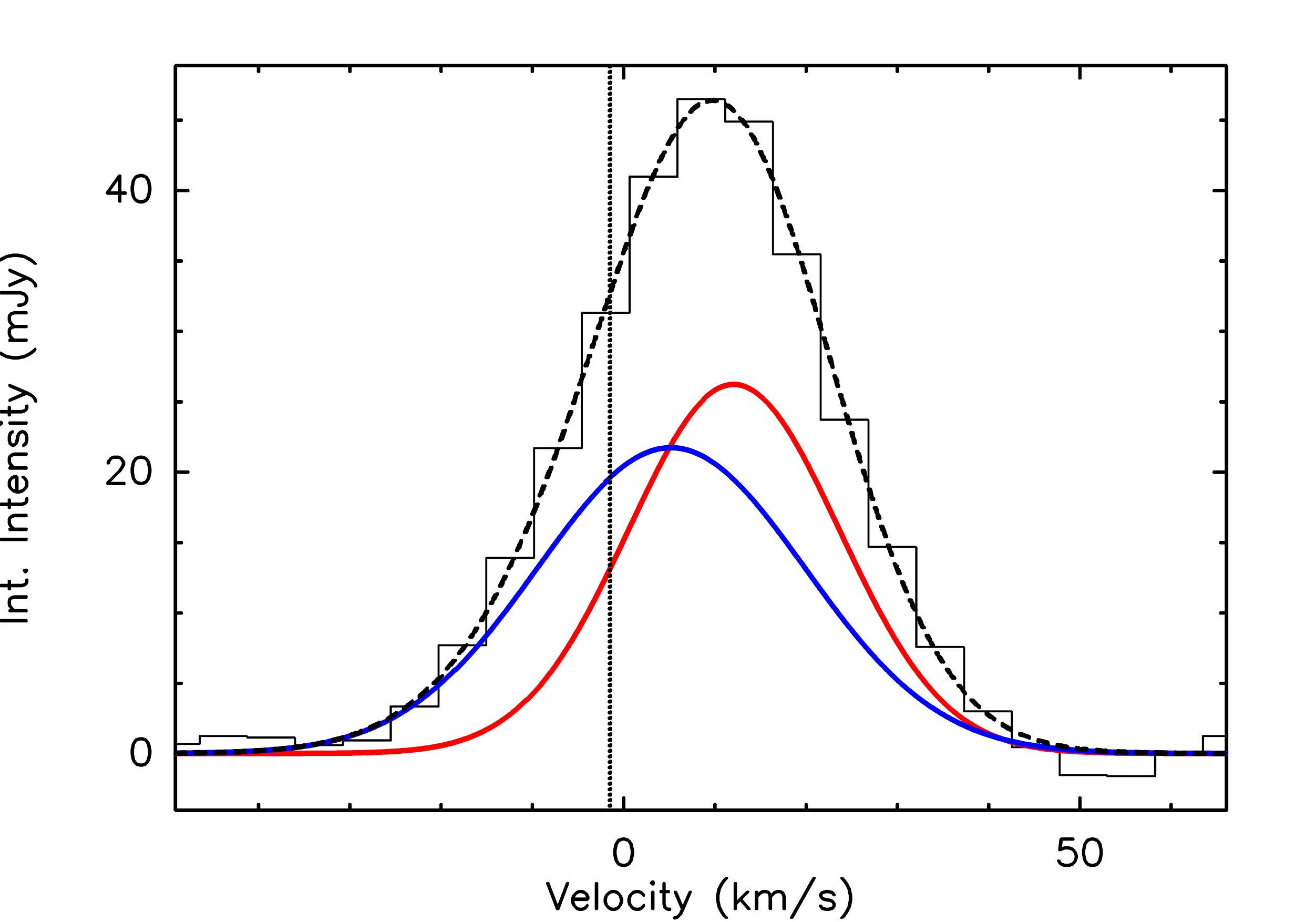}}
	\subfloat[POS24]{\includegraphics[angle=270,width=7.5cm]{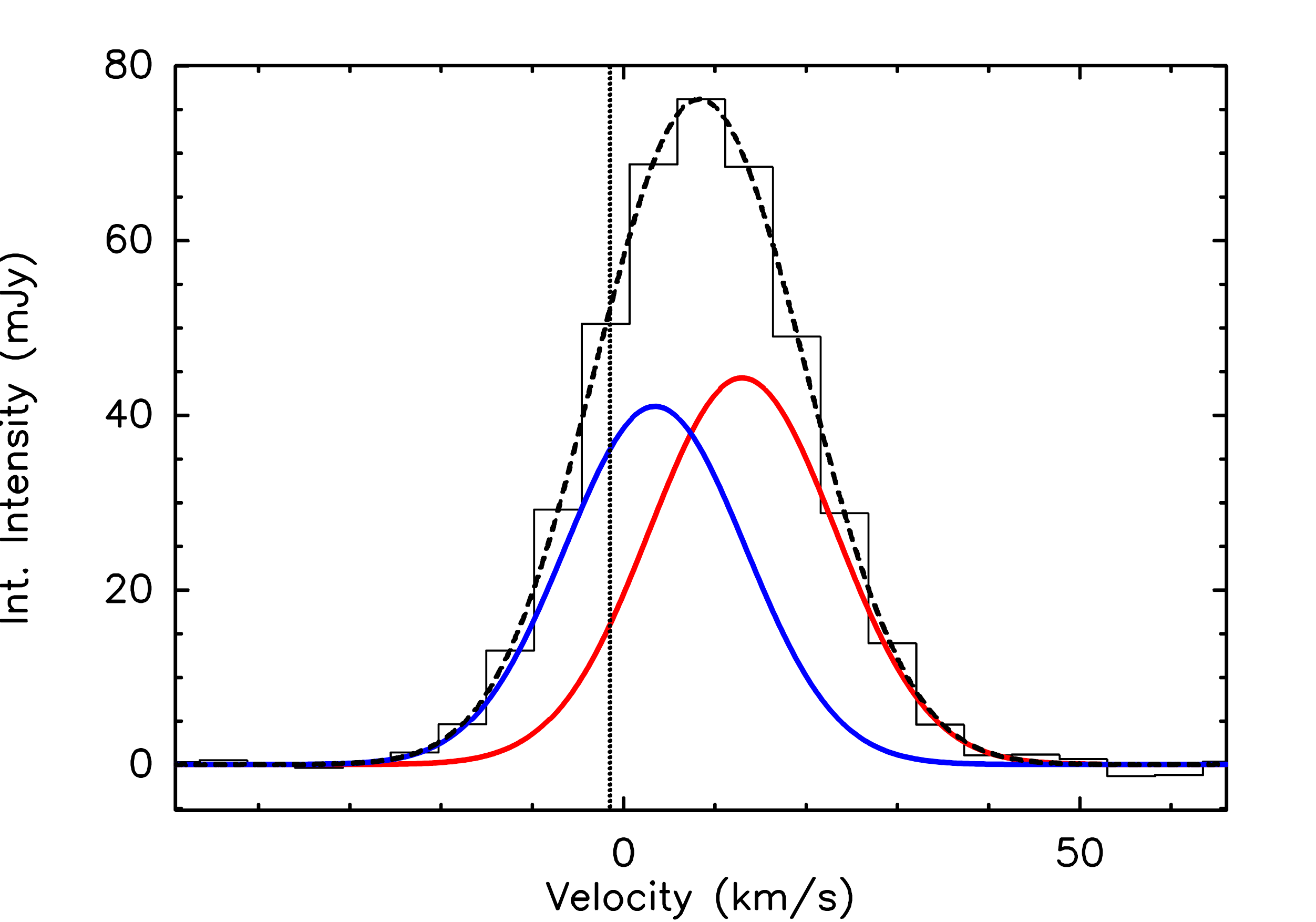}}
\end{figure*}

\addtocounter{figure}{-1}
\begin{figure*}
	\centering
	\caption{Continued}
	\subfloat[POS25]{\includegraphics[angle=270,width=7.5cm]{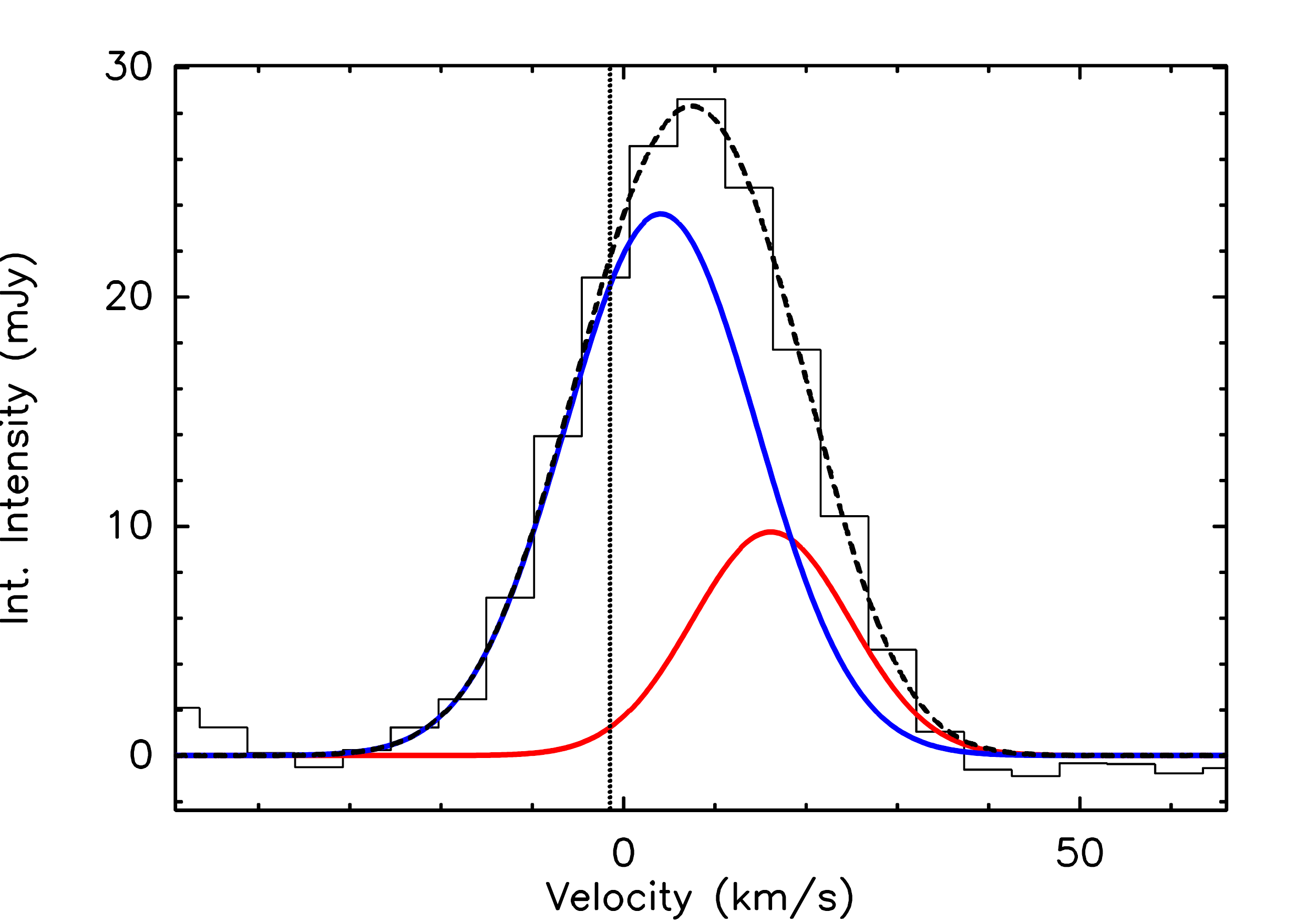}}	
	\subfloat[POS26]{\includegraphics[angle=270,width=7.5cm]{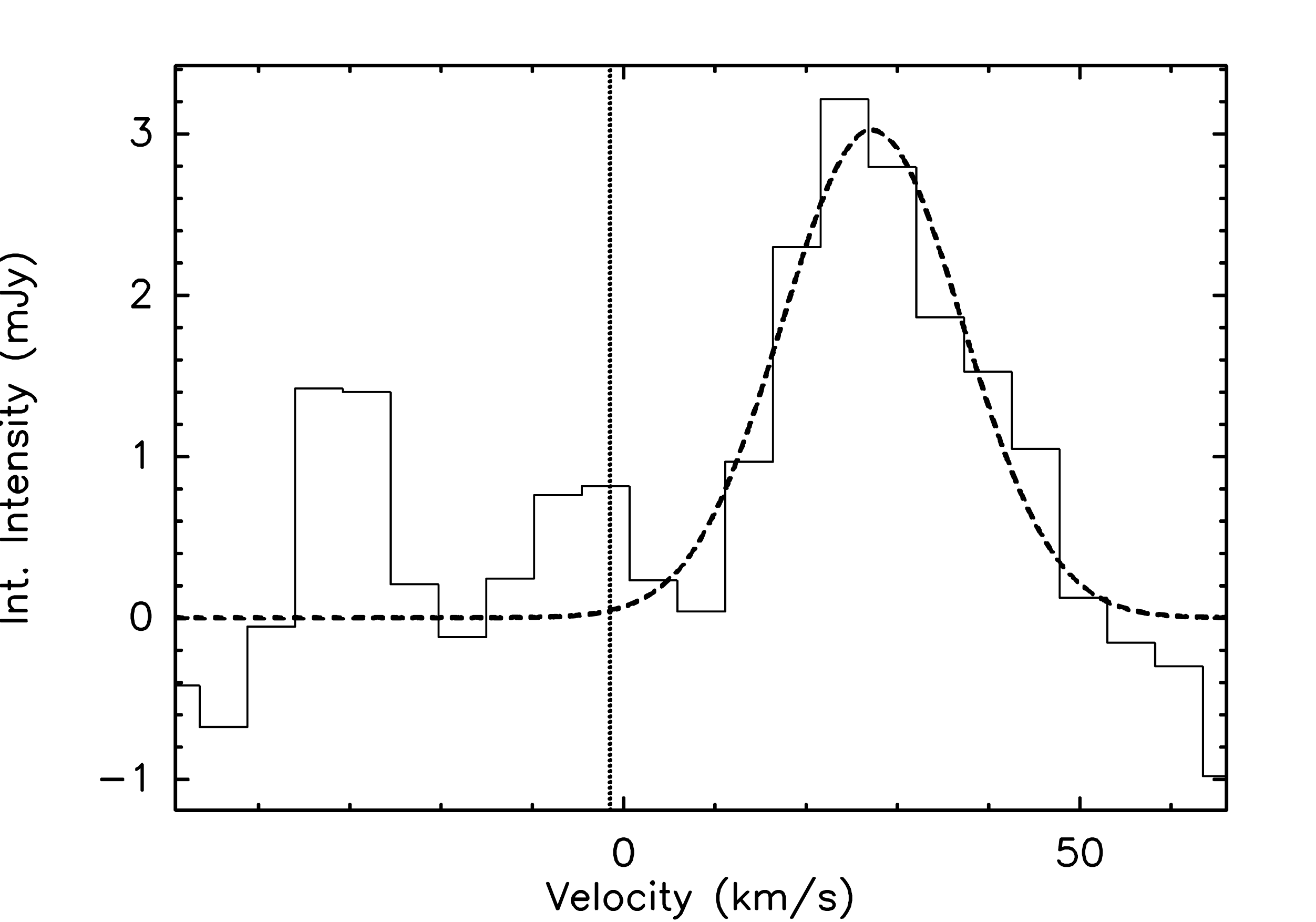}}\\
	\subfloat[POS27]{\includegraphics[angle=270,width=7.5cm]{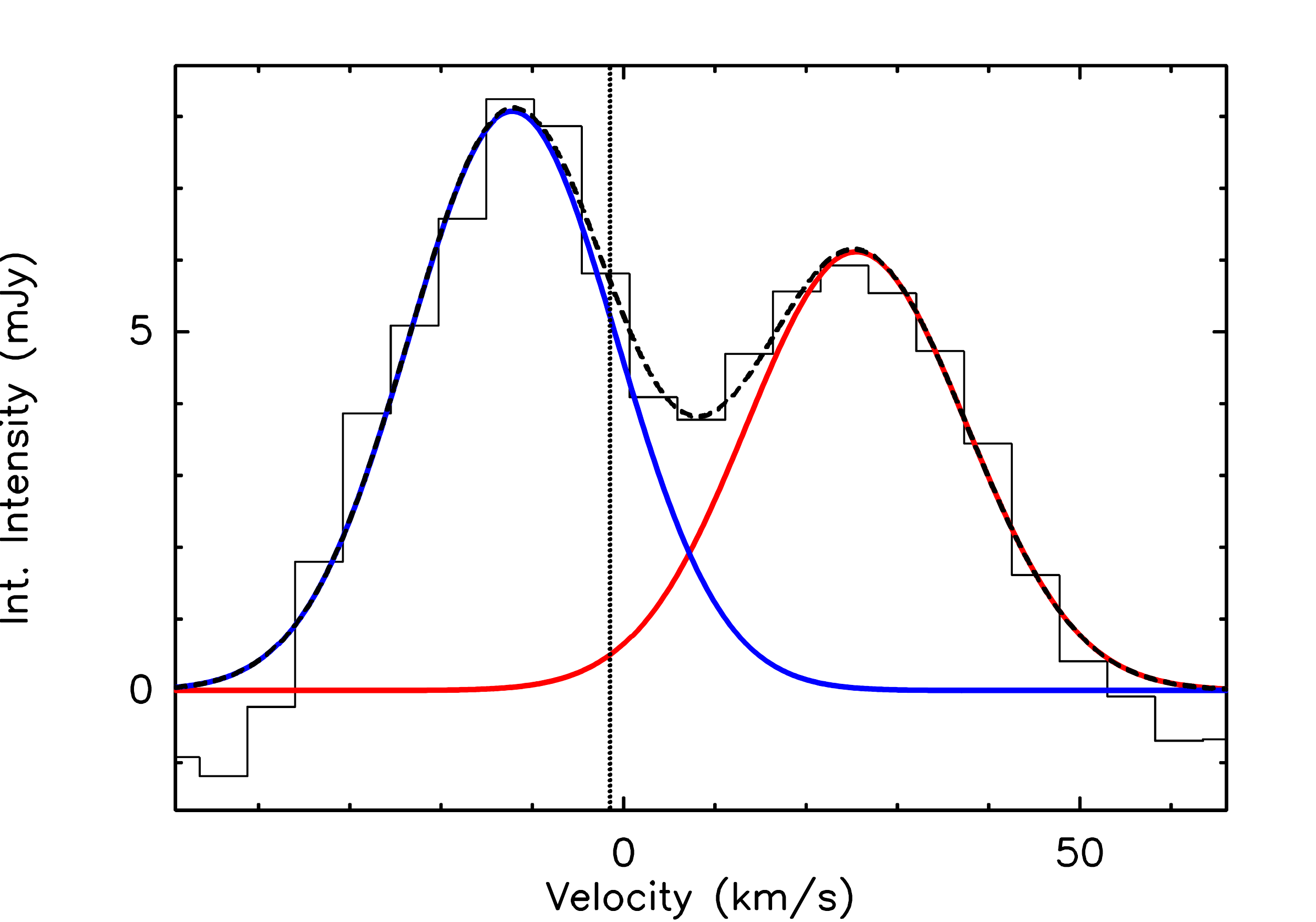}}
	\subfloat[POS28]{\includegraphics[angle=270,width=7.5cm]{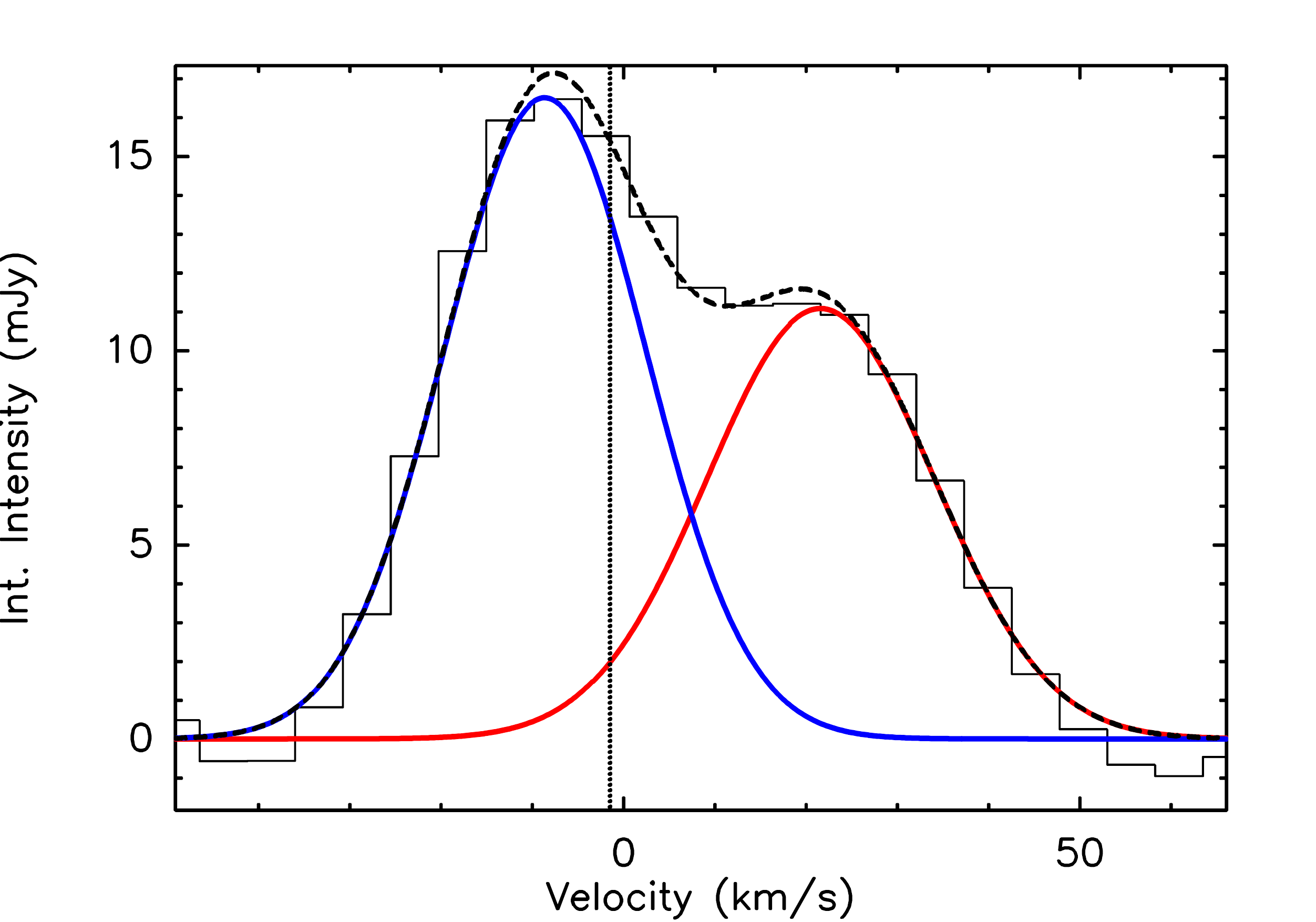}}\\	
	\subfloat[POS29]{\includegraphics[angle=270,width=7.5cm]{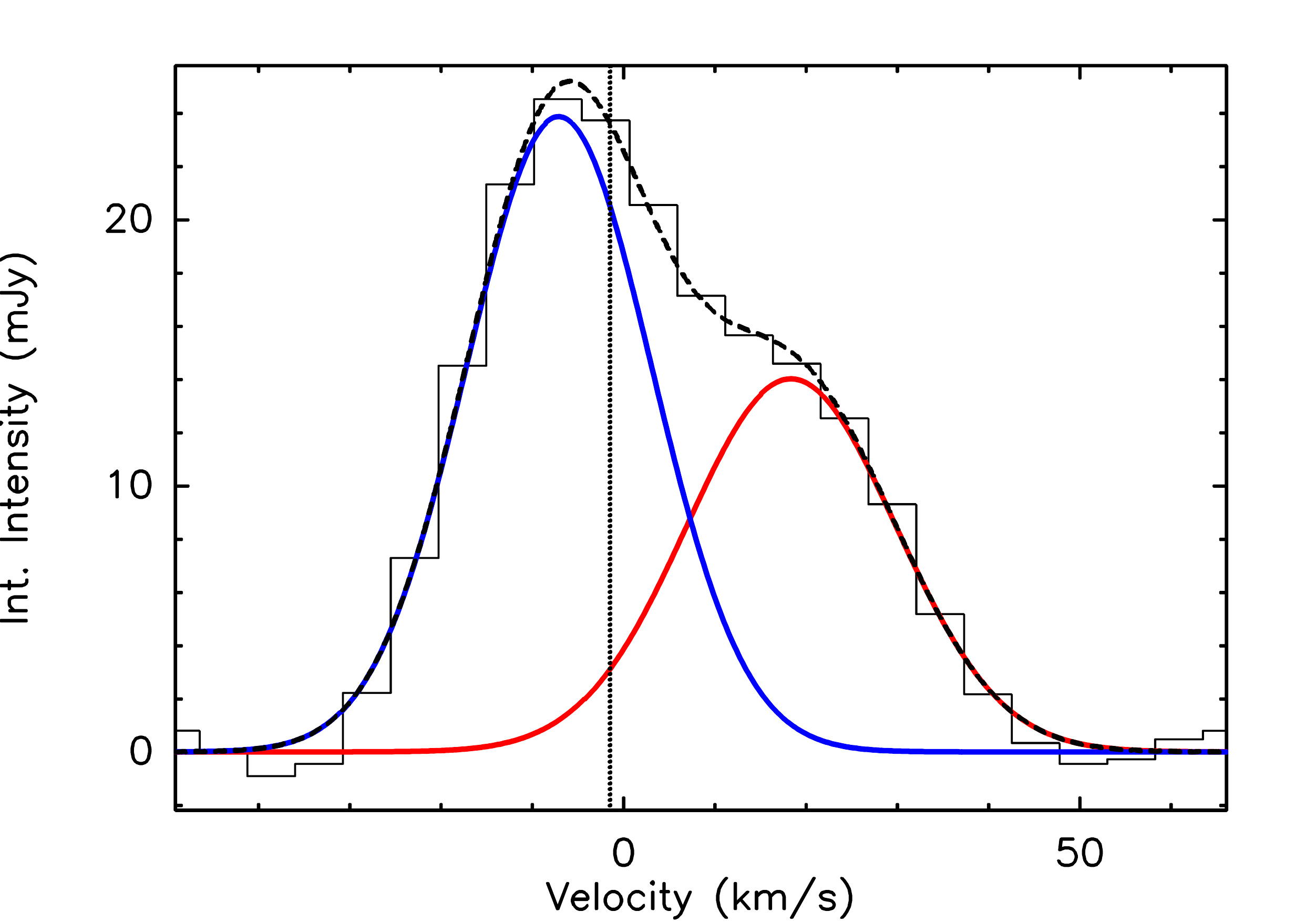}}
	\subfloat[POS30]{\includegraphics[angle=270,width=7.5cm]{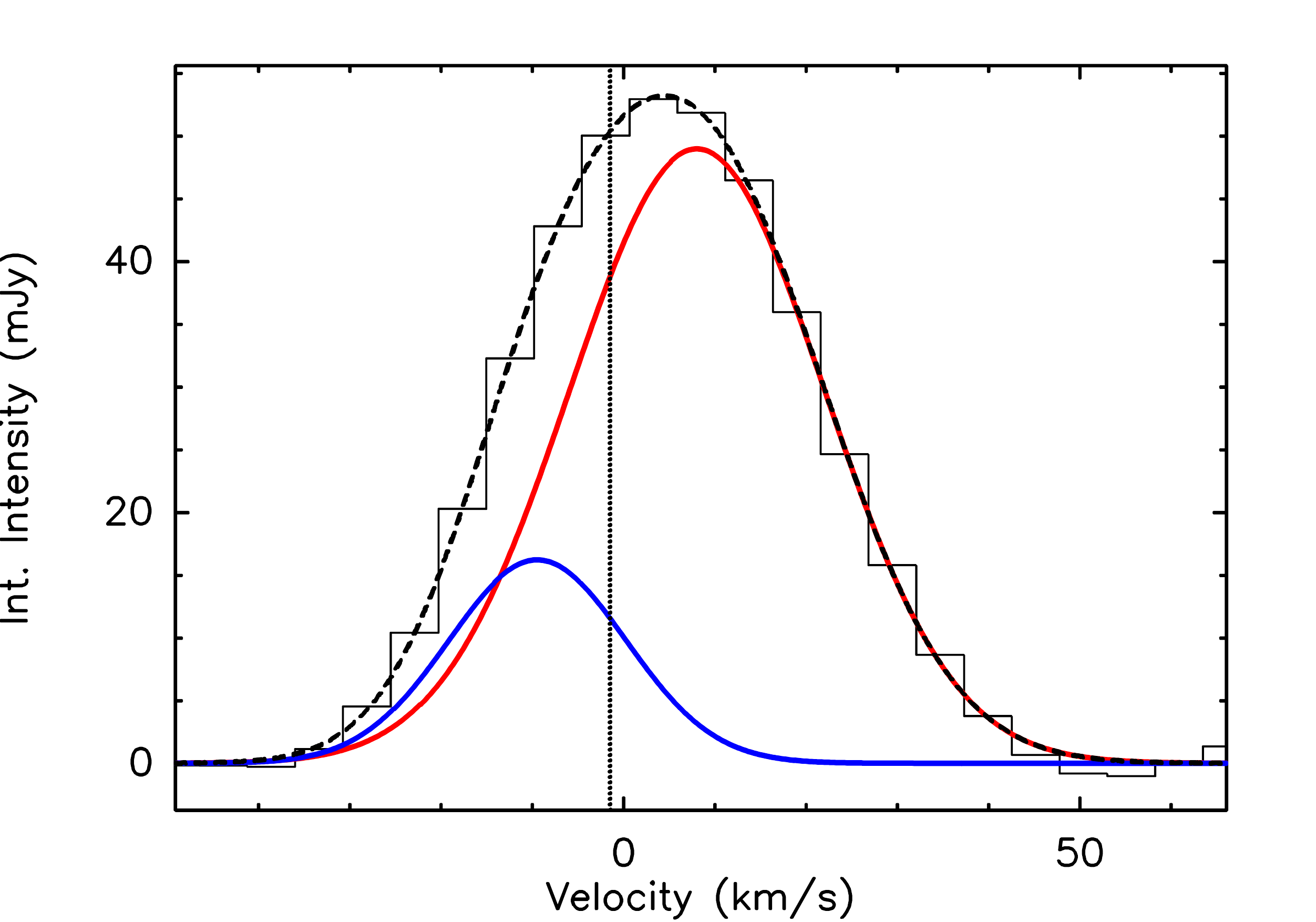}}\\	
	\subfloat[POS31]{\includegraphics[angle=270,width=7.5cm]{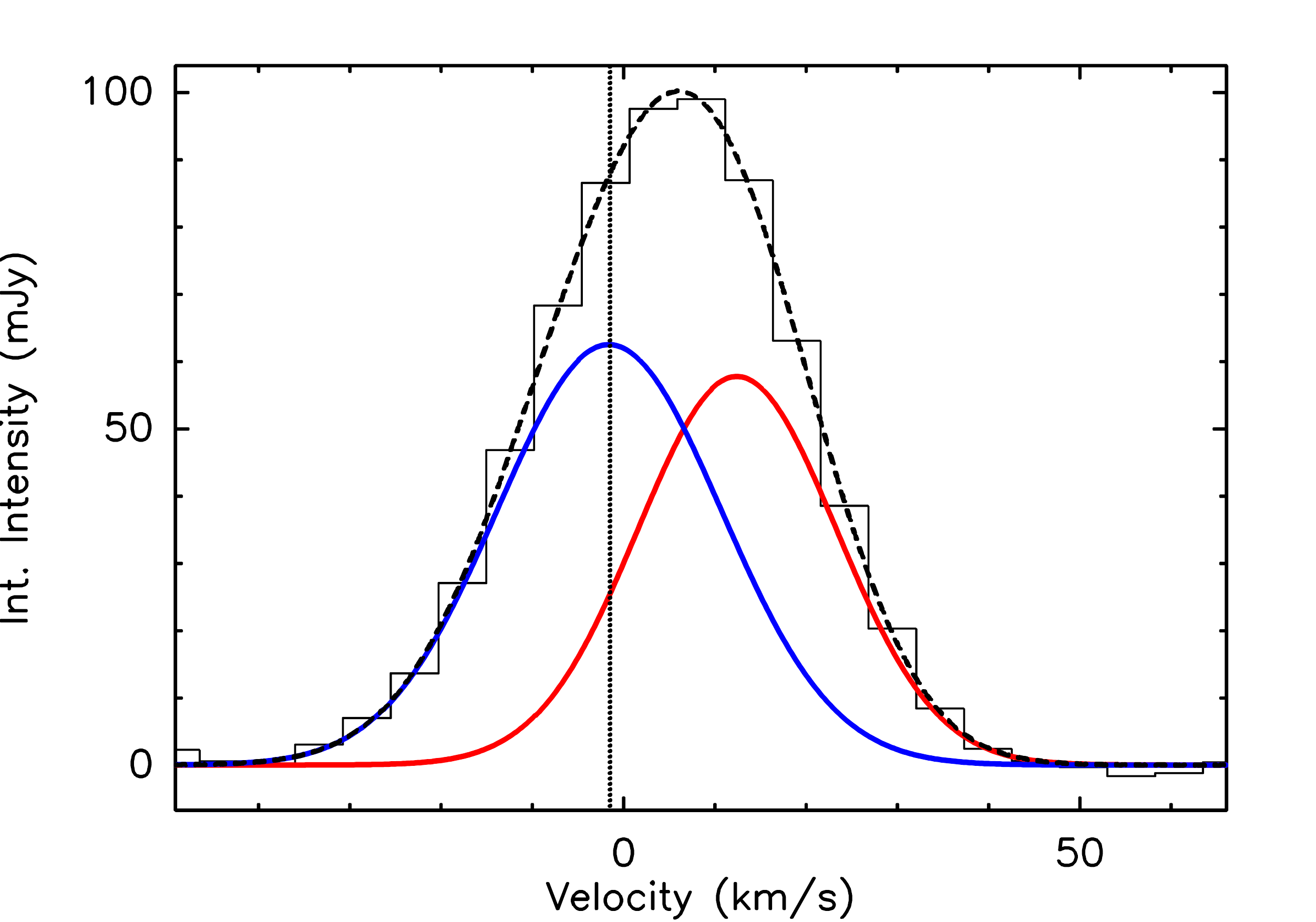}}
	\subfloat[POS32]{\includegraphics[angle=270,width=7.5cm]{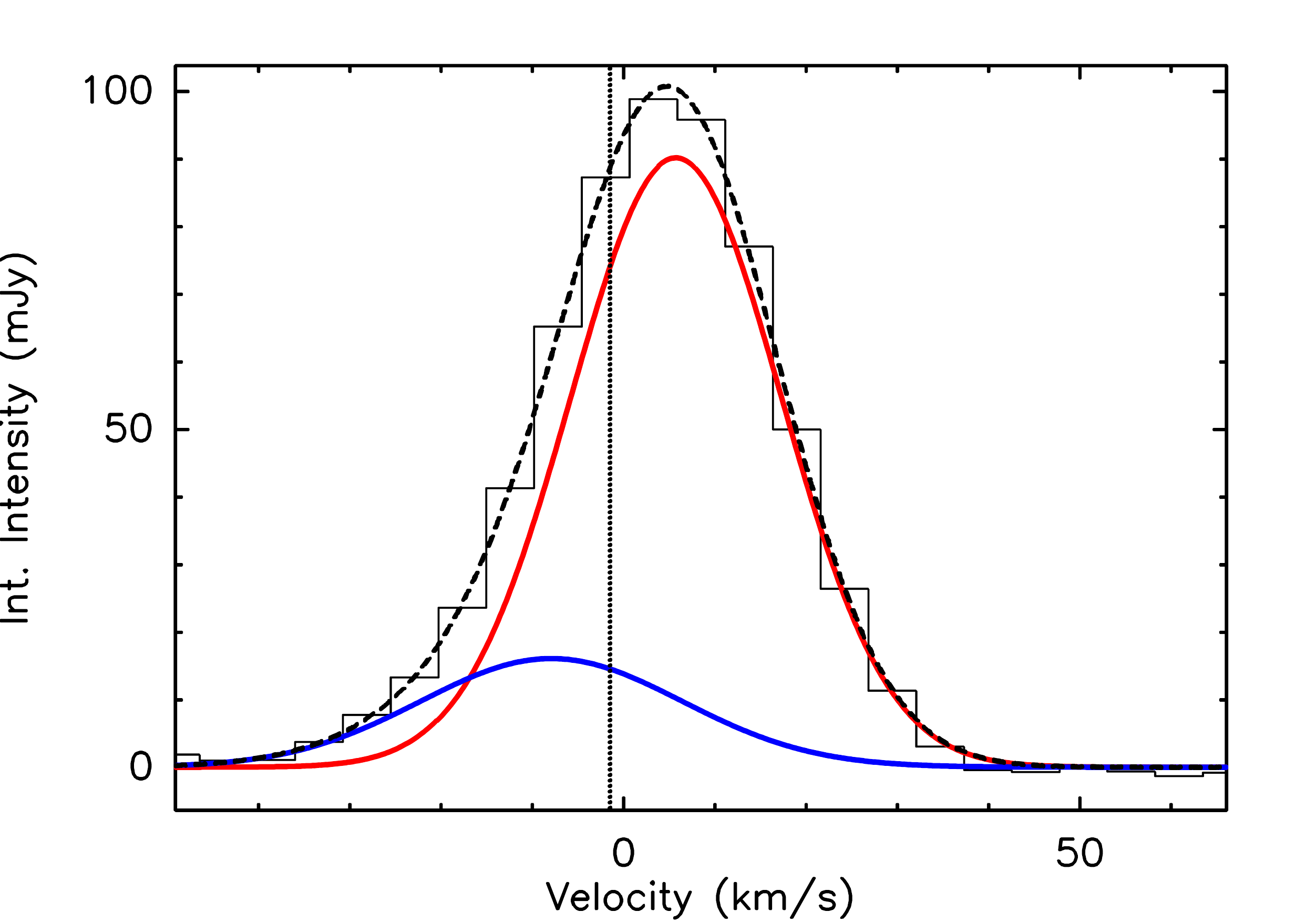}}
\end{figure*}

\addtocounter{figure}{-1}
\begin{figure*}
	\centering		
	\caption{Continued}
	\subfloat[POS33]{\includegraphics[angle=270,width=7.5cm]{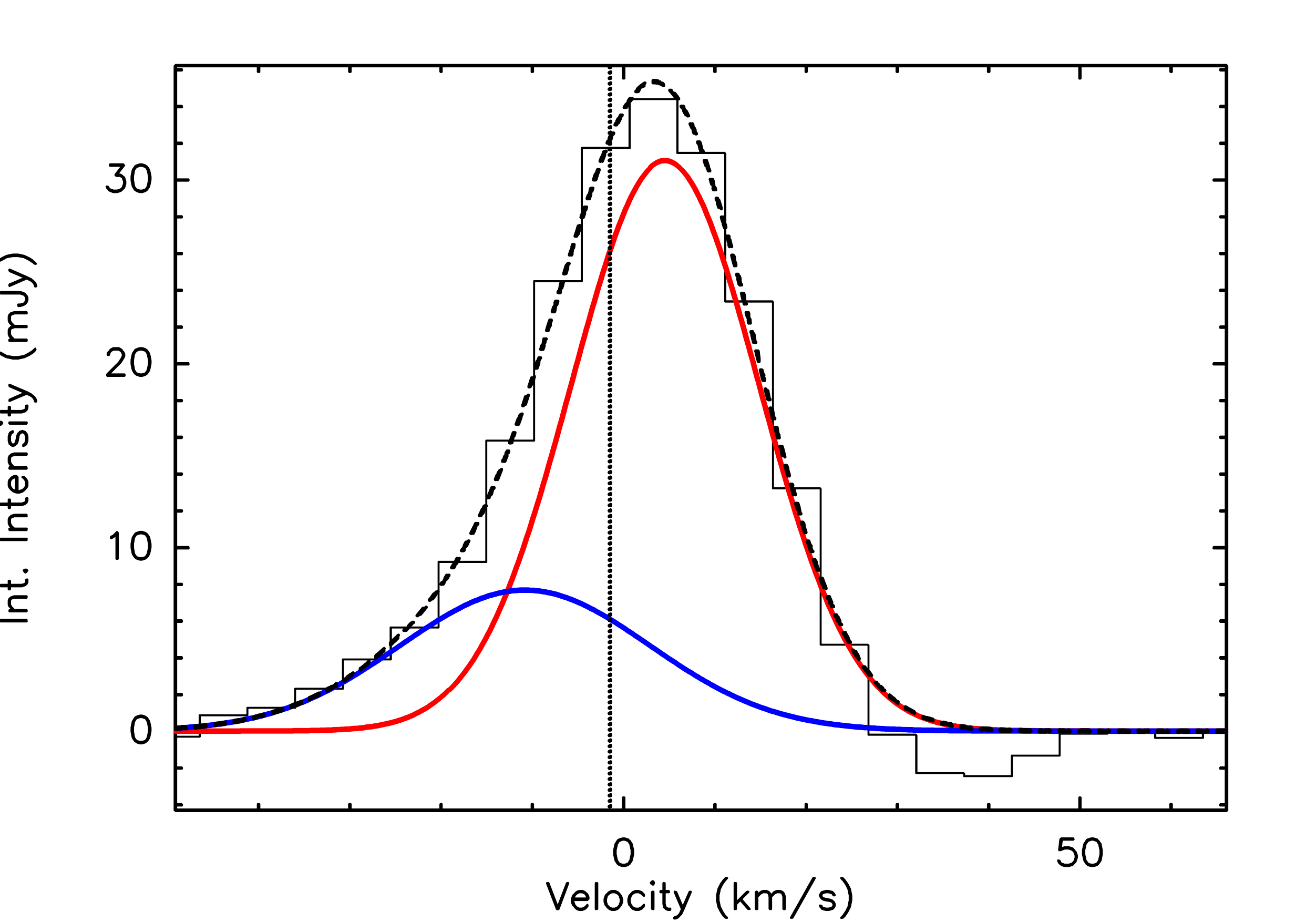}}	
	\subfloat[POS34]{\includegraphics[angle=270,width=7.5cm]{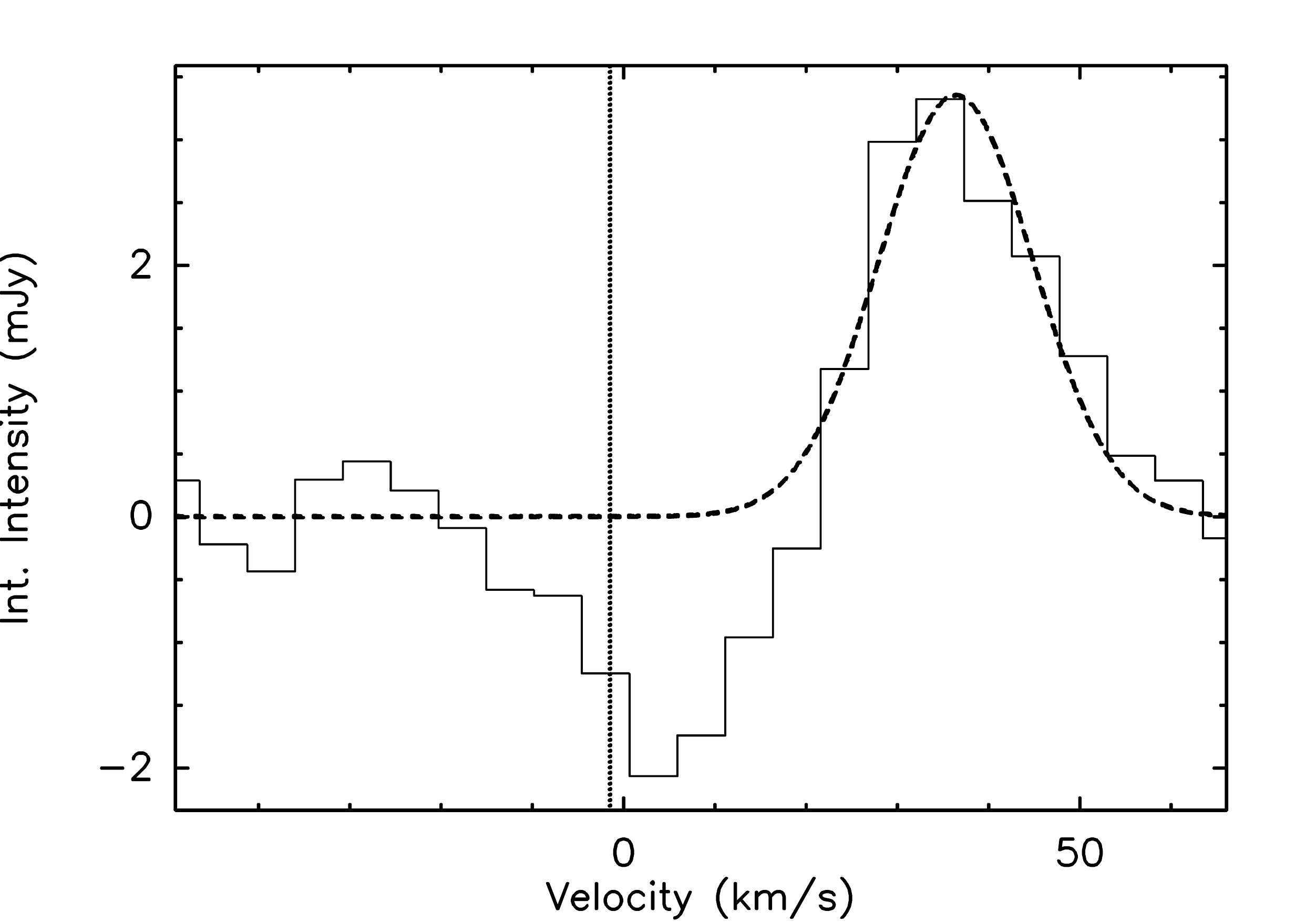}}\\
	\subfloat[POS35]{\includegraphics[angle=270,width=7.5cm]{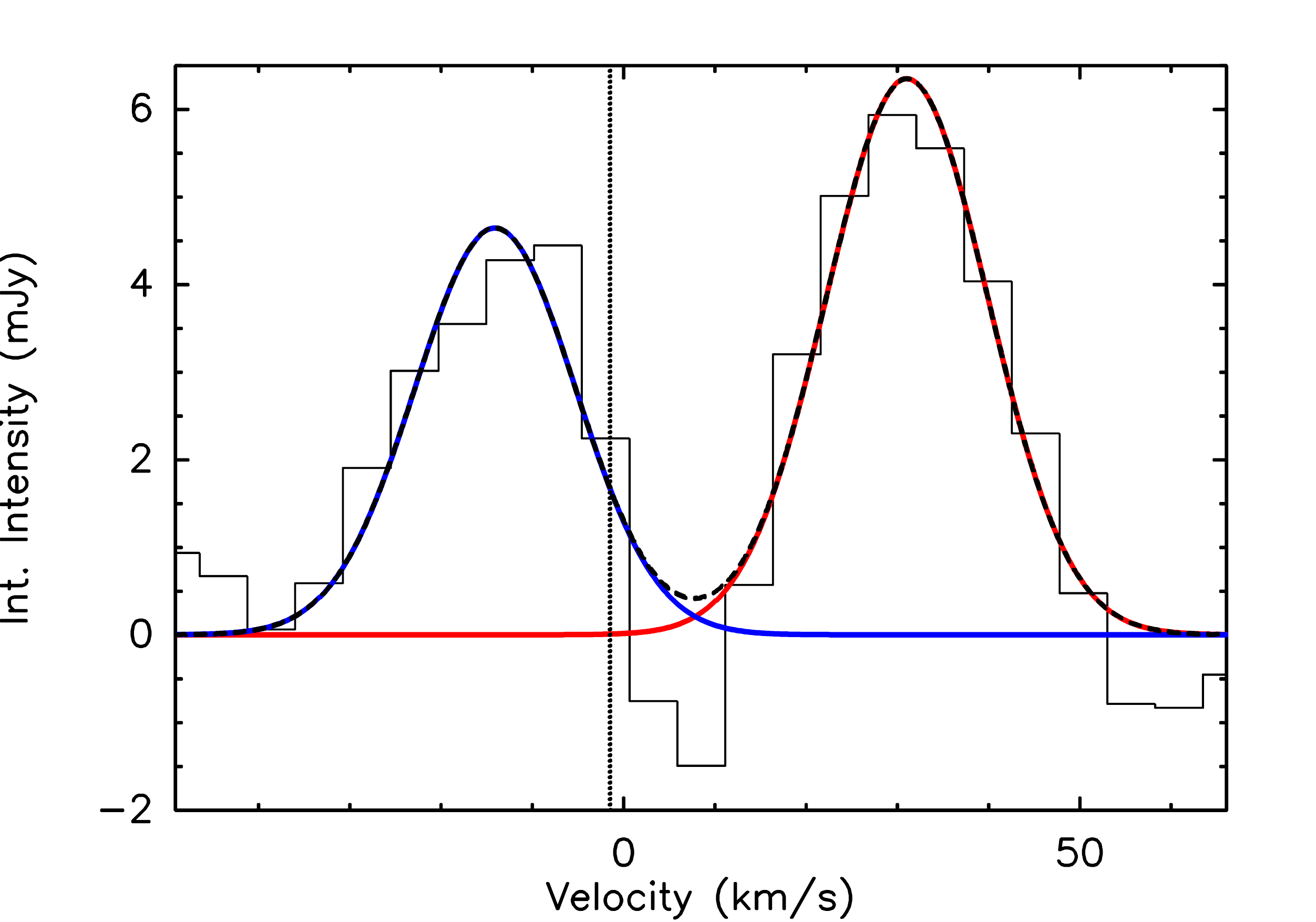}}
	\subfloat[POS36]{\includegraphics[angle=270,width=7.5cm]{POS36.pdf}}\\	
	\subfloat[POS37]{\includegraphics[angle=270,width=7.5cm]{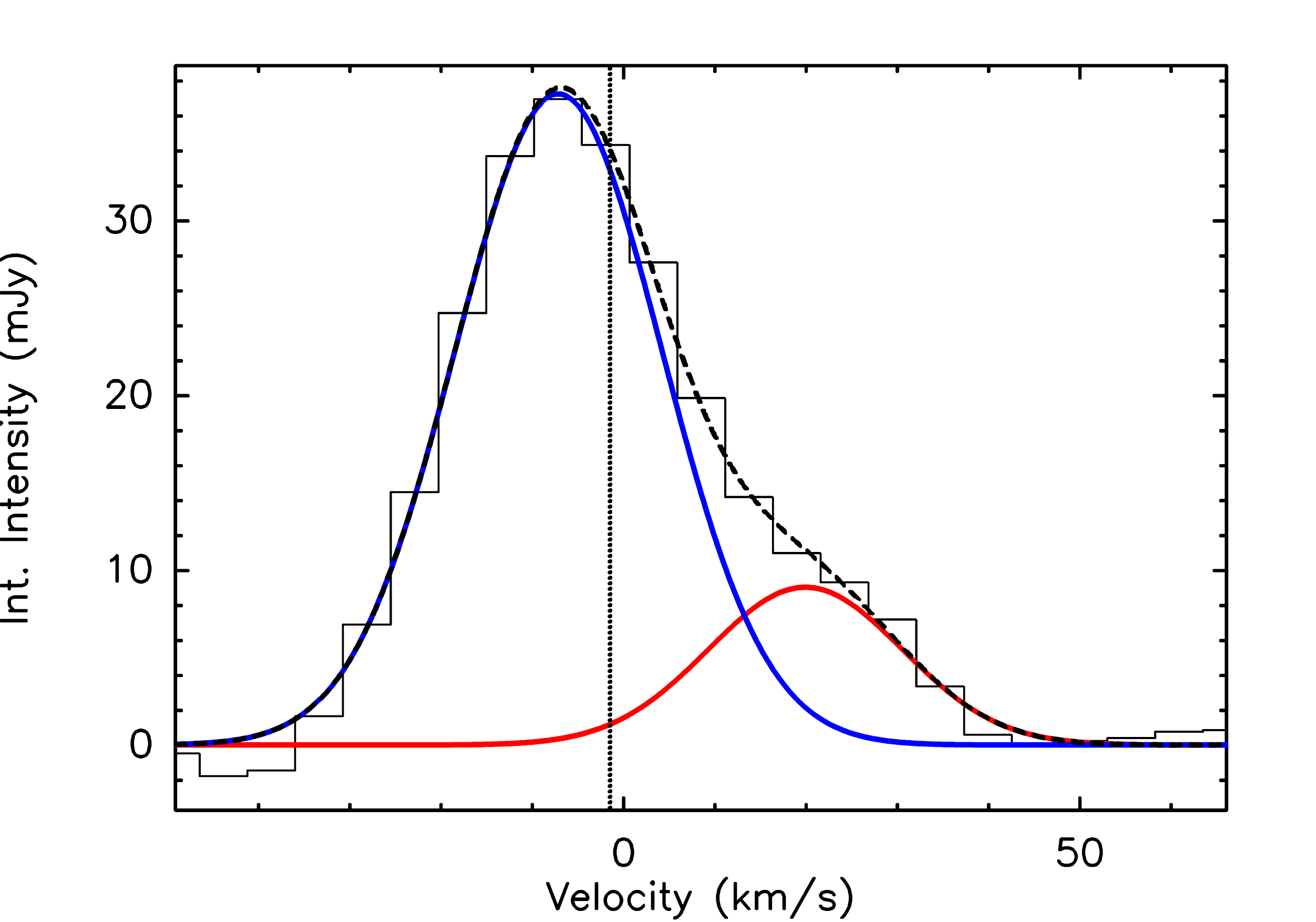}}
	\subfloat[POS38]{\includegraphics[angle=270,width=7.5cm]{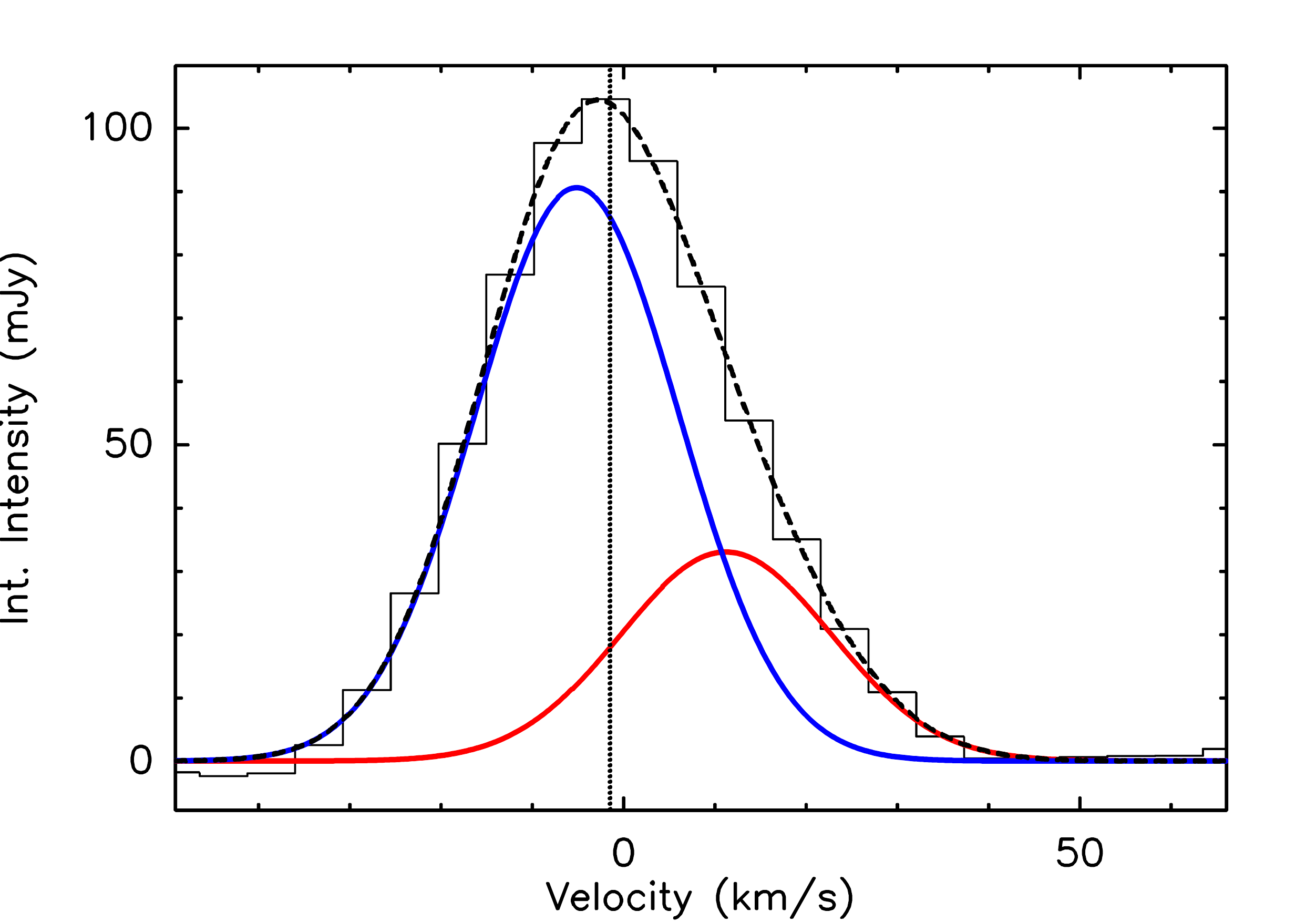}}\\	
	\subfloat[POS39]{\includegraphics[angle=270,width=7.5cm]{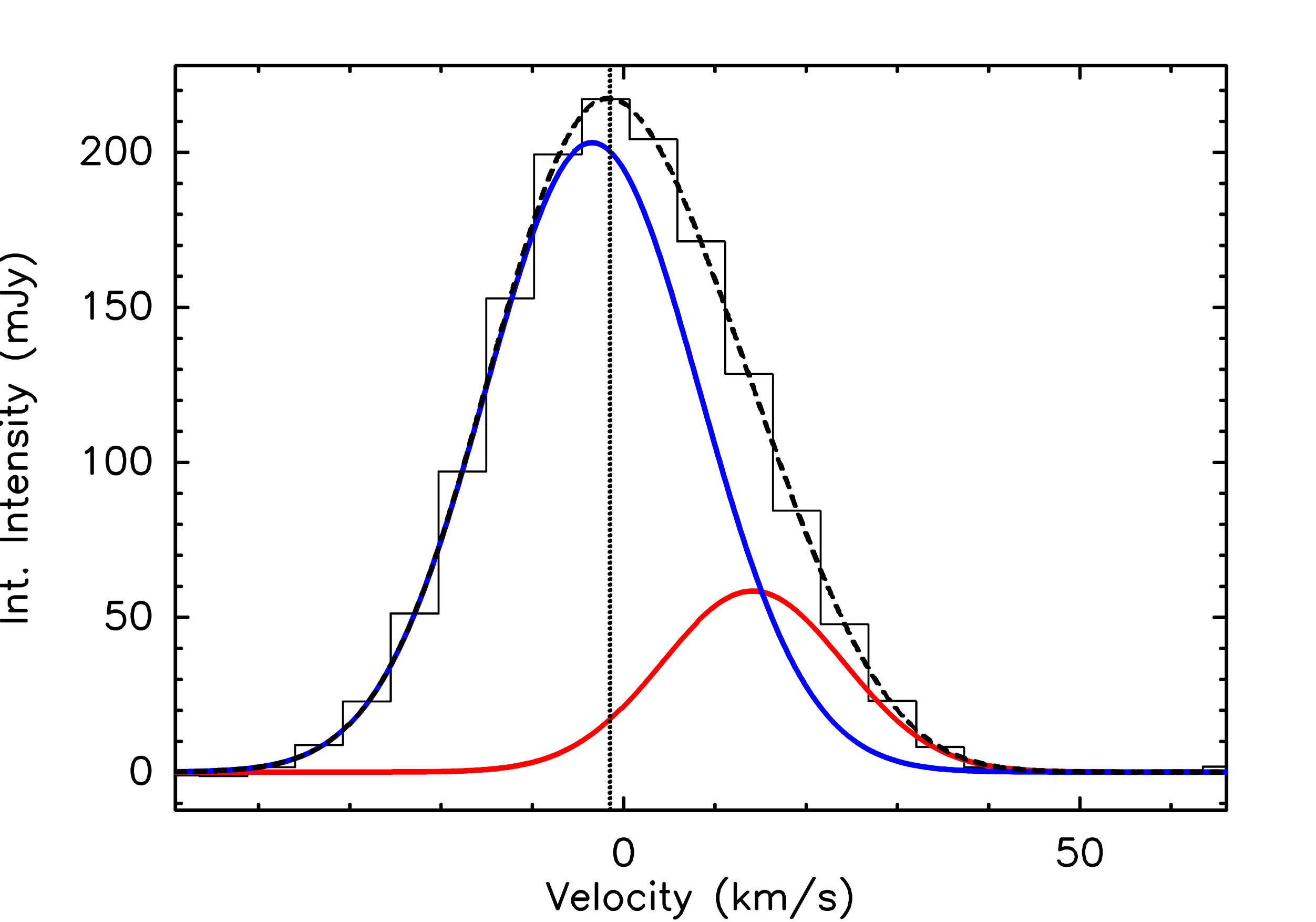}}
	\subfloat[POS40]{\includegraphics[angle=270,width=7.5cm]{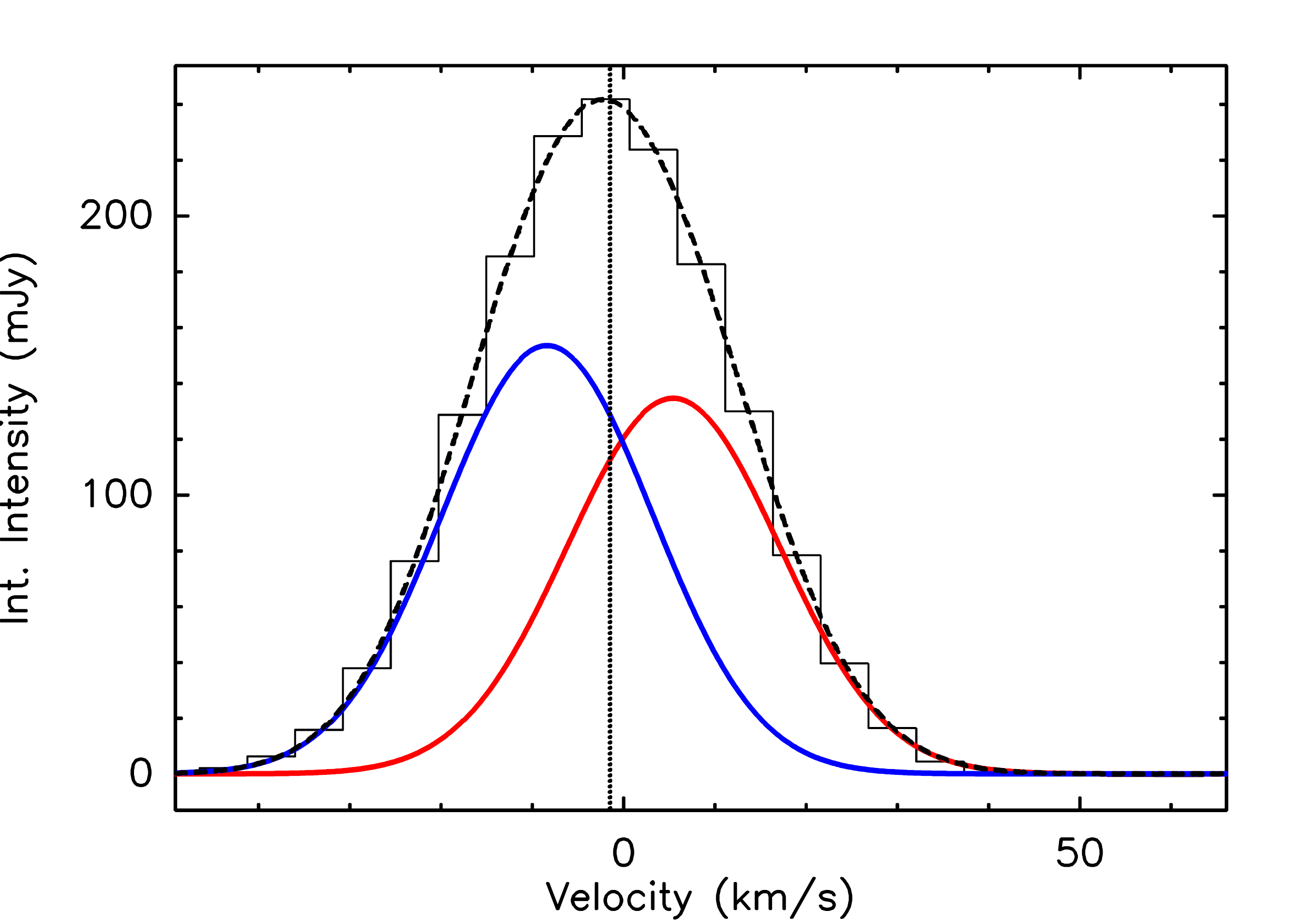}}
\end{figure*}

\addtocounter{figure}{-1}
\begin{figure*}
	\centering			
	\caption{Continued}
	\subfloat[POS41]{\includegraphics[angle=270,width=7.5cm]{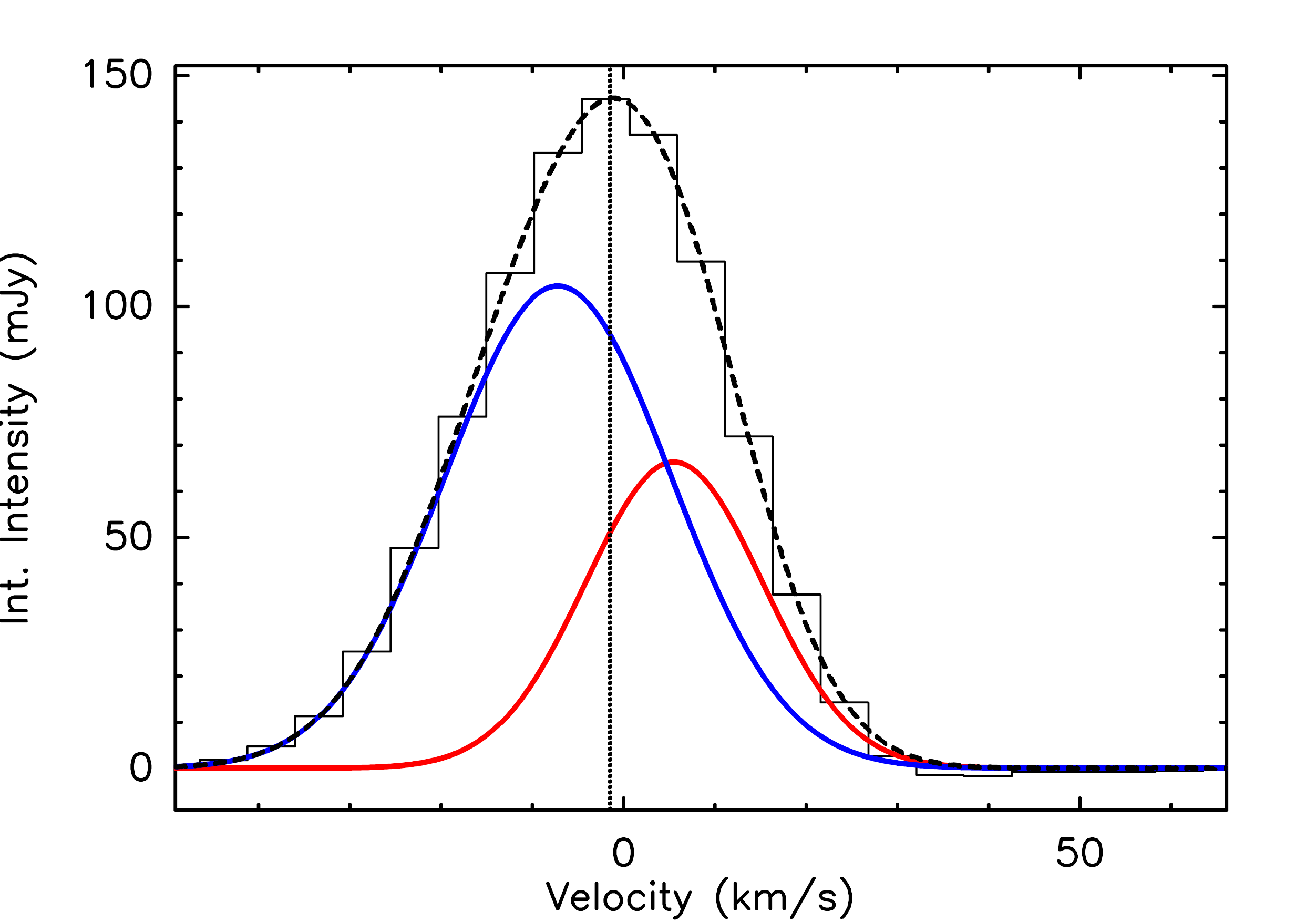}}	
	\subfloat[POS42]{\includegraphics[angle=270,width=7.5cm]{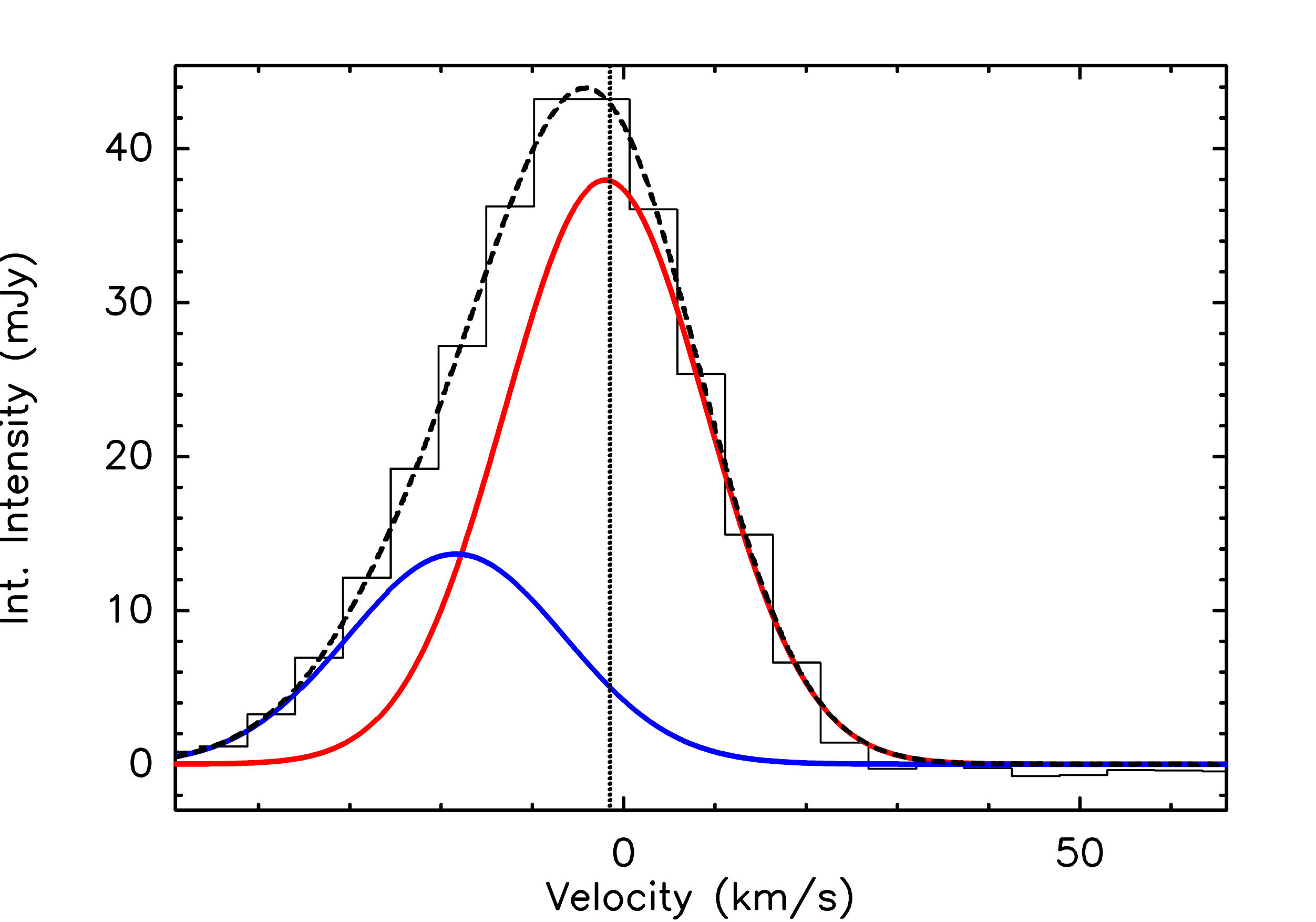}}\\
	\subfloat[POS43]{\includegraphics[angle=270,width=7.5cm]{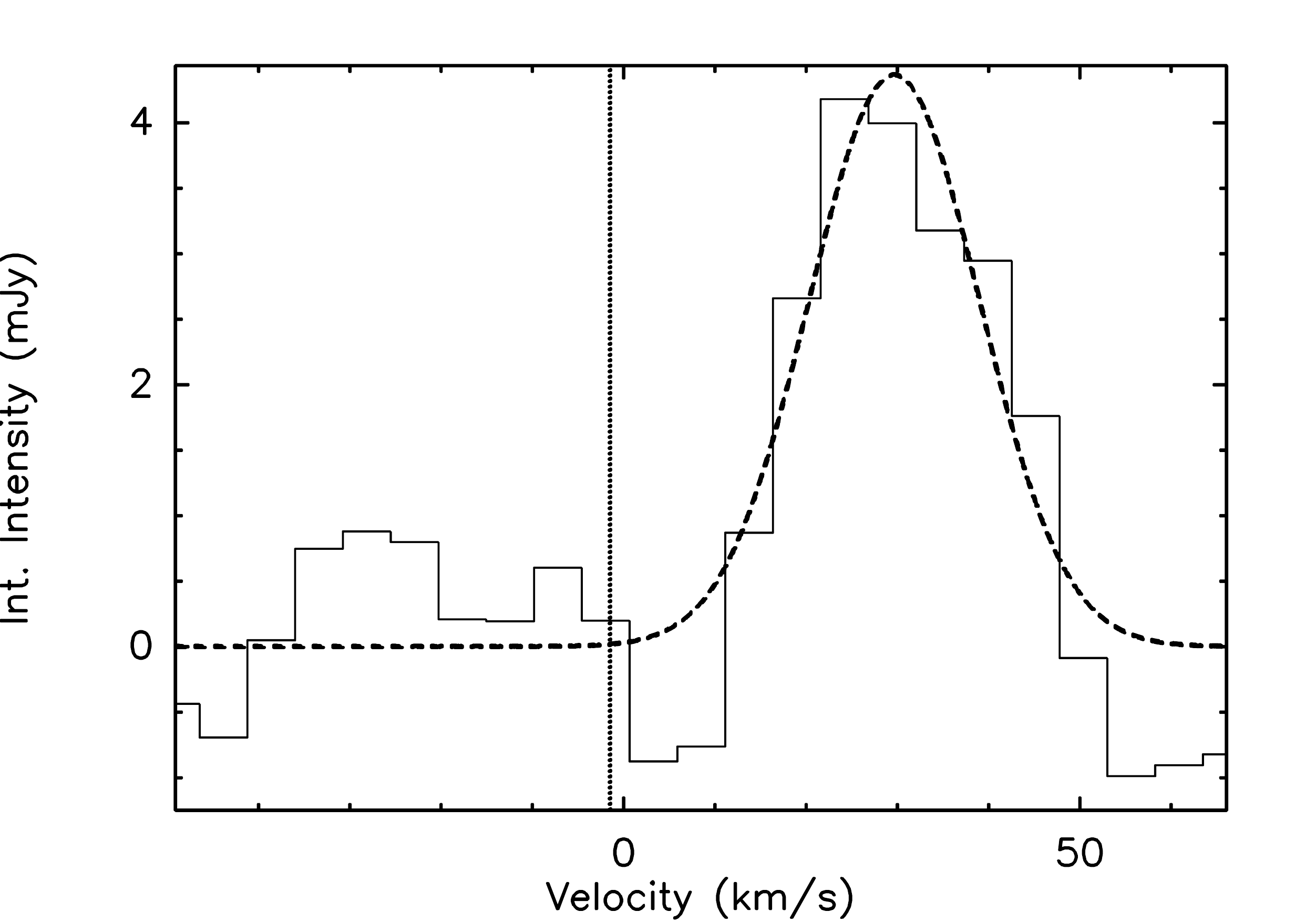}}
	\subfloat[POS44]{\includegraphics[angle=270,width=7.5cm]{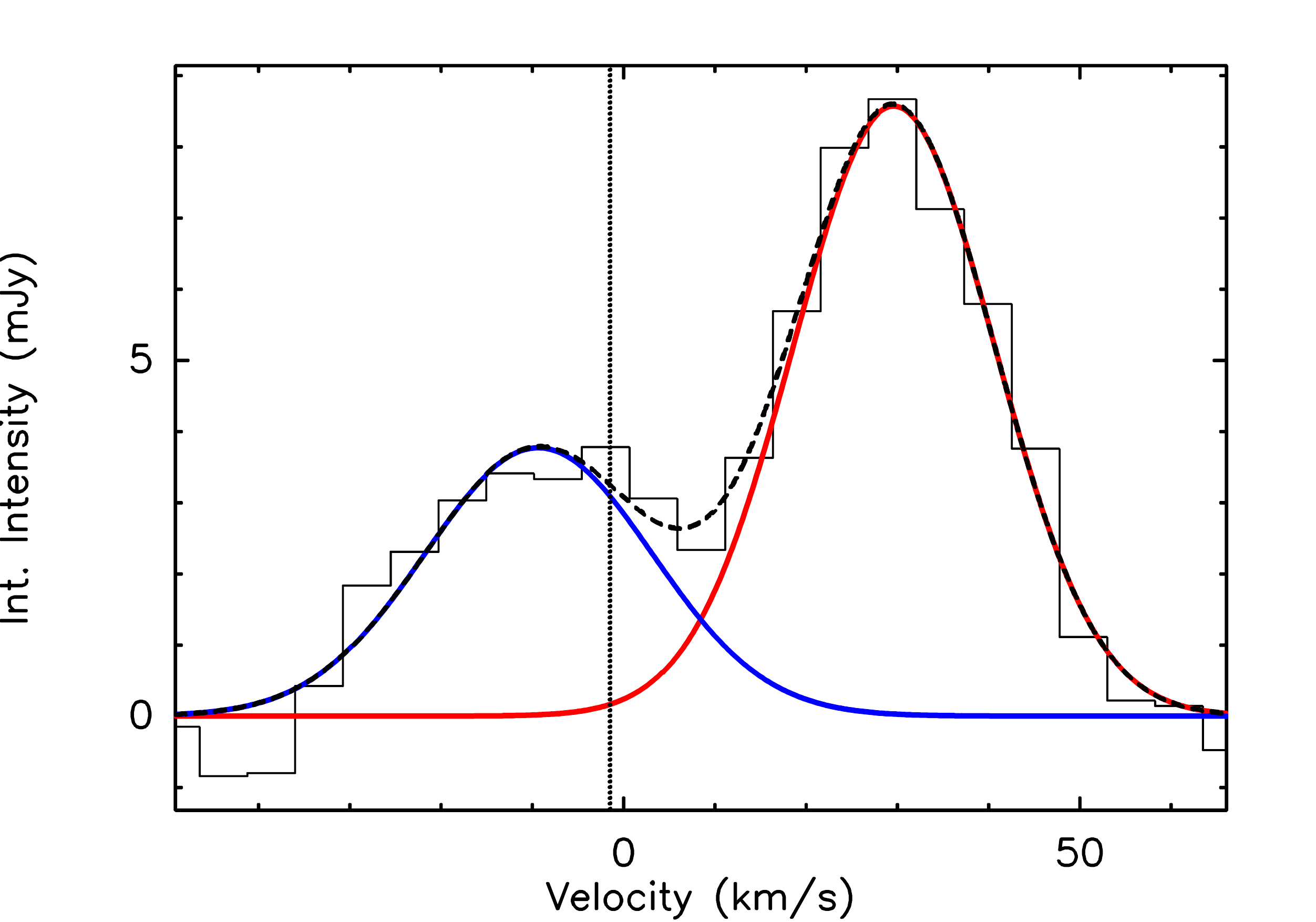}}\\	
	\subfloat[POS45]{\includegraphics[angle=270,width=7.5cm]{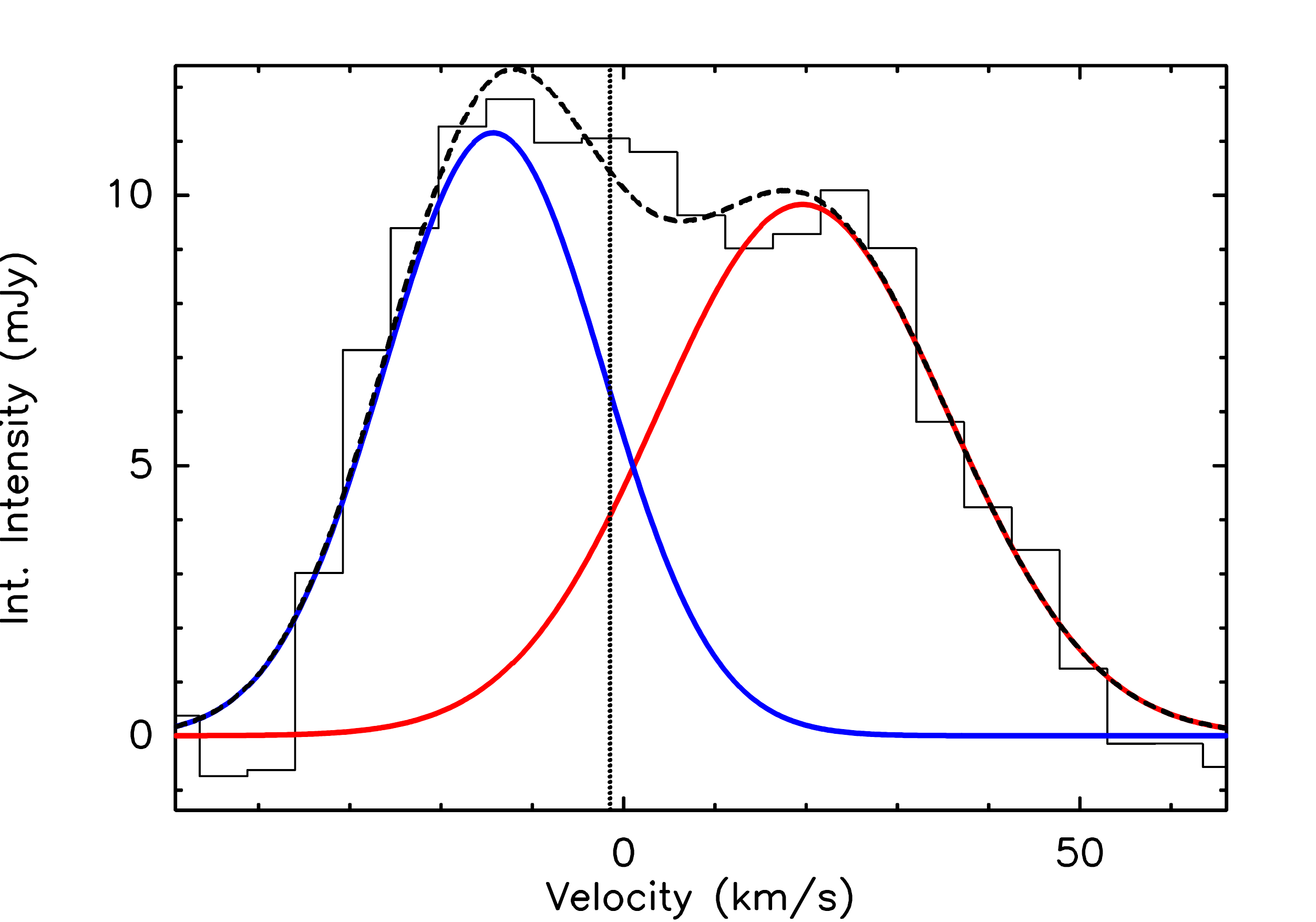}}
	\subfloat[POS46]{\includegraphics[angle=270,width=7.5cm]{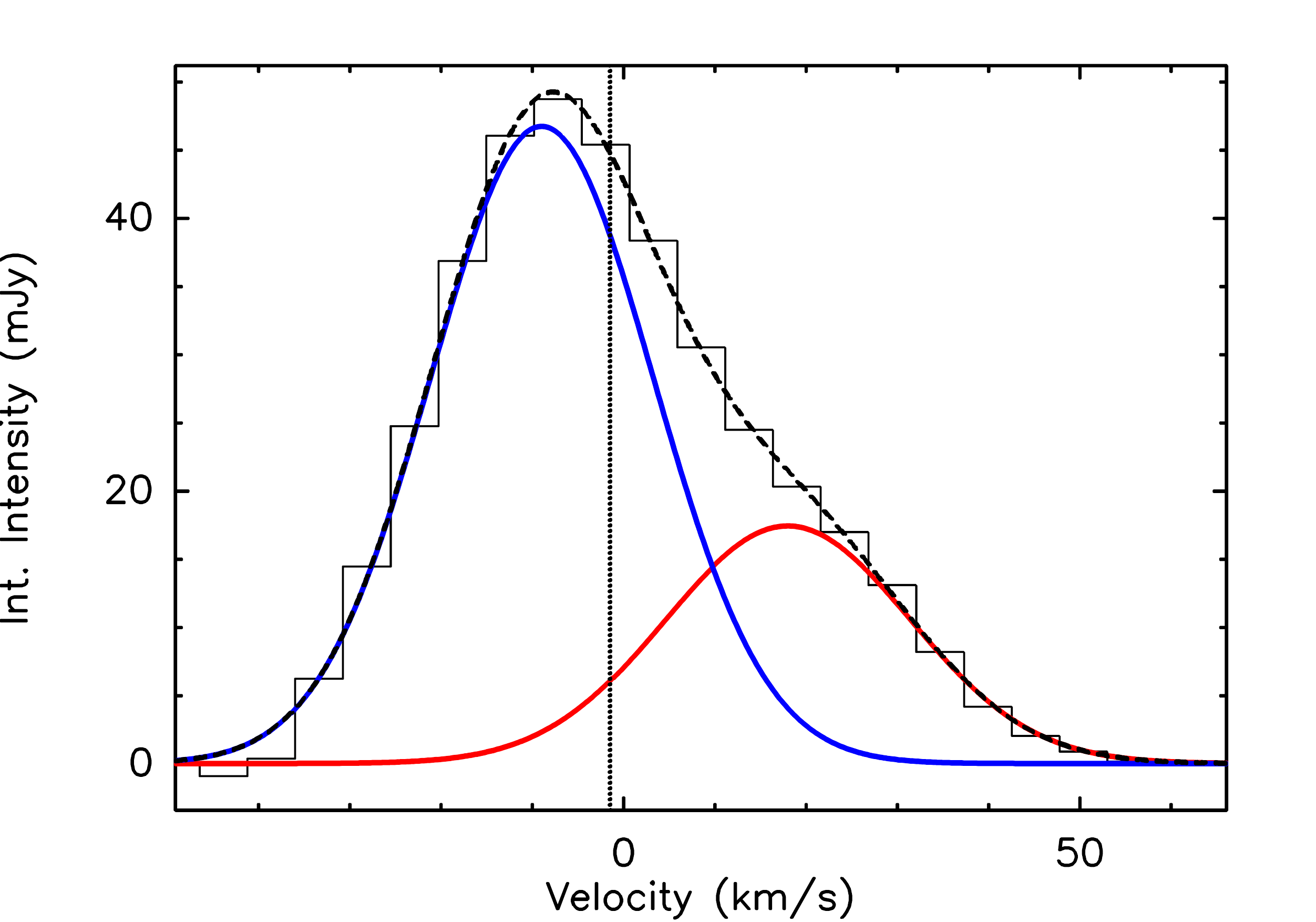}}\\	
	\subfloat[POS47]{\includegraphics[angle=270,width=7.5cm]{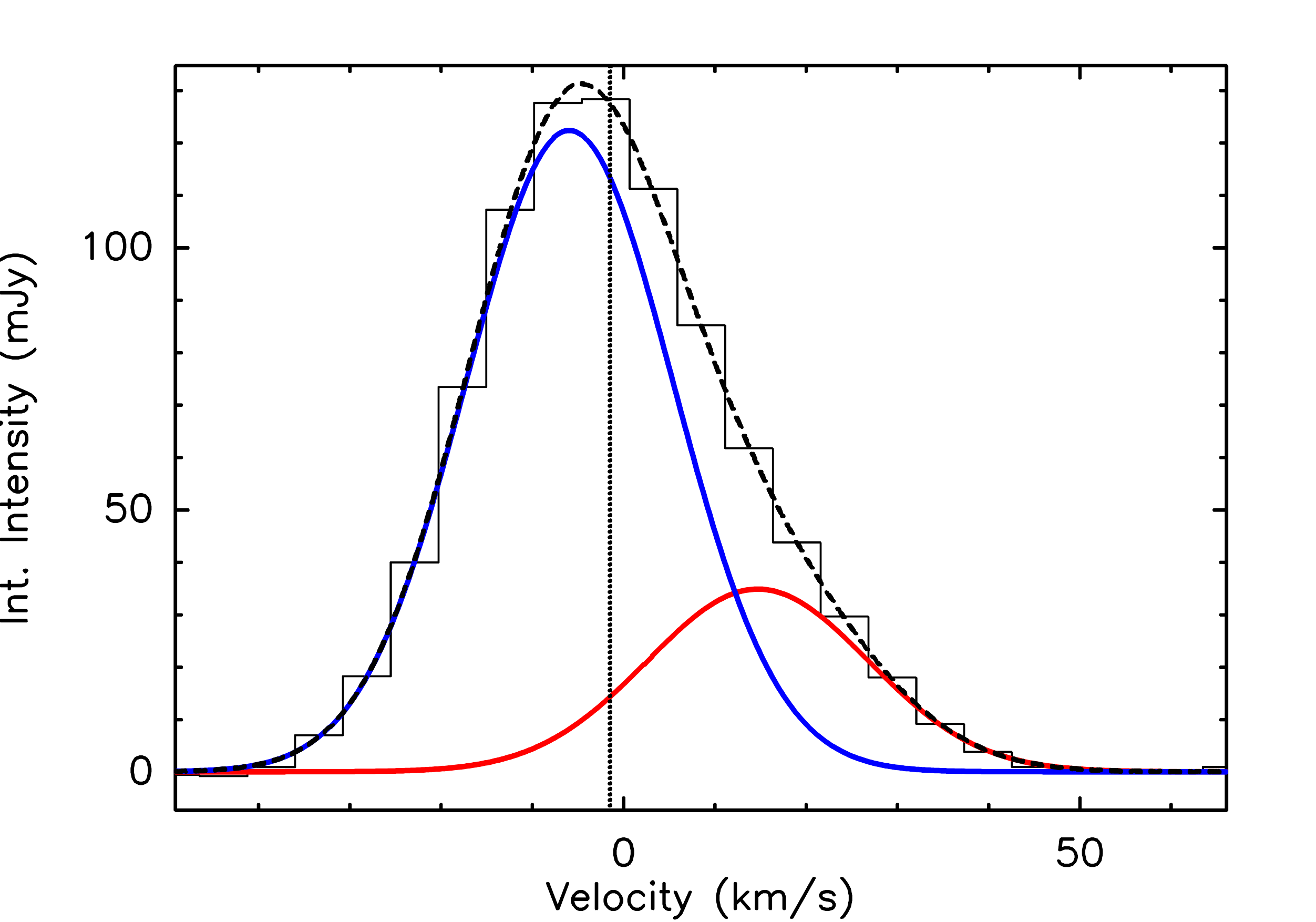}}
	\subfloat[POS48]{\includegraphics[angle=270,width=7.5cm]{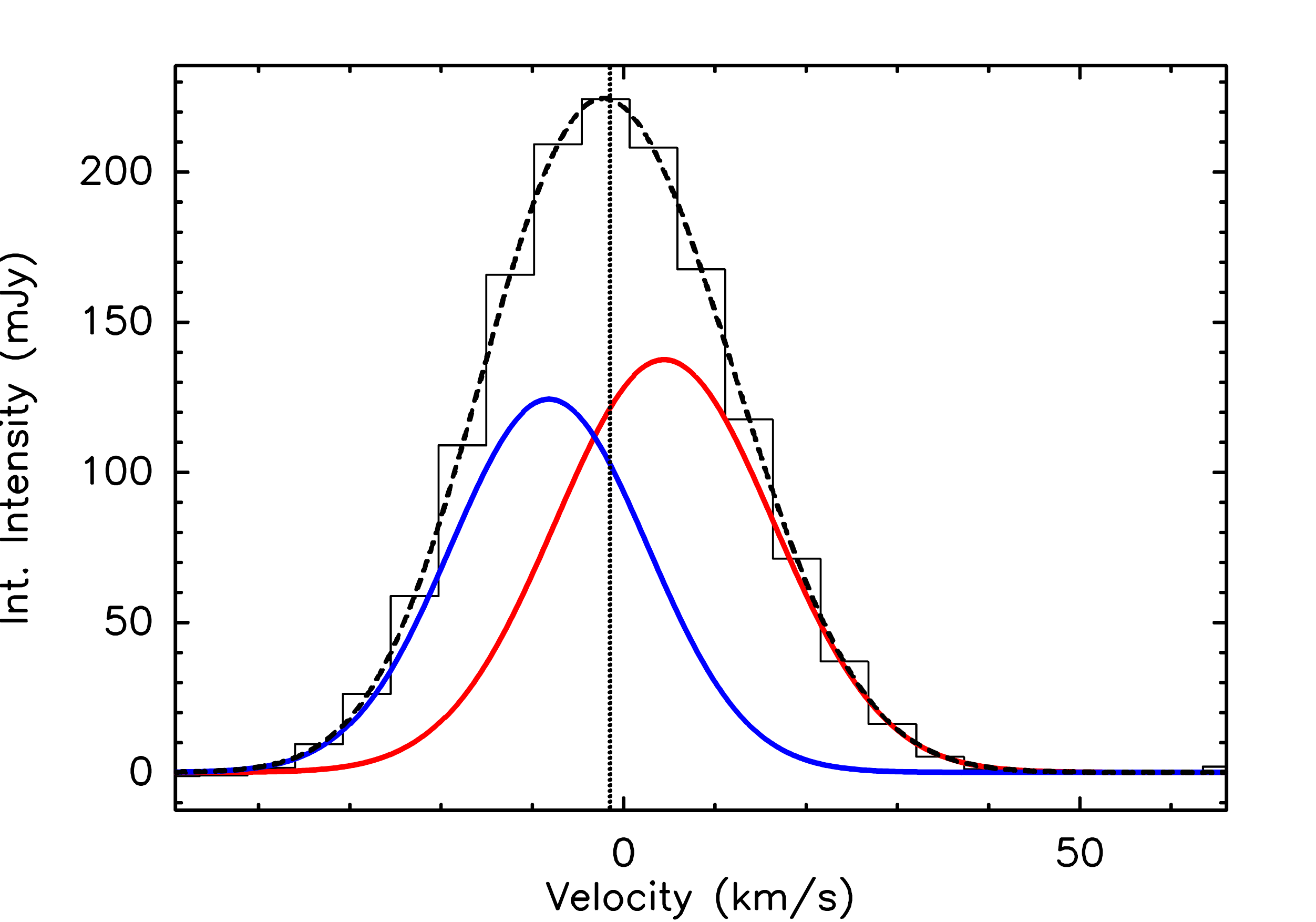}}
\end{figure*}

\addtocounter{figure}{-1}
\begin{figure*}
	\centering
	\caption{Continued}
	\subfloat[POS49]{\includegraphics[angle=270,width=7.5cm]{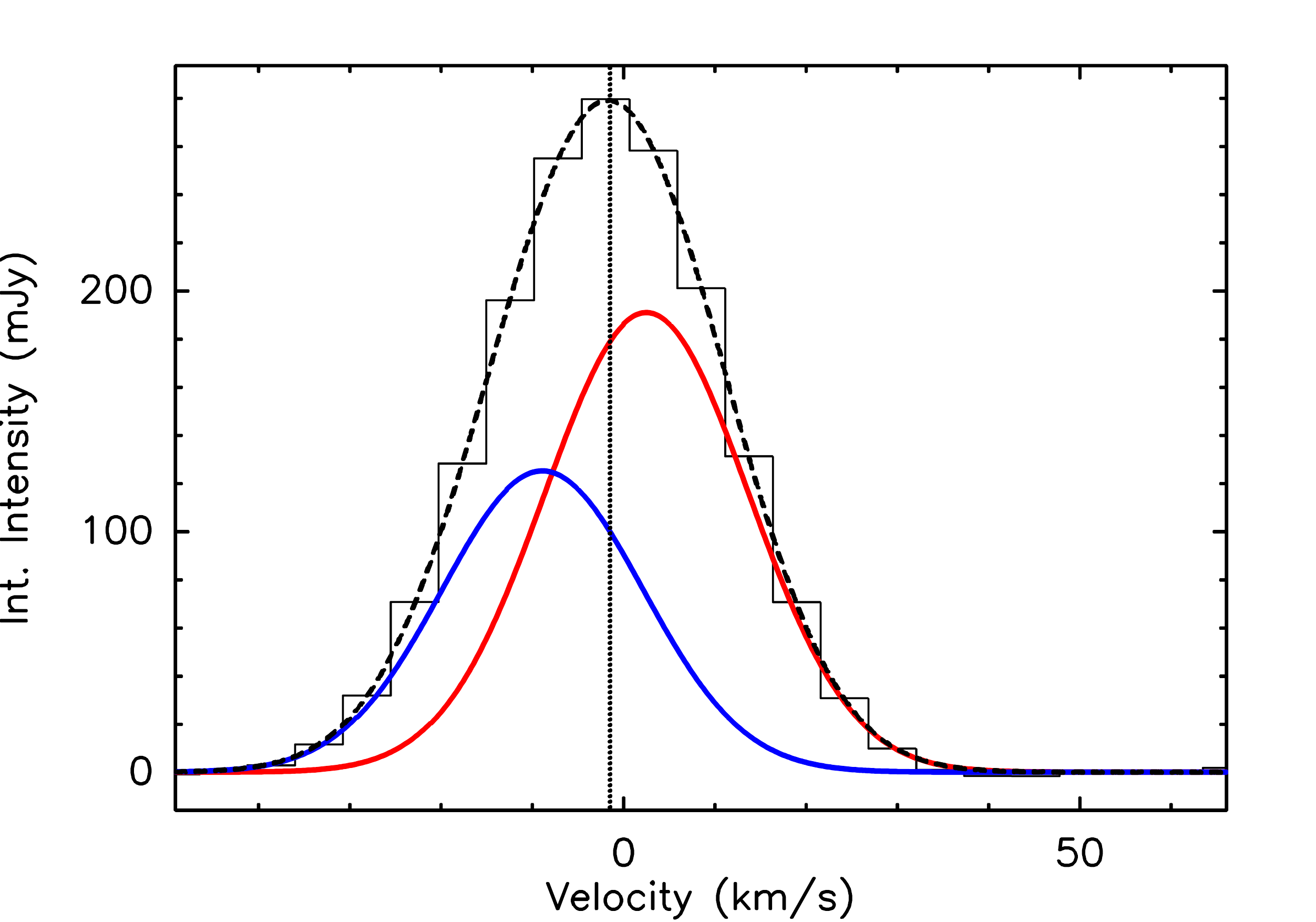}}	
	\subfloat[POS50]{\includegraphics[angle=270,width=7.5cm]{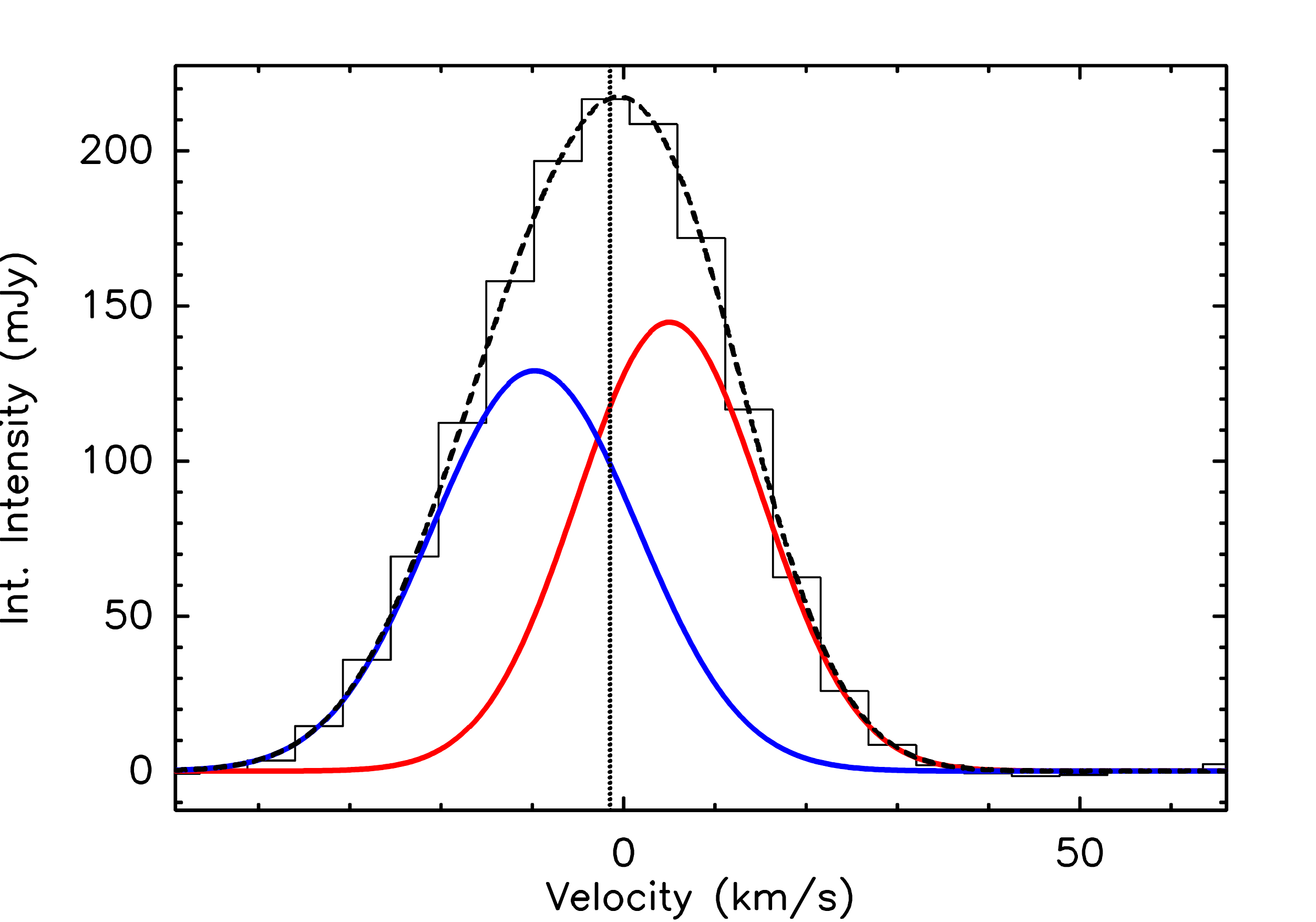}}\\
	\subfloat[POS51]{\includegraphics[angle=270,width=7.5cm]{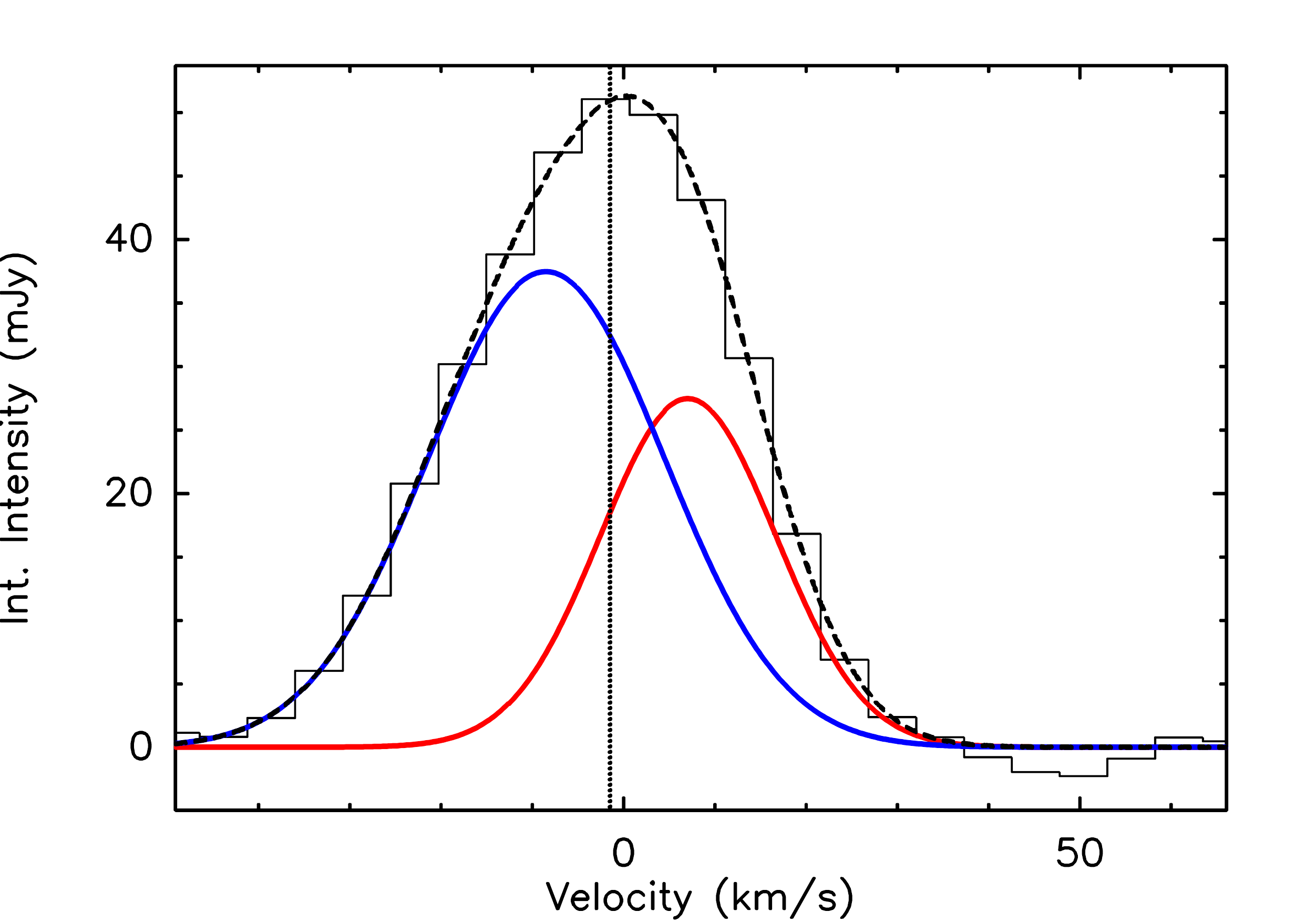}}
	\subfloat[POS52]{\includegraphics[angle=270,width=7.5cm]{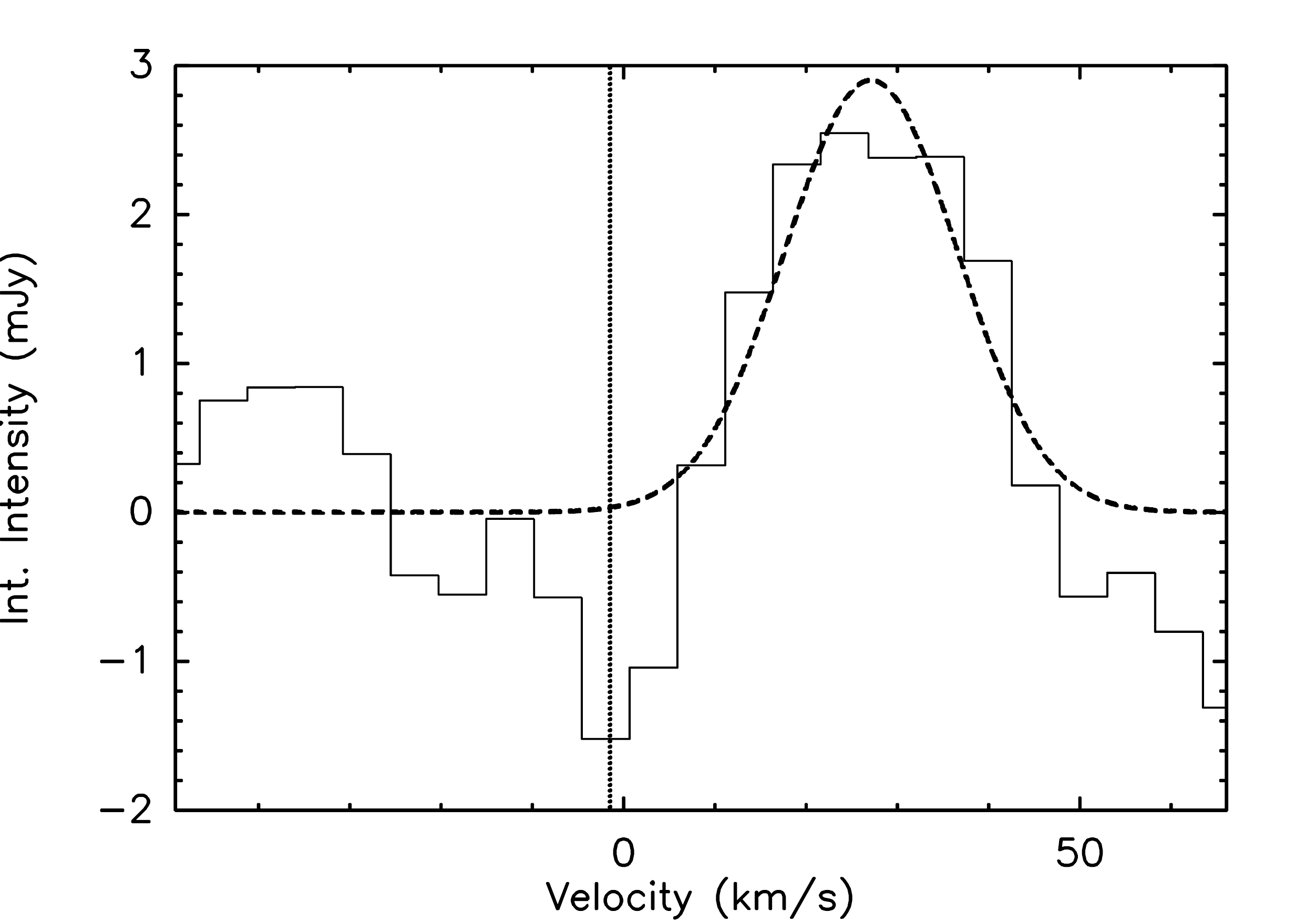}}\\	
	\subfloat[POS53]{\includegraphics[angle=270,width=7.5cm]{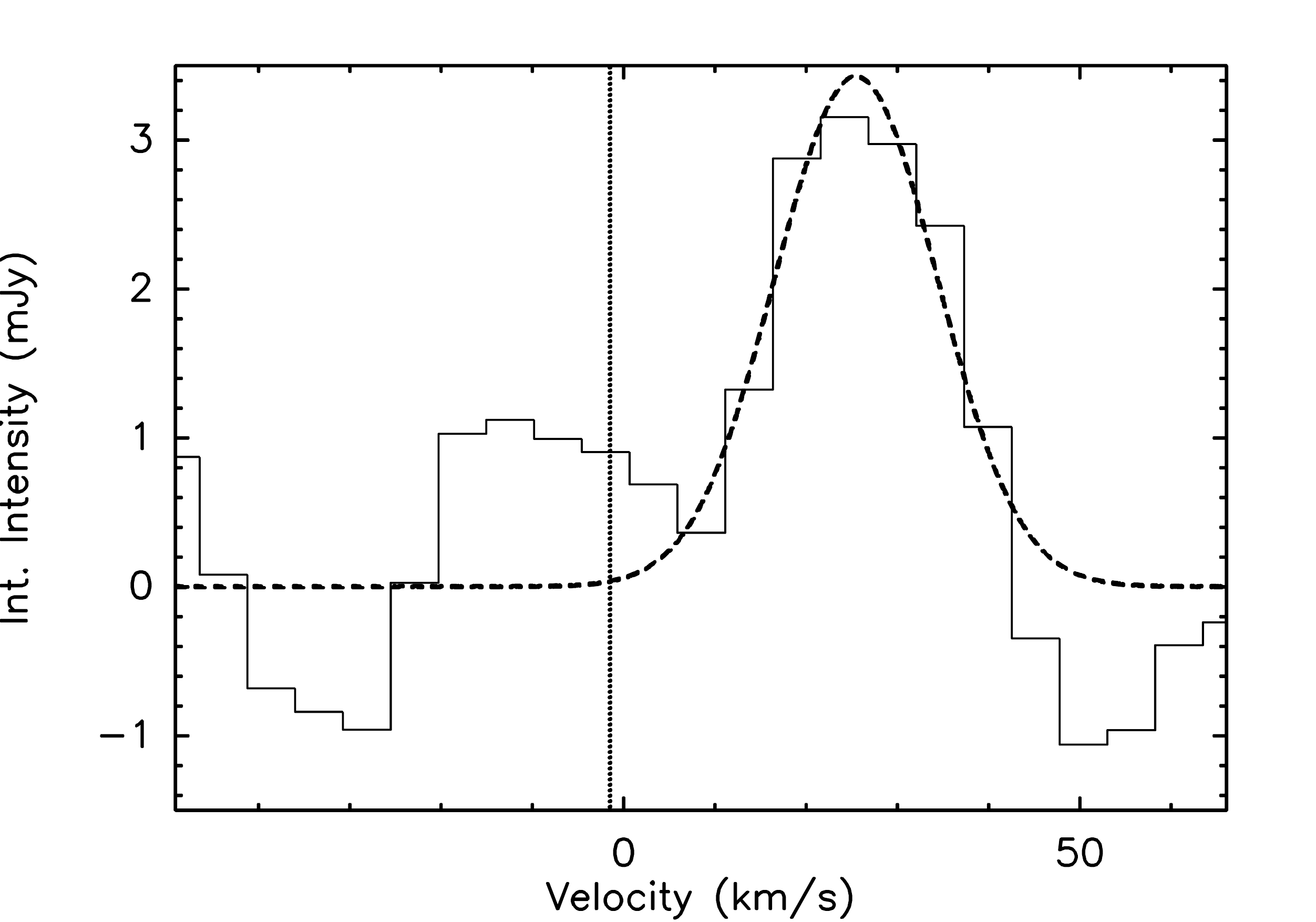}}
	\subfloat[POS54]{\includegraphics[angle=270,width=7.5cm]{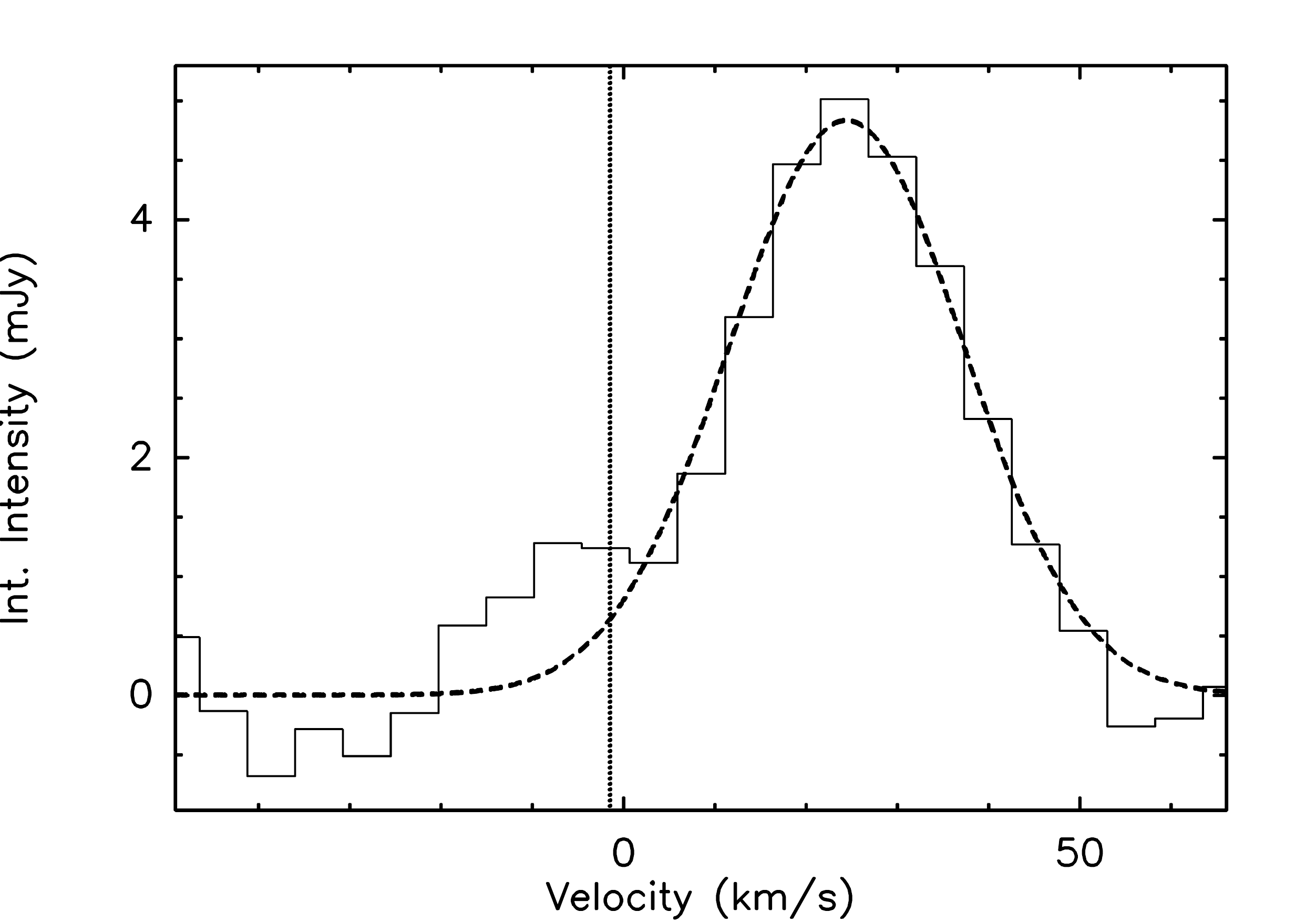}}\\	
	\subfloat[POS55]{\includegraphics[angle=270,width=7.5cm]{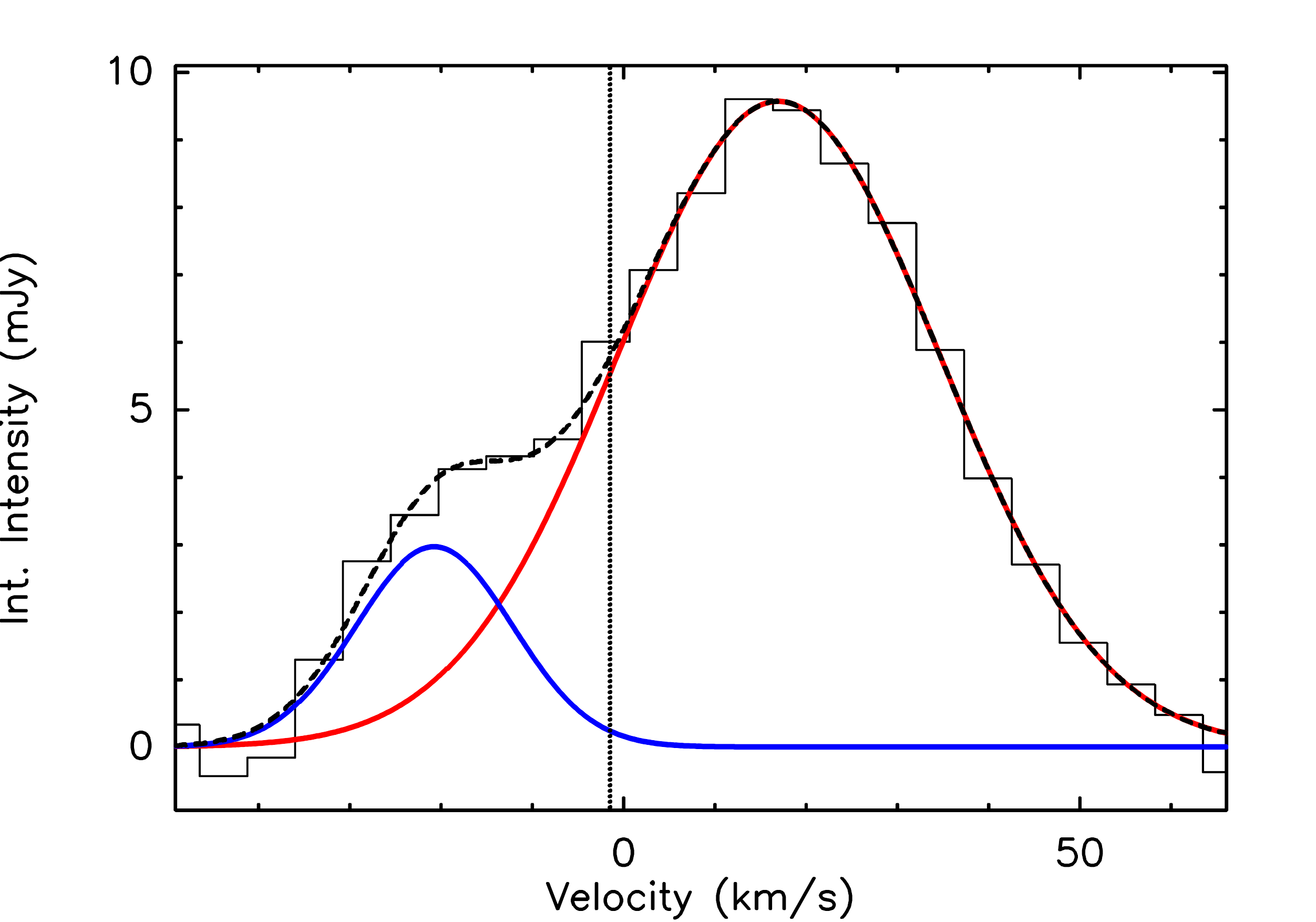}}
	\subfloat[POS56]{\includegraphics[angle=270,width=7.5cm]{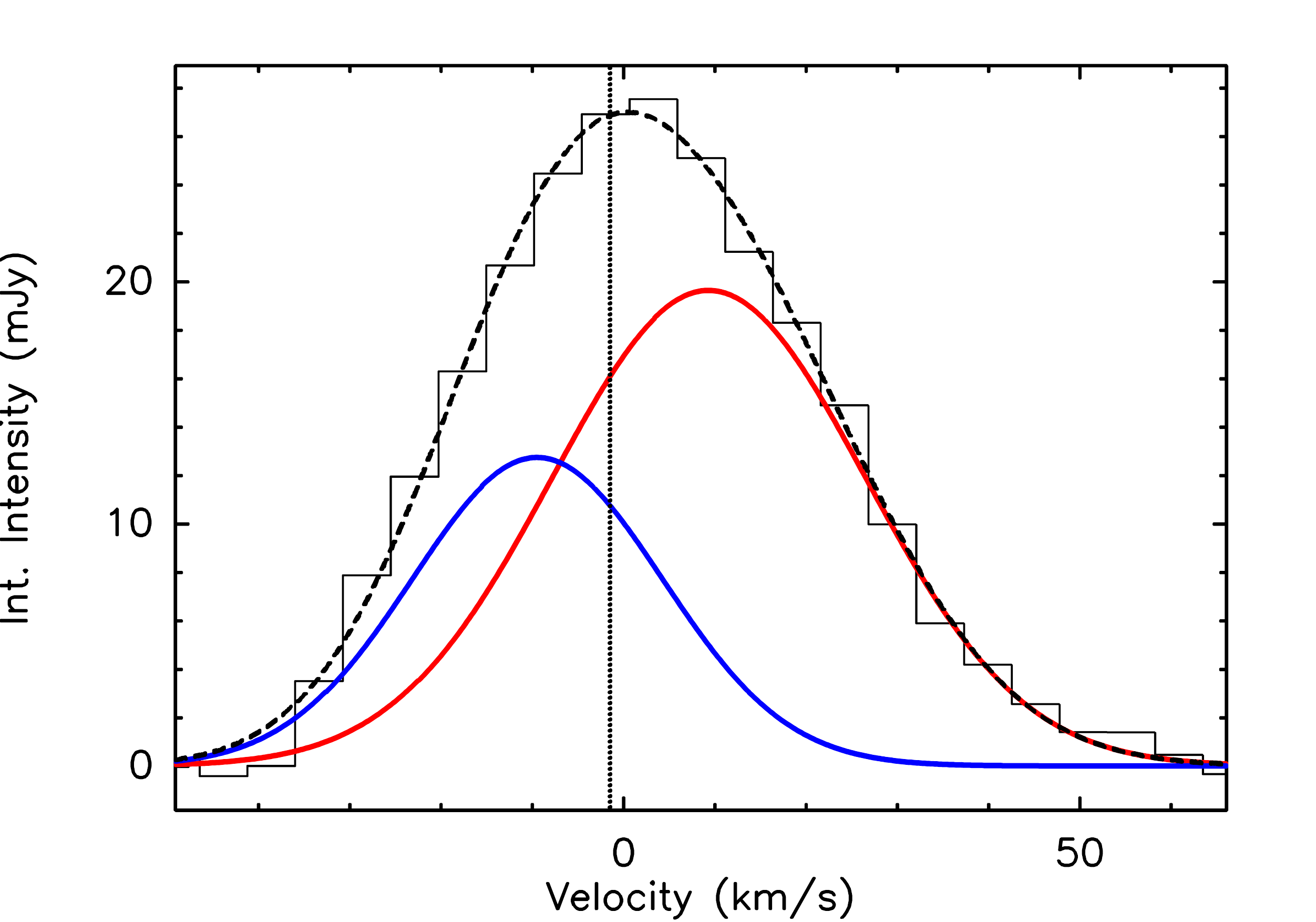}}
\end{figure*}

\addtocounter{figure}{-1}
\begin{figure*}
	\centering
	\caption{Continued}
	\subfloat[POS57]{\includegraphics[angle=270,width=7.5cm]{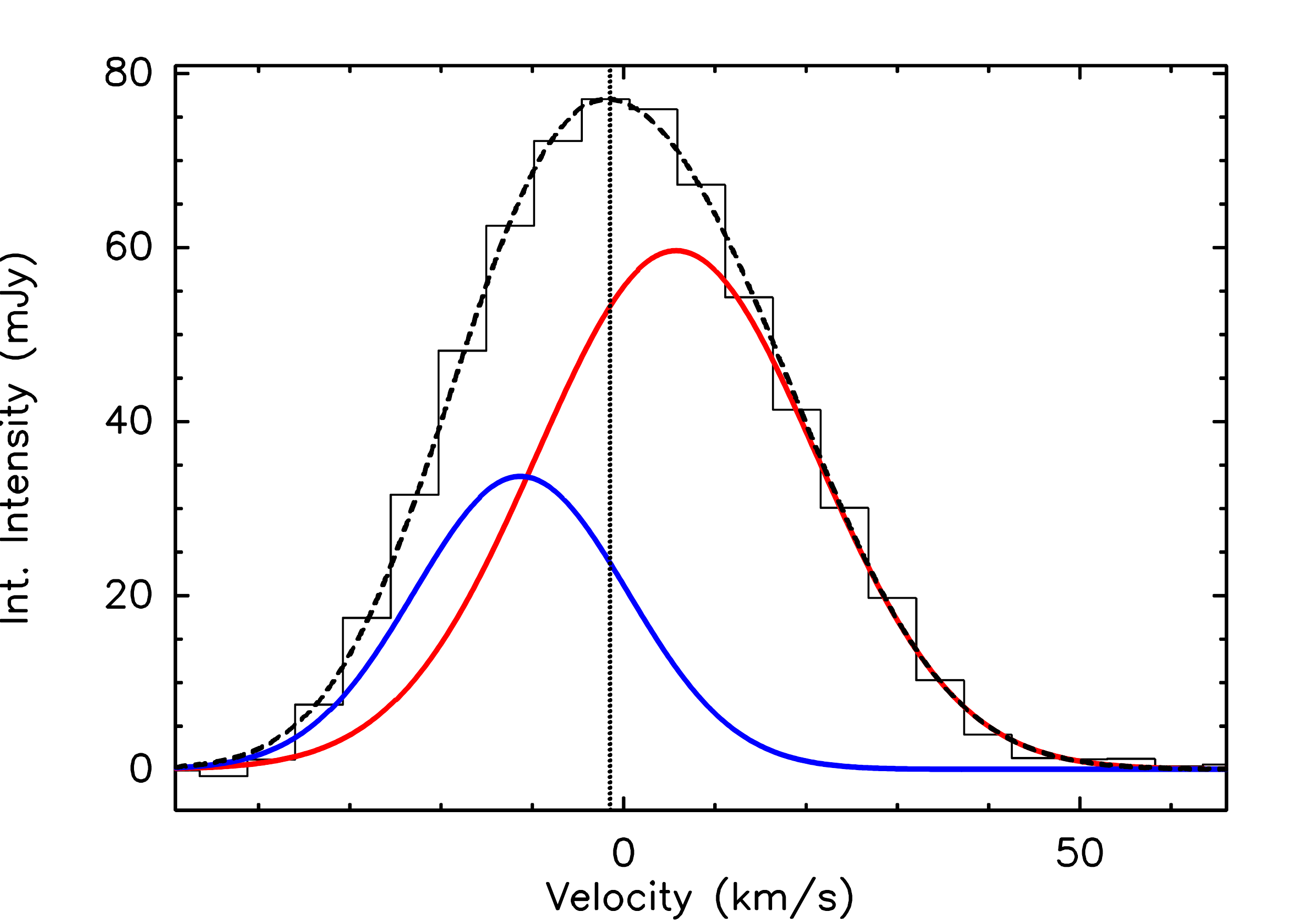}}	
	\subfloat[POS58]{\includegraphics[angle=270,width=7.5cm]{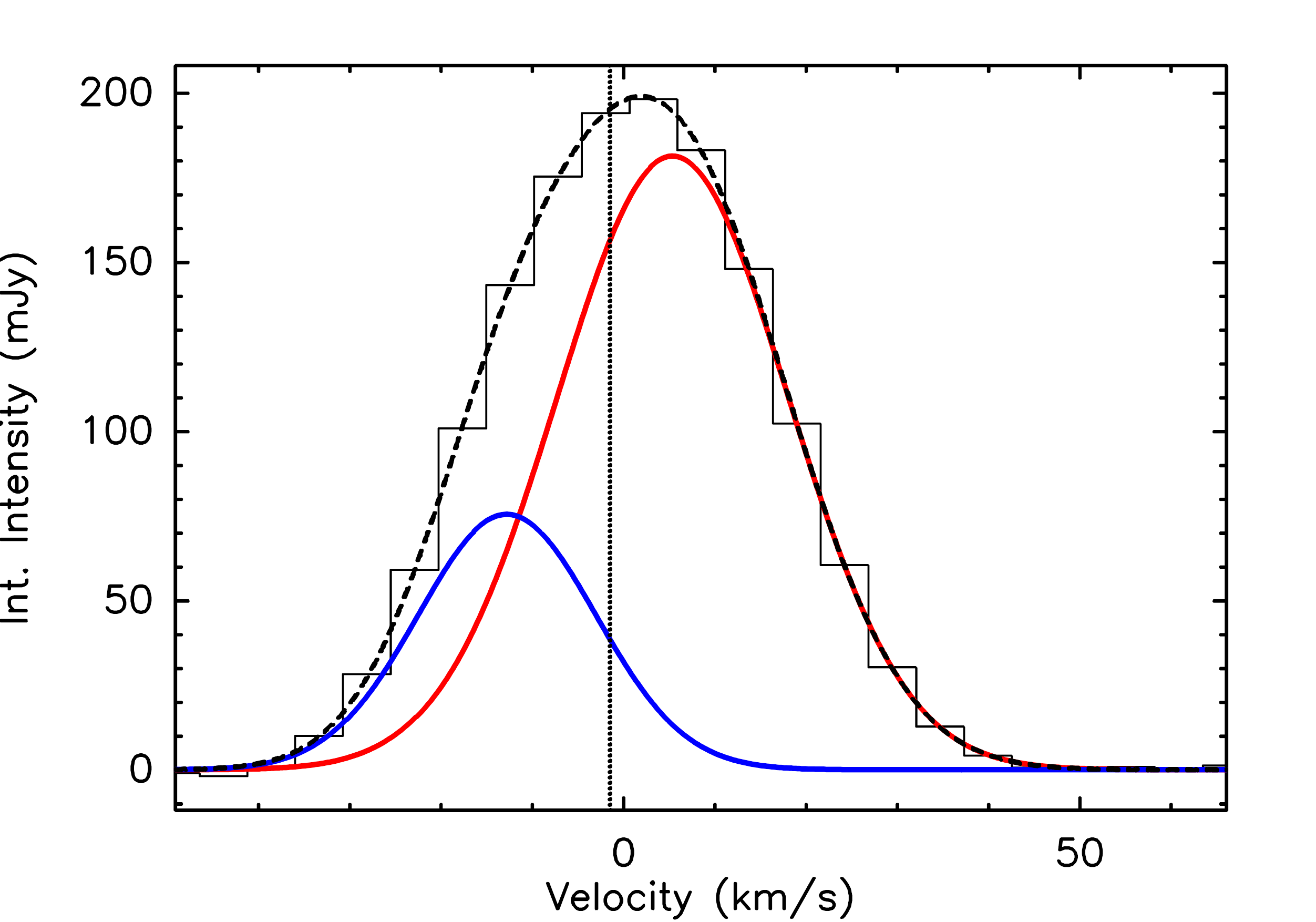}}\\
	\subfloat[POS59]{\includegraphics[angle=270,width=7.5cm]{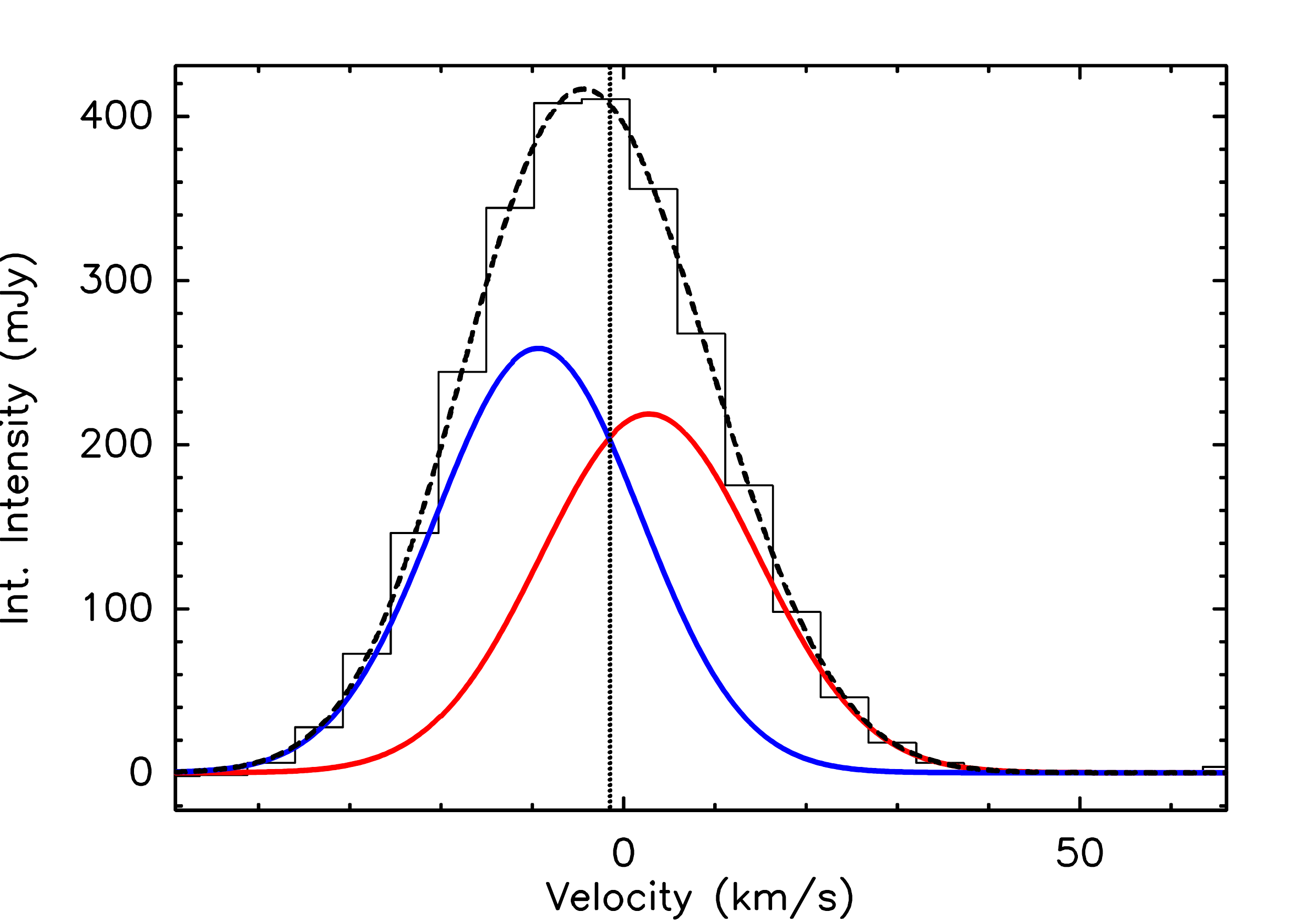}}
	\subfloat[POS60]{\includegraphics[angle=270,width=7.5cm]{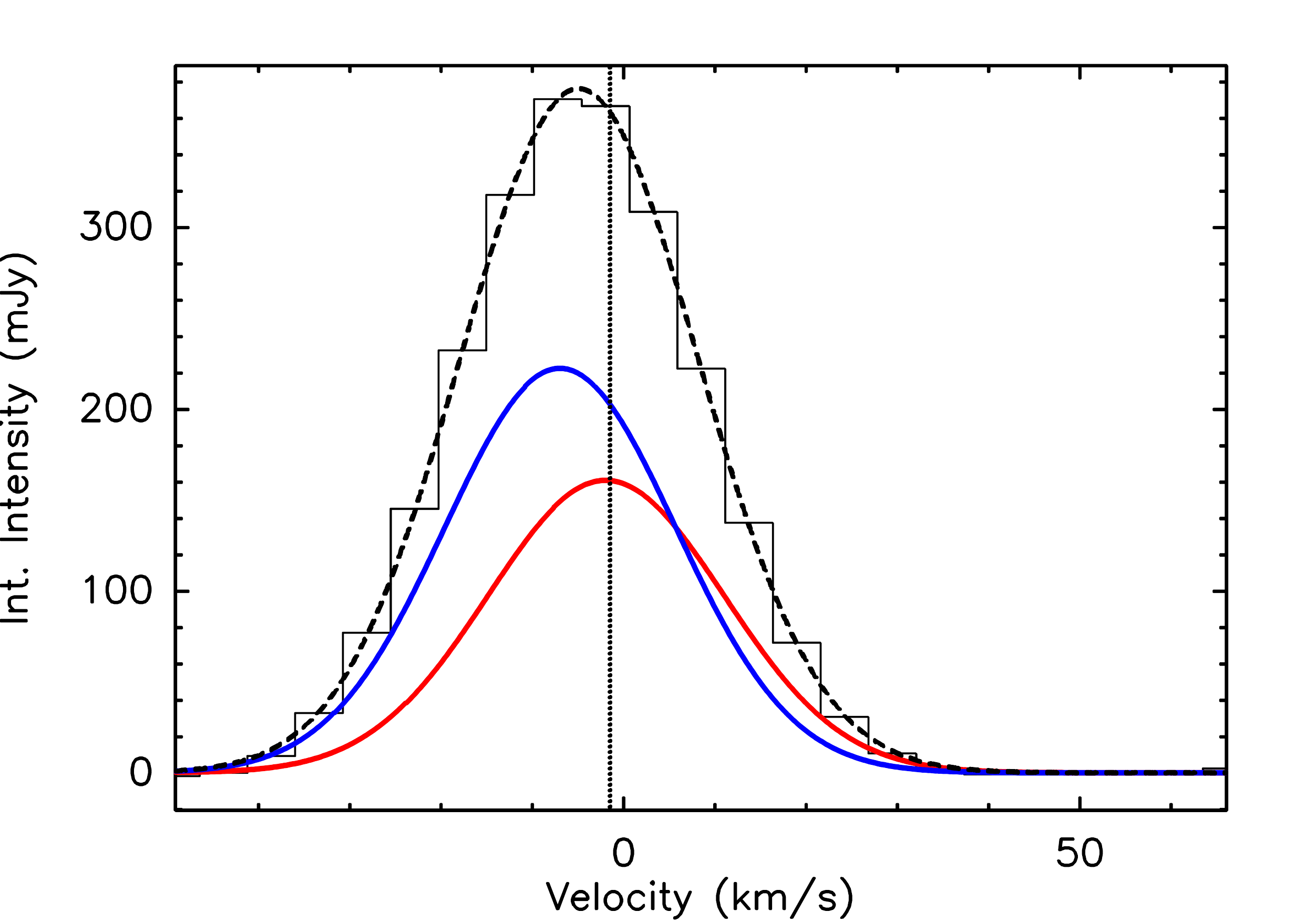}}\\	
	\subfloat[POS61]{\includegraphics[angle=270,width=7.5cm]{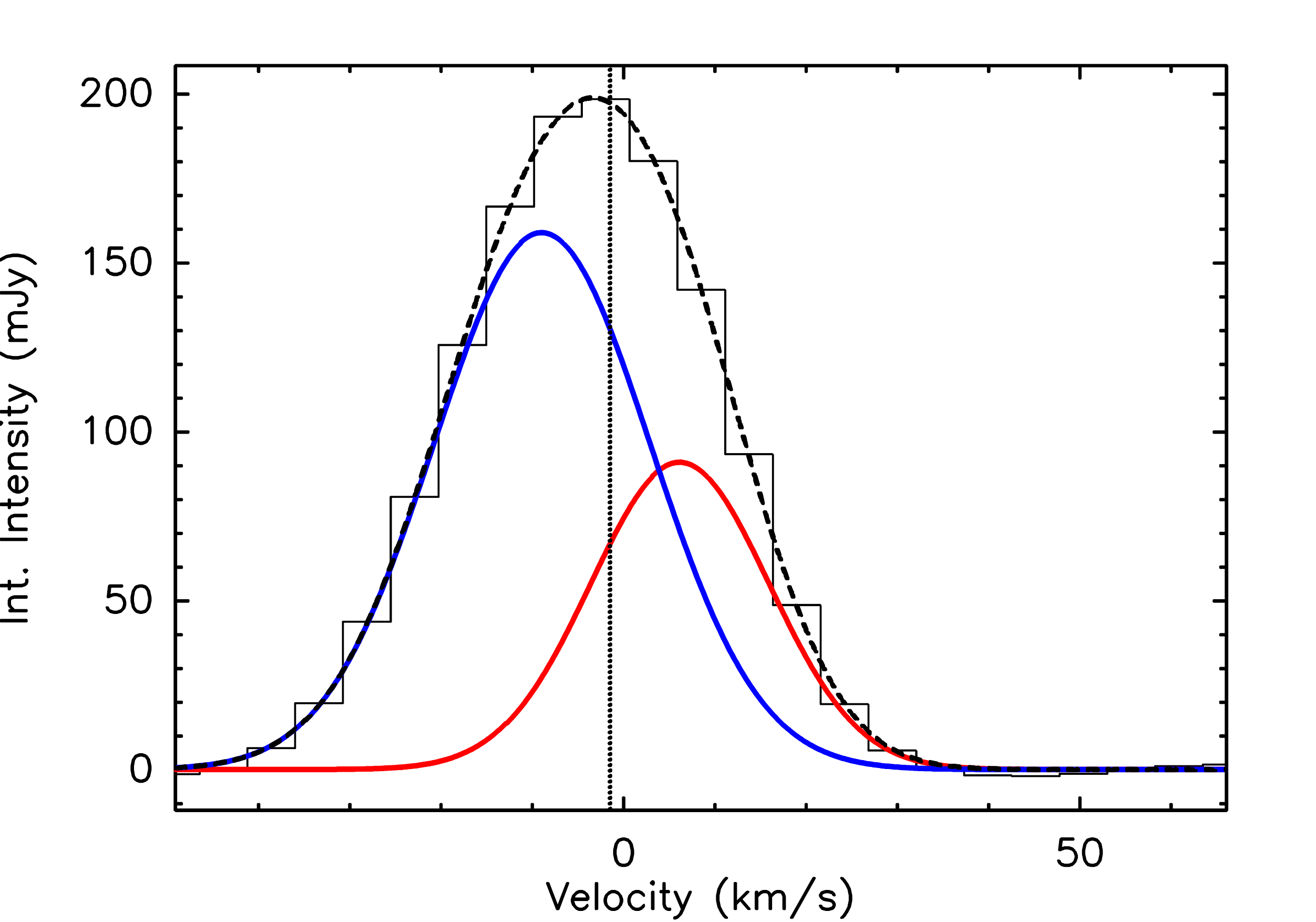}}
	\subfloat[POS62]{\includegraphics[angle=270,width=7.5cm]{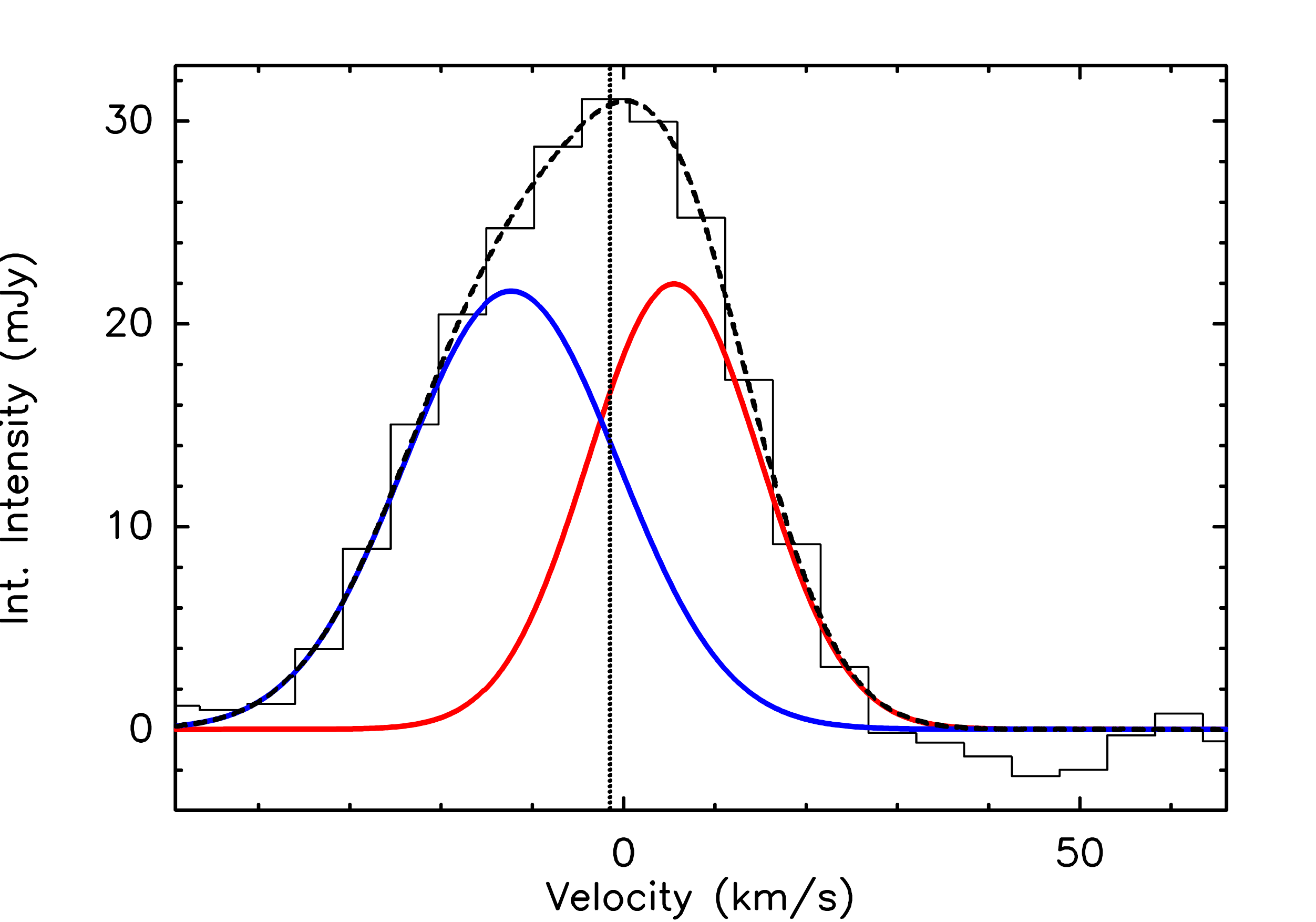}}\\	
	\subfloat[POS63]{\includegraphics[angle=270,width=7.5cm]{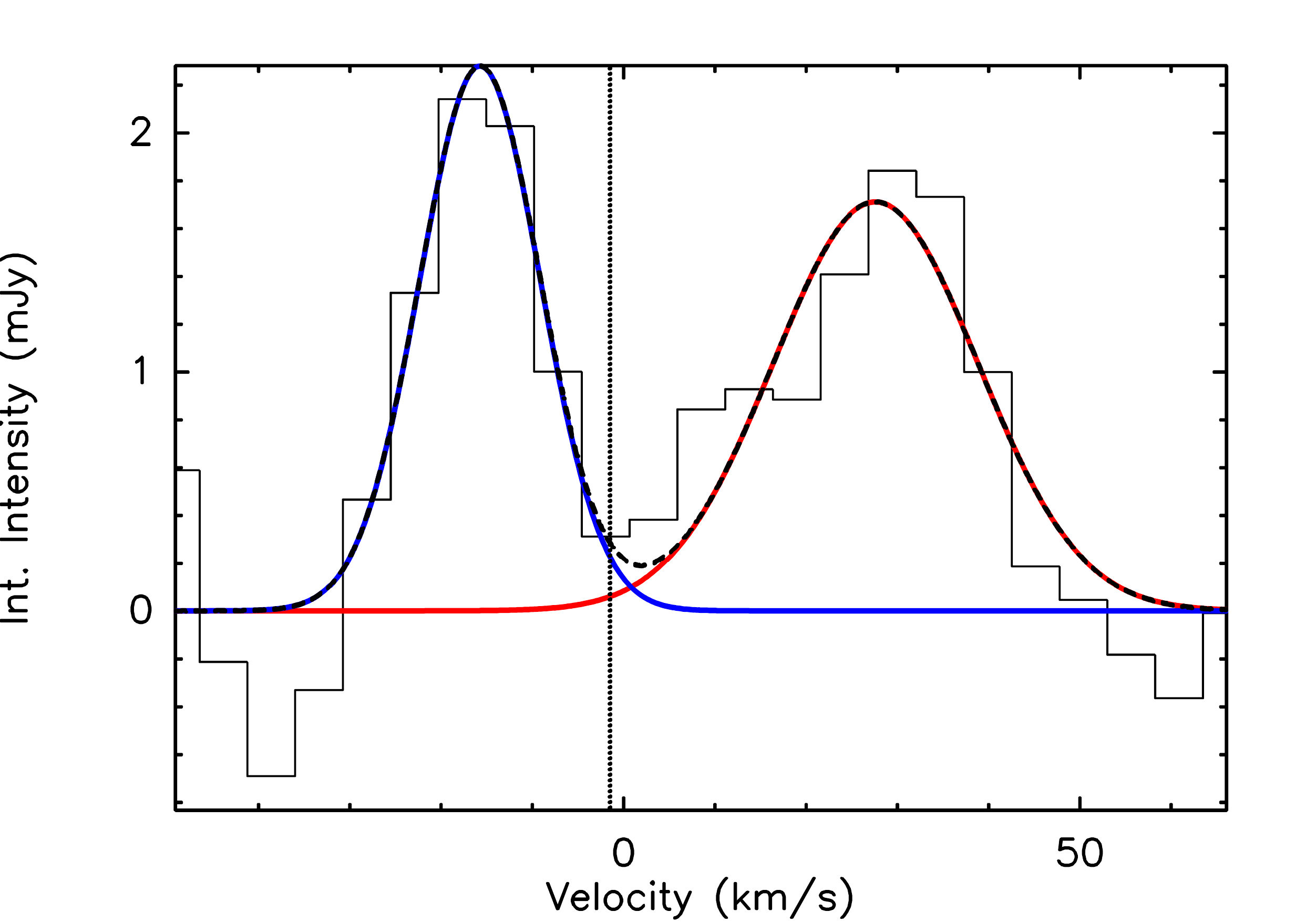}}
	\subfloat[POS64]{\includegraphics[angle=270,width=7.5cm]{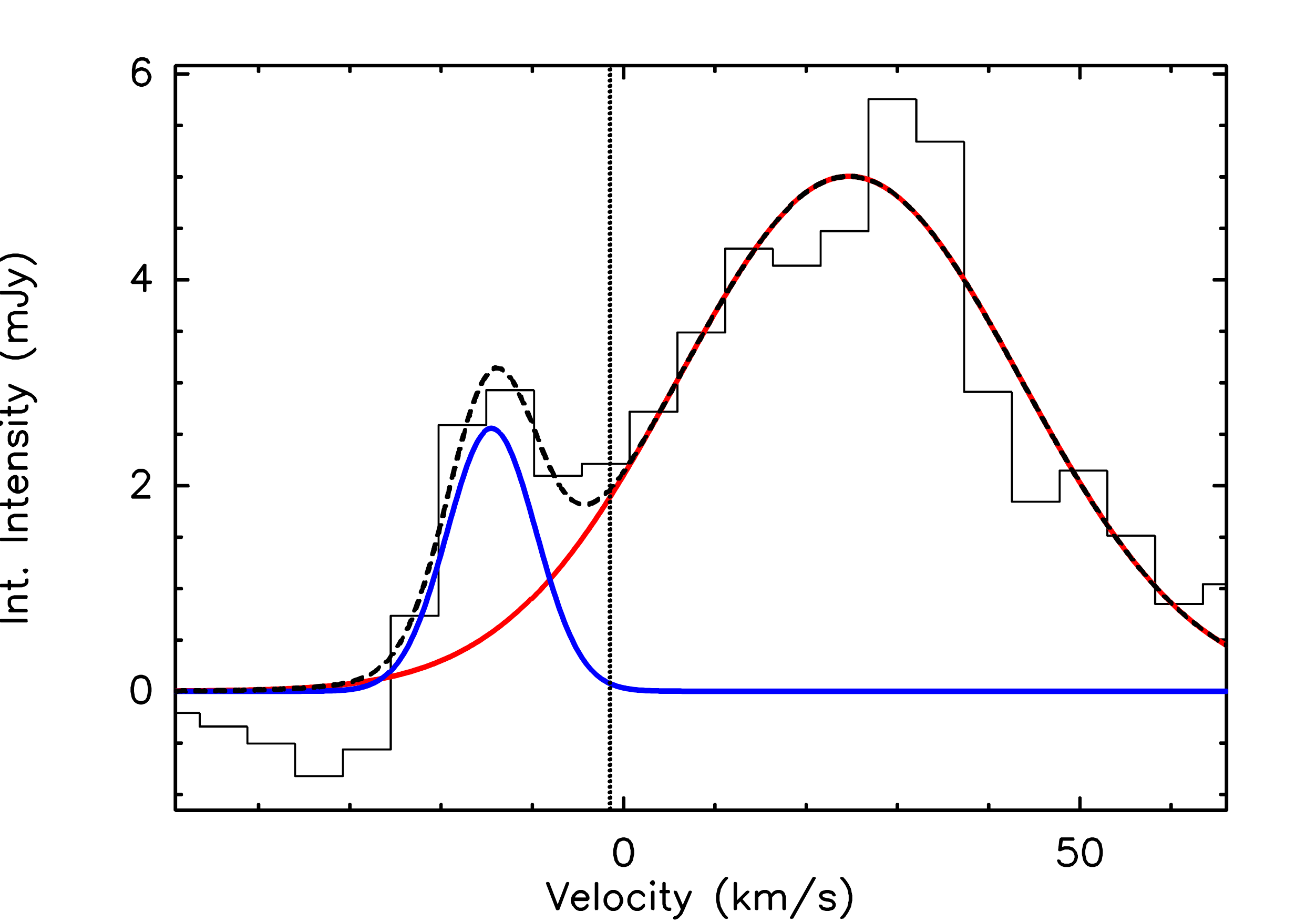}}
\end{figure*}

\addtocounter{figure}{-1}
\begin{figure*}
	\centering
	\caption{Continued}
	\subfloat[POS65]{\includegraphics[angle=270,width=7.5cm]{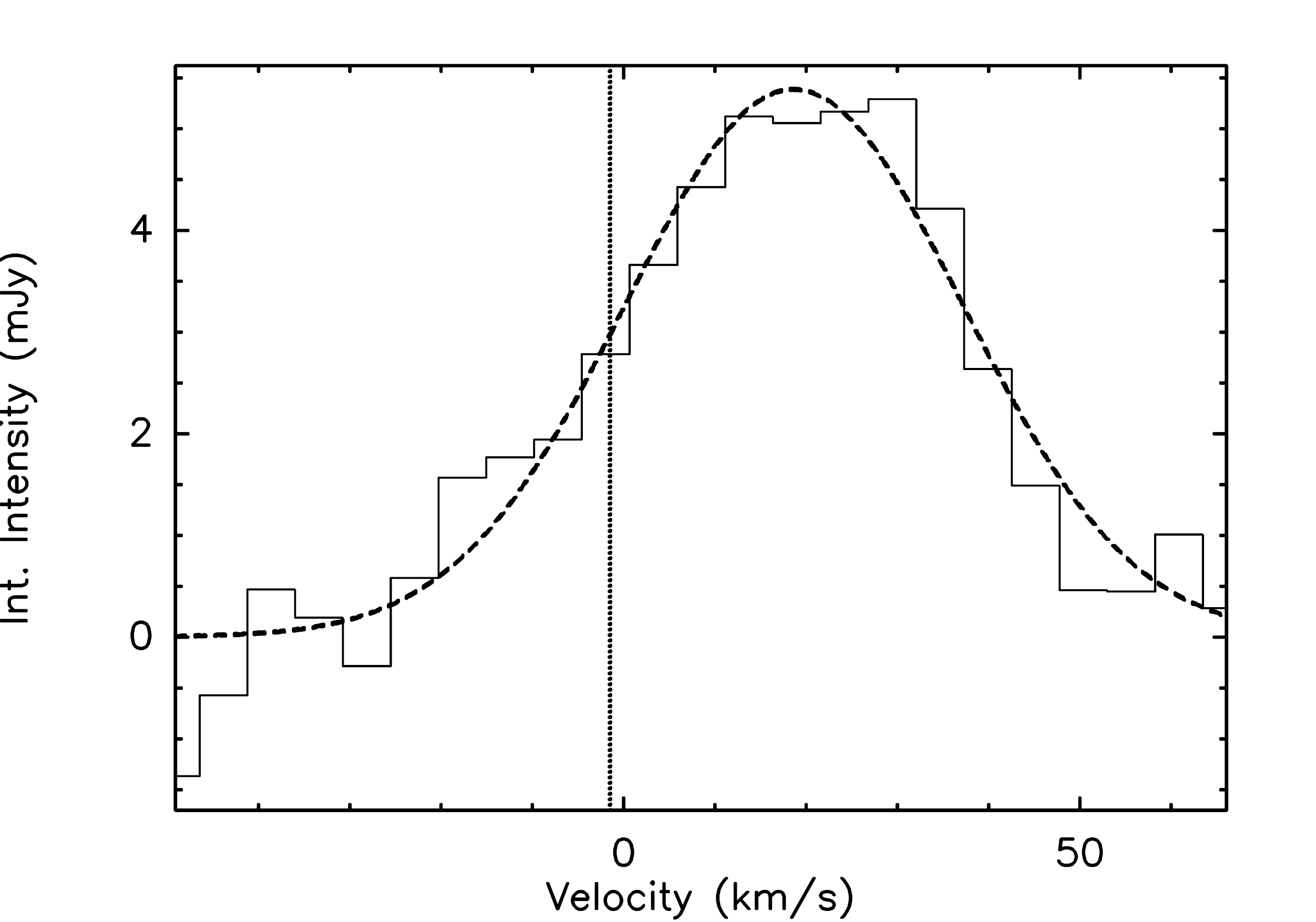}}	
	\subfloat[POS66]{\includegraphics[angle=270,width=7.5cm]{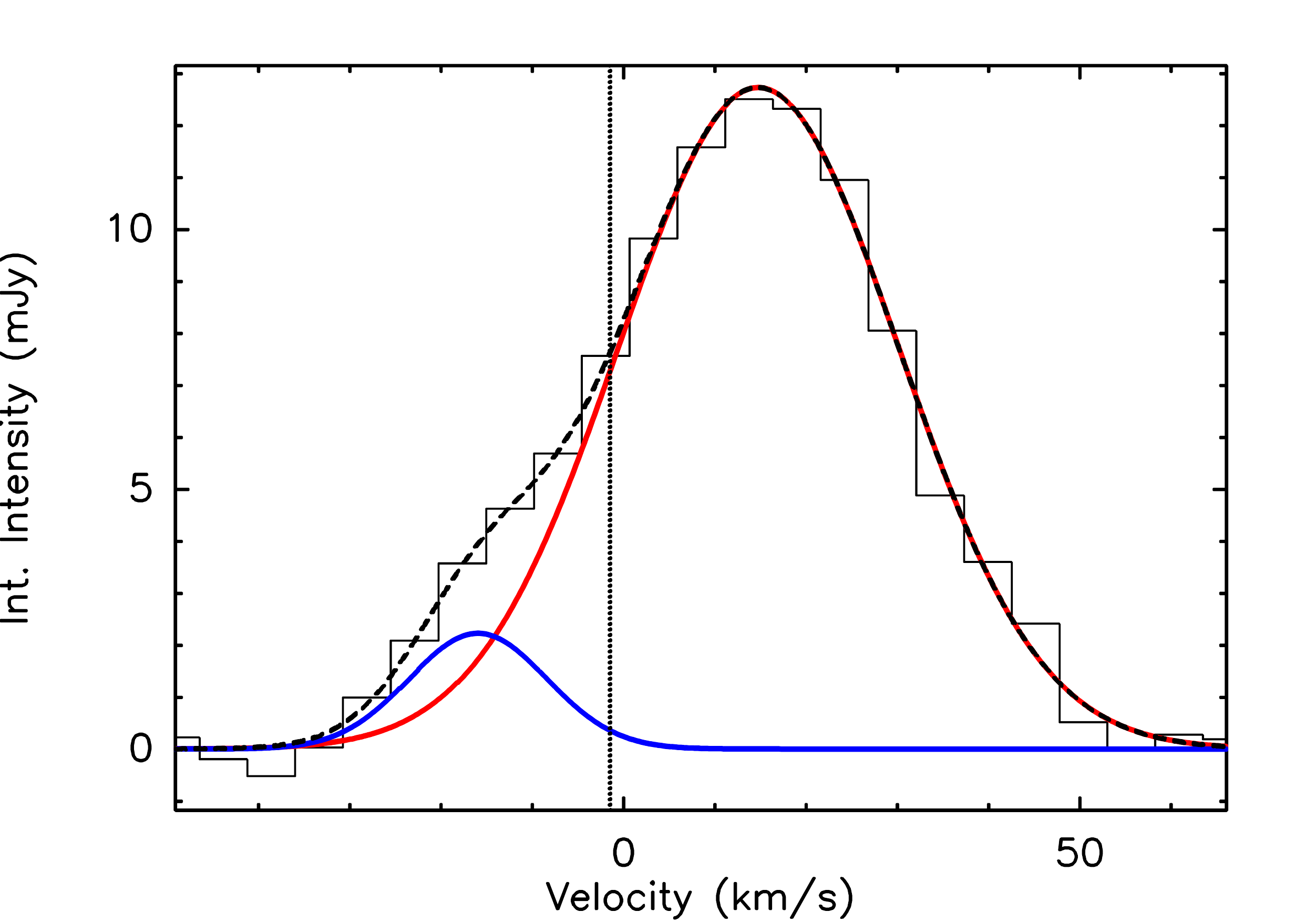}}\\
	\subfloat[POS67]{\includegraphics[angle=270,width=7.5cm]{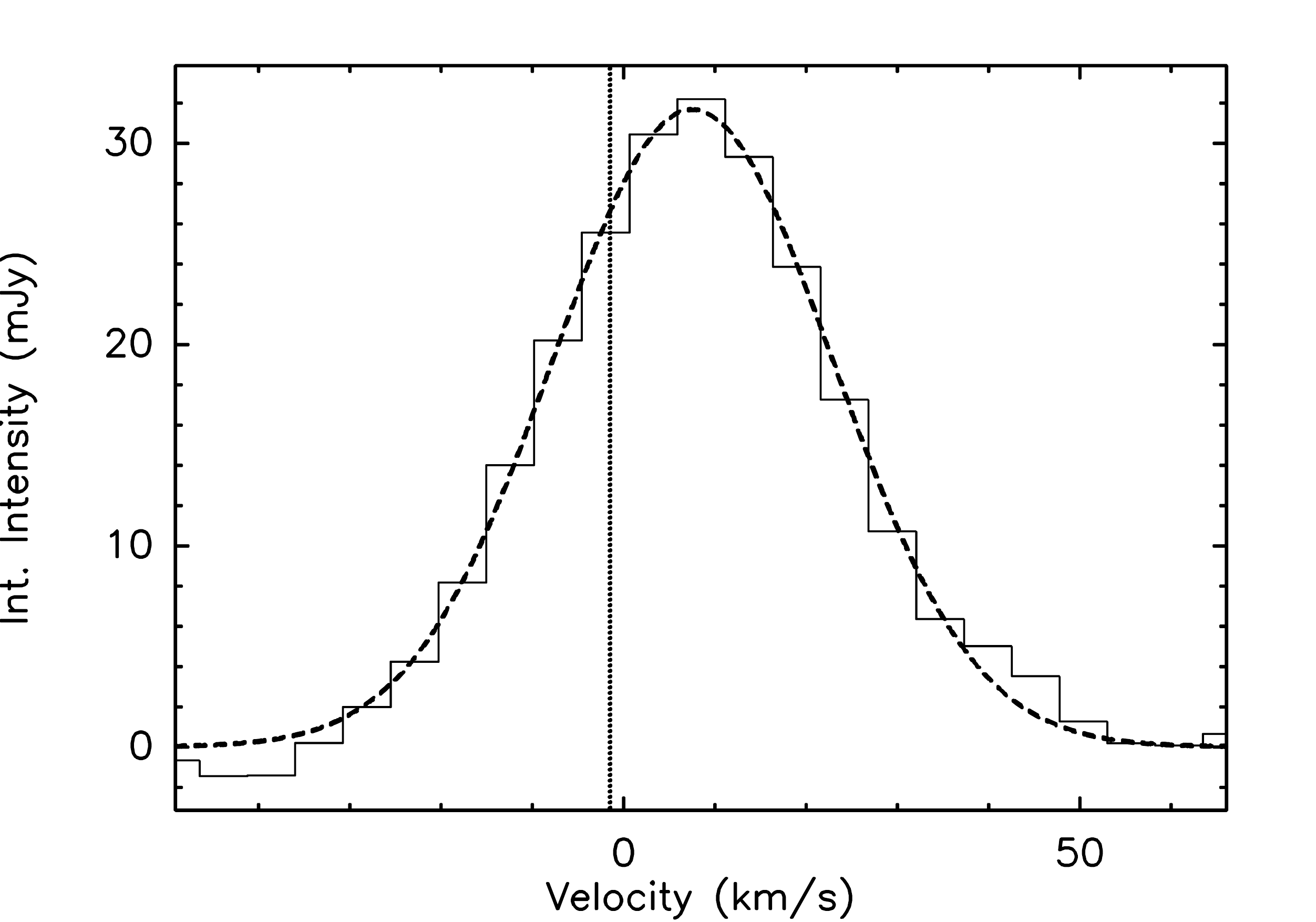}}
	\subfloat[POS68]{\includegraphics[angle=270,width=7.5cm]{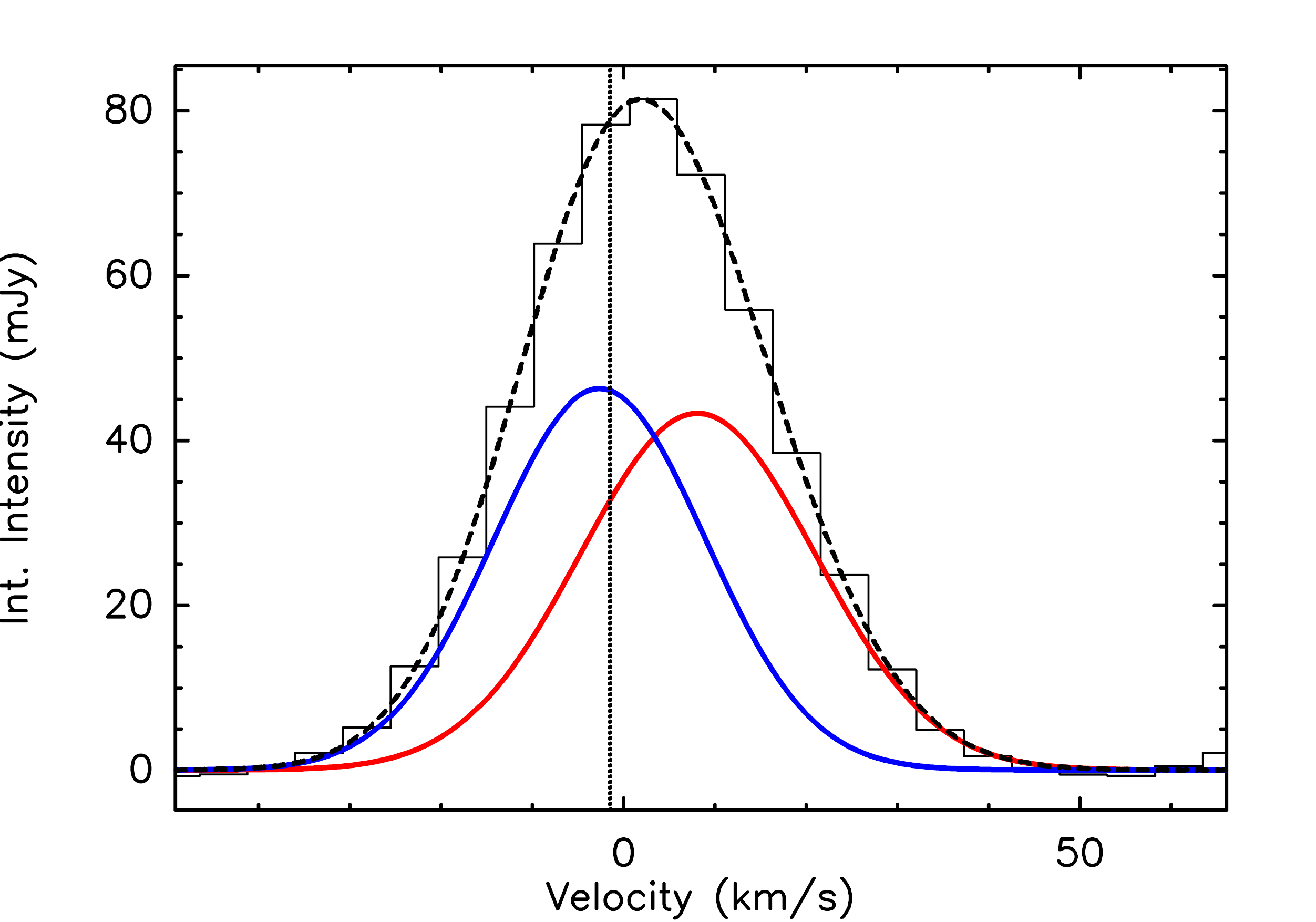}}\\	
	\subfloat[POS69]{\includegraphics[angle=270,width=7.5cm]{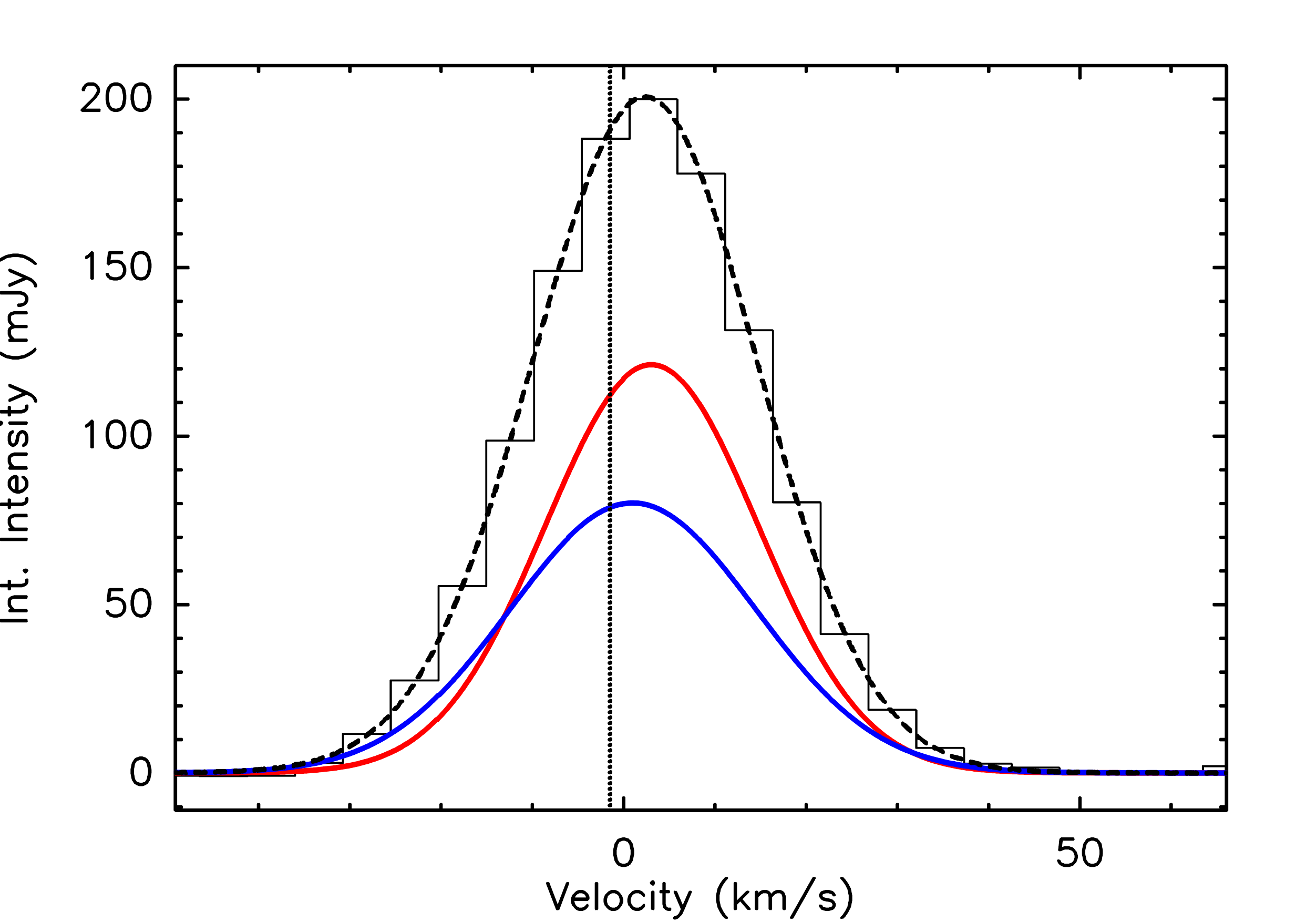}}
	\subfloat[POS70]{\includegraphics[angle=270,width=7.5cm]{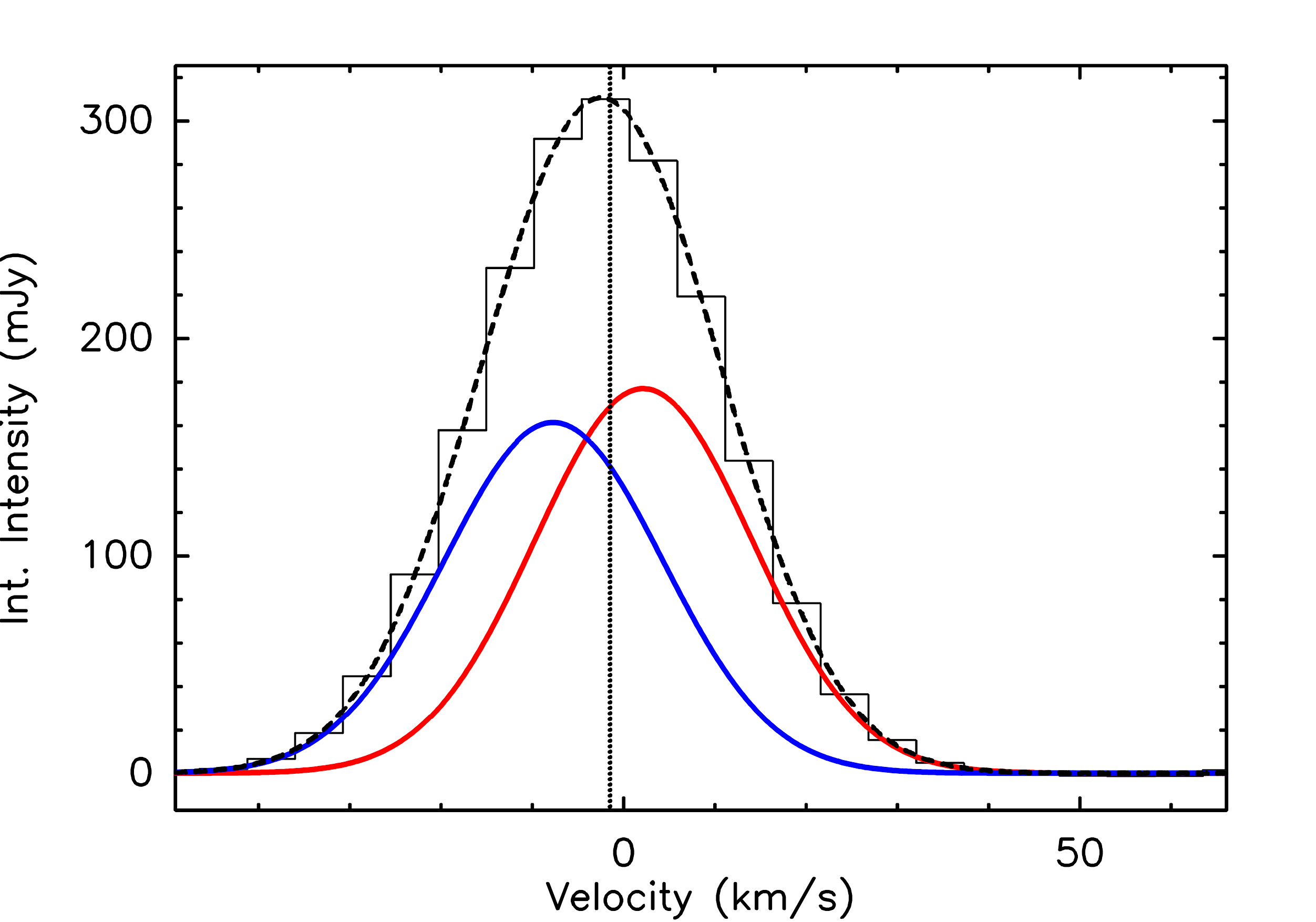}}\\	
	\subfloat[POS71]{\includegraphics[angle=270,width=7.5cm]{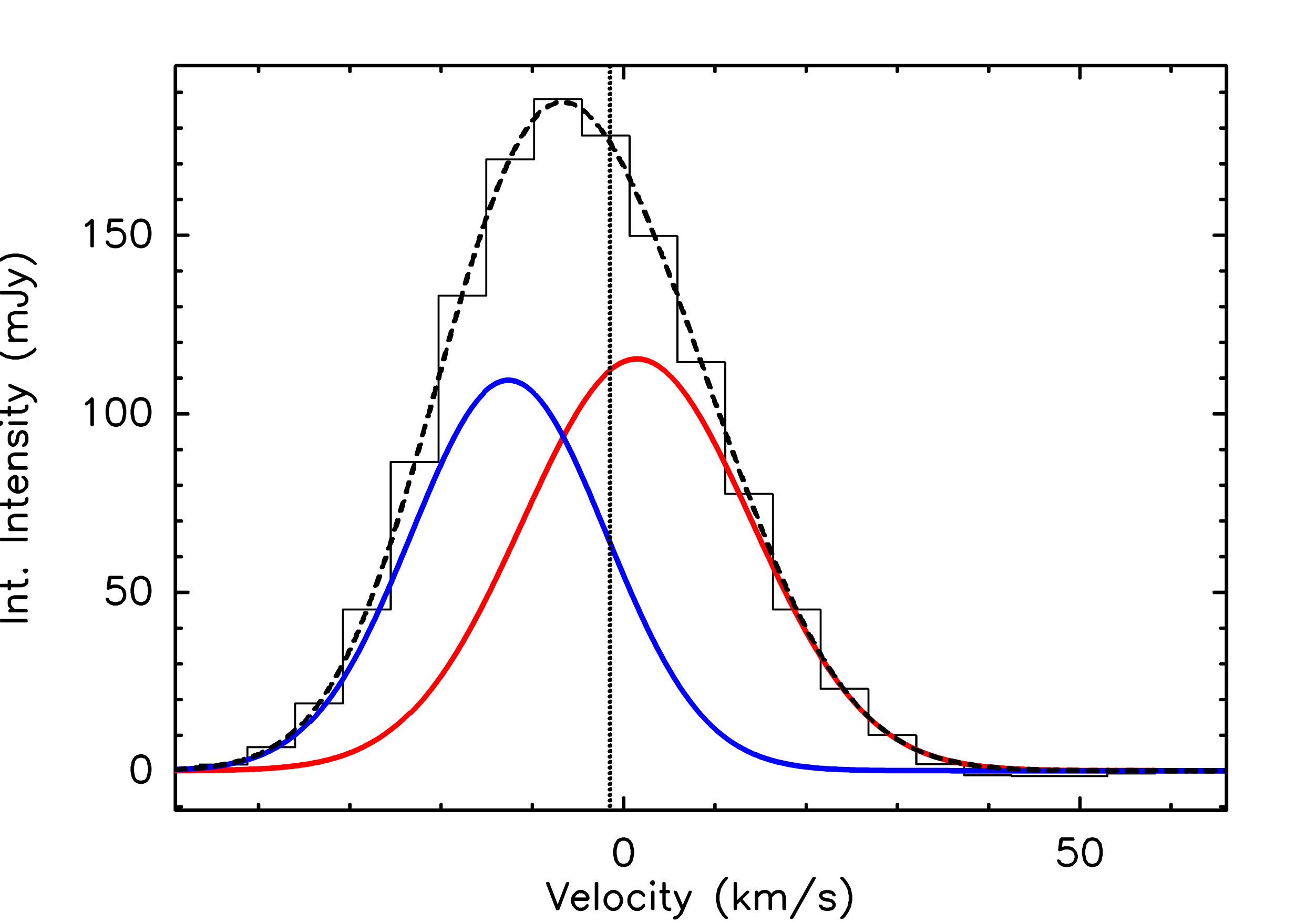}}
	\subfloat[POS72]{\includegraphics[angle=270,width=7.5cm]{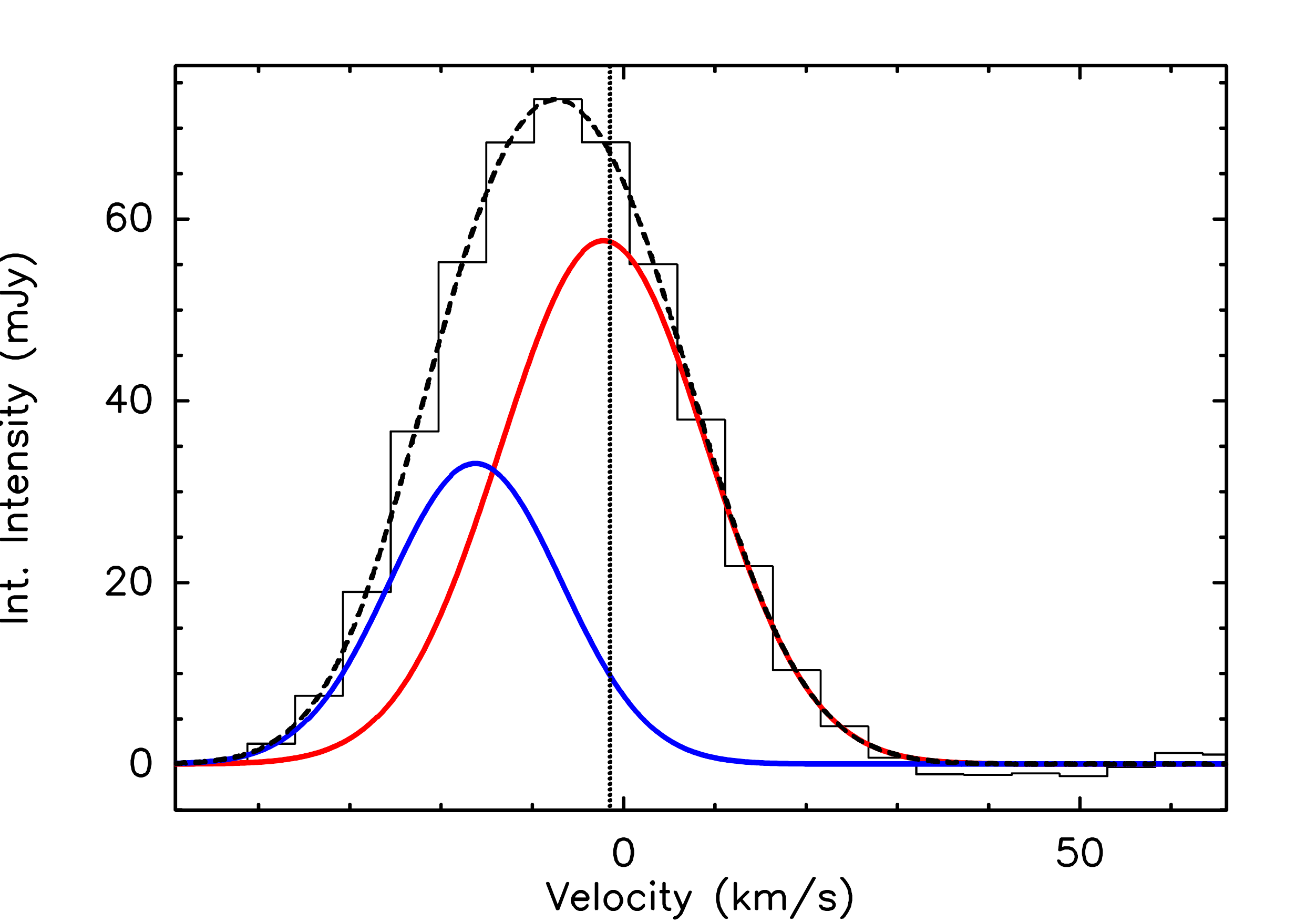}}
\end{figure*}			

\end{appendix}

\end{document}